\documentclass[a4paper]{article}
\pdfoutput=1
\usepackage{hyperref}

\usepackage{spconf,amsmath,epsfig}
\usepackage{booktabs} 
\newcommand{\Reffig}[1]{Figure~\ref{#1}}

\usepackage{multirow}
\usepackage{multicol}
\usepackage{enumitem} 
\setlist[itemize]{leftmargin=*} 
\usepackage[square,sort,comma,numbers]{natbib}
\usepackage{url}

\newif\iflong
\longtrue 
\newcommand{\sht}[2]{\iflong #1\else #2\fi}
\newcommand{\rem}[1]{\iflong #1\fi}

\rem{\usepackage{fancyhdr}
\usepackage{pdfpages}}

 \makeatletter
 \def\@name{ \emph{Mario Michael Krell$^{1,2,*}$, Julia Bernd$^{1,*}$, Yifan Li$^{2}$, Daniel Ma$^{2}$,
Jaeyoung Choi$^{1,3}$,}  \\ {\emph{\rem{Michael Ellsworth$^{1,2}$, }Damian Borth$^{4}$, Gerald Friedland$^{1,2,5}$}}}
 \makeatother

\begin{document}\sloppy

\sht{\title{Field Studies with Multimedia Big Data: \\
Opportunities and Challenges (Extended Version)}}{\title{Multimedia Big Data Field Studies:
Opportunities \& Challenges}}

\sht{
\address{$^{1}$International Computer Science Institute, Berkeley,
$^{2}$University of California, Berkeley,\\
$^{3}$Delft University of Technology, Delft,
$^{4}$German Research Center for Artificial Intelligence, Saarbr\"{u}cken,\\
$^{5}$Lawrence Livermore National Laboratory, Livermore.
*These authors contributed equally to this work.
}
}{
\name{Anonymous ICME submission}
\address{}
}

\rem{\onecolumn\includegraphics[width=\textwidth]{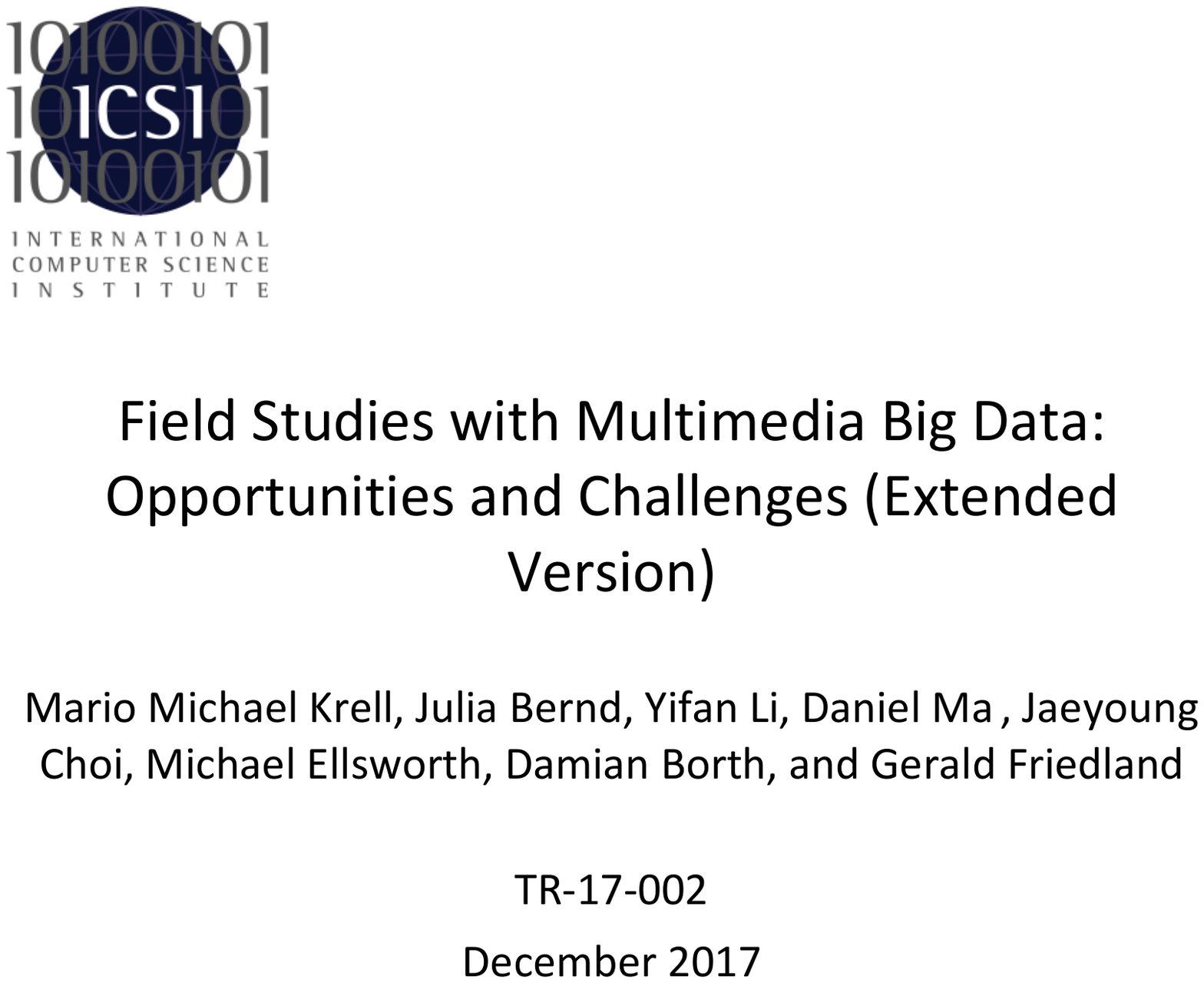}

\twocolumn
\setcounter{page}{1}
\pagestyle{fancyplain}
\cfoot[\textit{Field Studies with Multimedia Big Data // ICSI TR-17-002}]{\textit{Field Studies with Multimedia Big Data // ICSI TR-17-002}}
\rfoot[\thepage]{\thepage}
\lfoot[M. M. Krell, J. Bernd, et al.]{M. M. Krell, J. Bernd, et al.}
\lhead{}
\rhead{}
\renewcommand{\headrulewidth}{0pt}
}

\maketitle

\begin{abstract}
Social multimedia users \sht{are increasingly sharing}{share} 
all kinds of data about the world. 
\sht{They do this for their own reasons, 
not to provide data for field studies---but 
the trend presents a great opportunity for scientists.}{This presents a great opportunity for scientists.}
\rem{The Yahoo Flickr Creative Commons $100$ Million (YFCC100M) dataset 
comprises $99$ million images and nearly $800$ thousand videos from \sht{Flickr,
all shared under Creative Commons licenses.}{Flickr.}}
\sht{}{But there is a gap between what 
multimedia researchers have shown to be
possible with such data and the
follow-through of creating a \textit{comprehensive} 
framework that supports scientists\rem{ across disciplines}.} 
\sht{To enable \sht{scientists}{researchers} to leverage \sht{these media records}{such data}\rem{ for field studies},
we}{We} propose a new framework that extracts 
targeted subcorpora \sht{from the YFCC100M,}{from the Yahoo Flickr Creative Commons 
$100$ Million dataset (YFCC100M),} in a format usable by
researchers who are not experts in \sht{big data retrieval and processing.

This}{big data processing. The first phase of our framework interface is available as
Multimedia Commons Search. This}
paper discusses \rem{a number of }examples\sht{ from 
the literature---as well as some entirely new ideas---}{ }of natural and social science
field studies that could be piloted, supplemented, replicated, or conducted using
YFCC100M \sht{data. These}{photos and videos; these}
examples illustrate the need for
\sht{a}{the}\sht{ general}{} 
new open-source framework for Multimedia Big Data Field Studies.
\rem{There is currently a gap between the separate aspects of what 
multimedia researchers have shown to be
possible with consumer-produced big data and the
follow-through of creating a \textit{comprehensive} 
\rem{field study} framework that supports scientists across \rem{other} disciplines.

To 
bridge this gap, we must meet several challenges. For example, the framework
must handle unlabeled and noisily labeled data to produce a filtered dataset
for a scientist---who naturally wants it to be both as large and as clean as
possible.
This requires an iterative approach that provides access to
statistical summaries and refines the search by constructing new classifiers.}
\rem{The first phase of our framework is available as
Multimedia Commons  \sht{Search (\url{http://search.mmcommons.org}, MMCS),
an intuitive interface that enables complex 
search queries at a large scale.}{Search.}} After 
outlining \rem{our proposal for }the \rem{general }framework 
\sht{and discussing the potential example studies,}{and its challenges,} 
this paper describes \sht{and evaluates a}{its} practical application
to \sht{the study of}{studying} origami.\sht{}{\footnote{A longer version of this paper, with additional details 
about the framework and case study and extensive discussion of specific possible studies, may be found at... 
\textit{[arXiv URL will be provided in final version].}}}
\end{abstract}

\section{Introduction}

The basis of science is quite often data.
Consequently, data science \sht{and machine learning are very hot topics,}{is a very hot topic,}
with new applications and insights coming out every hour.
Unfortunately, it can take a lot of time to \rem{record or }gather that data, 
and to \sht{add the necessary annotations and metadata for machine learning.}{annotate it for machine learning.}

For scientific field studies generally, a large proportion
of researchers' time is often spent in administrative tasks associated with
data-gathering. For example, they must seek funding, get approval from
institutional ethics \sht{committees or other relevant
committees,}{committees,} and, if children are involved, obtain parental
consent. Additionally, the data recording itself might take a lot of \sht{time,
if many samples are needed to confirm or invalidate a hypothesis.}{time.} In
particular, automatic analysis of data using machine learning requires quite
large datasets. Last but not least, the sampling variety a scientist can
achieve with a given dataset, in terms of geographical and cultural diversity,
is generally constrained by the time and money required for travel.

The Yahoo Flickr Creative Commons $100$ Million (YFCC100M) 
is the largest publicly available multimedia dataset \cite{Thomee2016}.
This user-generated content (UGC) corpus comprises $99.2$ million images and $800,000$ videos 
from Flickr\sht{, shared under 
Creative Commons \sht{copyright} licenses.}{.}
New extensions and subsets are frequently added, as part of the Multimedia Commons initiative\rem{ \cite{Bernd2015}}.
An important advantage of the YFCC100M for scientific studies is that access is open and simple.
\sht{Because}{Because all} the media are Creative Commons, 
they can be used for almost any type of research---and 
the standardized licenses make any restrictions clear\sht{ (for example, on commercial applications).
In \Reffig{f:examples} some example images and licenses are shown.}{.}

\rem{
\begin{figure*}
  \centering
  \includegraphics[height=.31\textwidth]{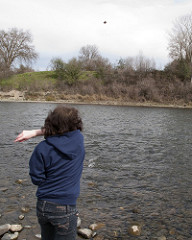} 
  \includegraphics[height=.31\textwidth]{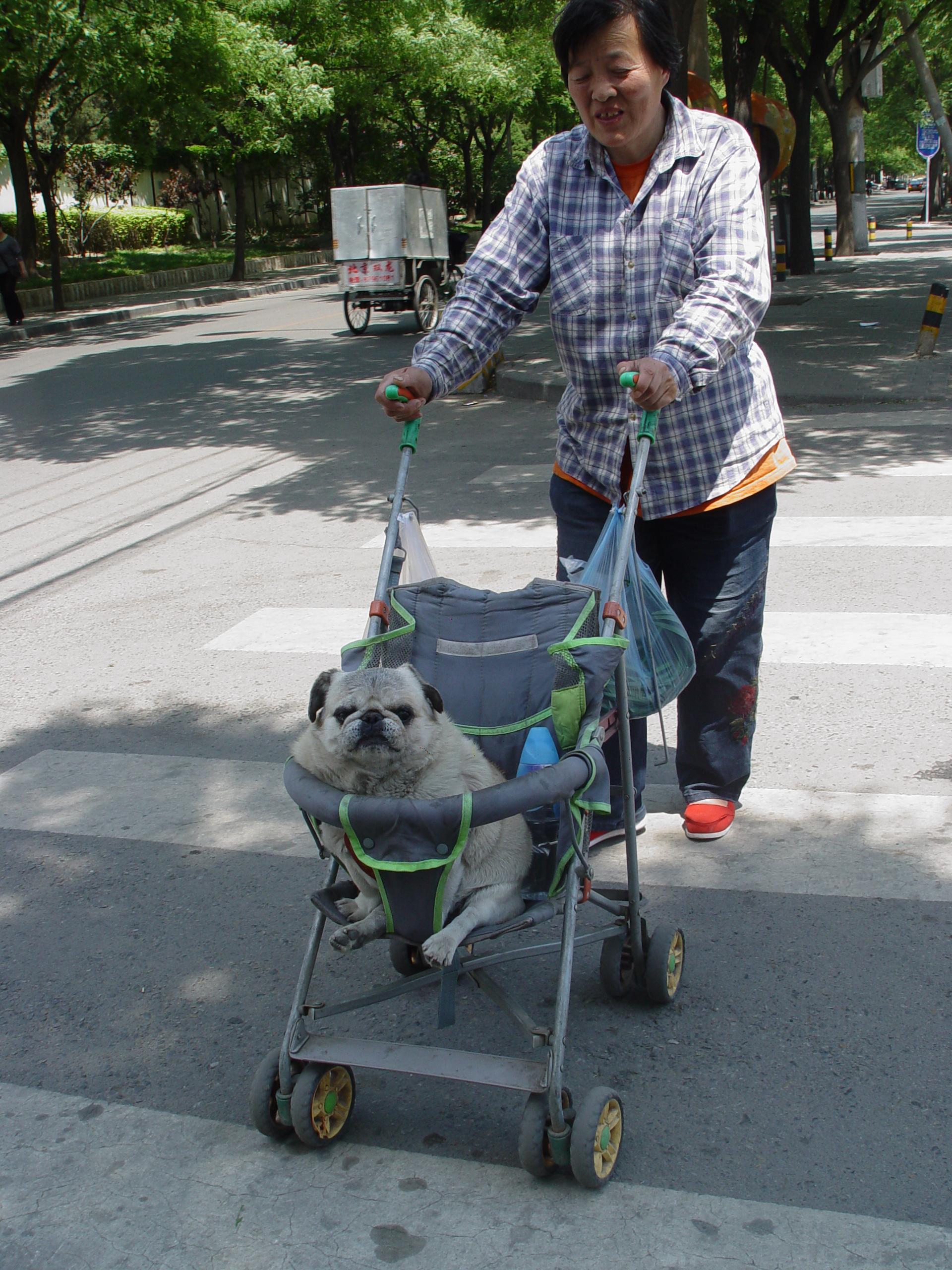}
  \includegraphics[height=.31\textwidth]{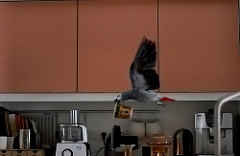}
  \caption{YFCC100M images extracted by using the pre-existing data browser
  ({http://YFCC100M.appspot.com/about})
  \cite{Kalkowski2015} to search for `throw', `dog baby', and `parrot'
  (from left to right: [https://www.flickr.com/photos/29233640@N07/3318234428, license: {CC BY 2.0}],
  [https://www.flickr.com/photos/7300793@N06/5725224597, license: CC BY-NC 2.0],
  [https://www.flickr.com/photos/69221340@N04/6912680630, license: CC BY 2.0]).}
  \label{f:examples}
\end{figure*}
}

We are proposing a framework to extend the existing YFCC100M ecosystem to enable
scientists with no expertise in big data retrieval and processing
to filter out data that is relevant for their research question,
and to process that data to begin answering it.
Depending on the topic, the framework might be used to collect and analyze the 
central dataset for a new study, or it might be used to pilot or prepare for on-the-ground data collection. 
In addition, the framework can provide a
way to replicate or extend existing studies, where the original 
recordings are not publicly available\sht{ due to administrative restrictions or ethics rules.}{.}

The contributions of this paper are in introducing the  
multimedia big data studies (MMBDS) concept and our search engine for the YFCC100M, 
discussing the benefits and limitations of MMBDS, 
\rem{providing many elaborated examples of potential MMBDS from different disciplines,}
and
analyzing some of the requirements for implementing those studies
in a comprehensive general framework\sht{ (including existing tools that could be leveraged).}{.}
For the requirements analysis, we apply our concept
to the concrete \rem{and entirely novel} example of origami \sht{studies.

Overall,}{studies. Overall,}
we hope the new possibilities offered by this framework
will inspire scientists to come up with new research ideas
that leverage the Multimedia Commons,
and to contribute new tools and data resources to \sht{the framework.}{it.}
We thus can bridge the gap 
between the \textit{potential} applications of multimedia research 
and its \textit{actual} application in the sciences and humanities.

In Section~\ref{s:data}, 
we describe the YFCC100M \rem{dataset} ecosystem\sht{ in more detail, with reference to}{ and} its potential for MMBDS. 
In Section~\ref{s:app}, we describe the basic structure of the framework.
Next, we give examples of studies that could be extended, 
new research that could be done in future,
and new applications that could be pursued \rem{using 
the YFCC100M or similar datasets in an MMBDS framework} 
(Sections~\ref{sec:sciexx} and \ref{sec:applexx}).
We describe a practical case study on origami in Section \ref{s:origami}.
Section~\ref{s:con} provides an outlook for the future.

\section{The YFCC100M Dataset}
\label{s:data}

\rem{This section describes the YFCC100M ecosystem, including the dataset, associated
extensions, and tools, focusing on the
advantages and limitations for MMBDS.}

While we eventually intend to incorporate additional data sources\sht{ into the MMBDS framework,}{,}
we began with the YFCC100M because of its size, its availability, and other 
suitable characteristics.

\rem{
\subsection{Purpose-Built vs.\ General Datasets}
} 

If a researcher can get all the data required for a study from an existing
purpose-built dataset, that is usually the best choice. But quite often,
existing datasets might not be diverse enough in terms of location, language,
etc.\ to suit their needs, or they might not have the right data type (text,
image, audio, \sht{video).

Scraping}{video). Scraping} data from the web presents a different set of
difficulties. Existing search engines \rem{(whether general, like Google, or
service-specific, like Flickr or YouTube) }are not built for data-gathering, and
are only partially able to handle unlabeled data. Constant updates to the \rem{search }engines and the content mean that the search that led to the dataset will not
necessarily be \emph{reproducible}. 
Providing the actual data, so other researchers can
\emph{replicate} the study, may not be possible due to restrictive or unclear
licensing, and maintaining the data in a public repository requires long-term
\sht{resources.

In}{resources. In} contrast, the YFCC100M can be accessed at any time, the data \rem{in it }remains the same, and
a subset can easily be shared as a list of IDs.

\subsection{YFCC100M Characteristics}
\label{s:yfccchar}

The YFCC100M is comprised of images and videos
\rem{that were} uploaded to Flickr under Creative Commons licenses
between $2004$ and $2014$.
This comparatively long time frame---and the enormous amount of data---make 
it particularly suited to scientific studies.
\sht{}{The average video length is
$39$ seconds. The distribution is displayed at \Reffig{f:hist},
using $80\%$ of the data.
There is one peak at $14$ seconds, indicating that there is a large
group of rather very short videos and, and another peak at $90$ seconds,
due to limitations from Flickr.
There are only very few videos after this peak.
}

\rem{
\begin{figure}
  \centering
  \includegraphics[width=0.49\textwidth]{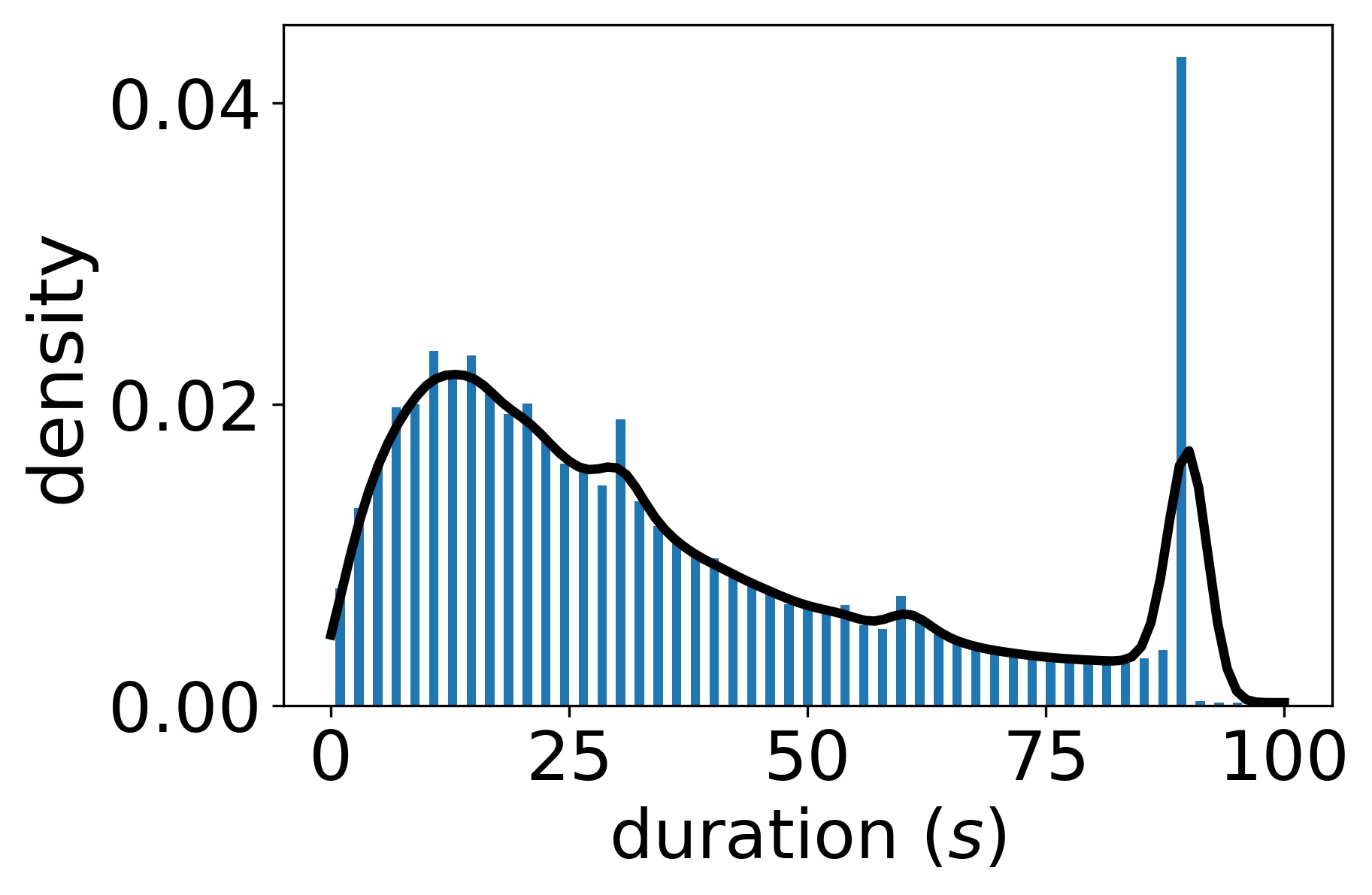}
  \caption{Duration distribution of YFCC100M videos.}
  \label{f:hist}
\end{figure}
}

Classification results show that the 
YFCC100M is very diverse in terms of subject matter \cite{Thomee2016}.
\sht{
And again, even if a researcher cannot find a sufficiently large set 
of high-quality images or videos for a full study on their topic, the YFCC100M 
can be very helpful for exploratory or preliminary studies. 
\sht{Such pre-studies can help a researcher
decide which factors to vary or explore in an in-depth, controlled study. 
They can also be quite useful in illuminating potential difficulties 
or confounds. This can help researchers avoid mistakes and delays during
targeted data acquisition.}{Such pre-studies can help a researcher decide which factors to explore in a controlled study, and can illuminate potential difficulties or confounds.}
}{
So even if there is not enough data for a full study, the YFCC100M can often support a preliminary study.
Such pre-studies can help a researcher decide which factors to explore in a controlled study, and can illuminate potential difficulties or confounds.
}

\rem{The average length of videos in the YFCC100M is
$39$ seconds. Some are, of course, much longer---but for studies 
that require longer continuous recordings, there might not be enough long examples.
However, Sections \ref{sec:sciexx} and \ref{sec:applexx} provide
a number of examples where short videos would suffice, 
such as analyzing human body movements.}

In addition to the raw data, the YFCC100M dataset includes 
metadata such as 
user-supplied tags and descriptions, locations (GPS coordinates and 
corresponding place names), \rem{recording} timestamps, and
camera types, for some or all of the media.
\rem{This metadata can be also used in MMBDS.}
In particular, location (available for about half the media) could be highly relevant when
a study requires data from a specific region 
or when location is a factor in the study, e.g., for analysis of environmental 
or cultural differences between regions.
Timestamps can be used when changes over time are of interest, such as changes 
in snow cover.\rem{\footnote{GPS coordinates and timestamps are not always accurate, 
but inaccurate data is usually easy to identify and discard.}}
\rem{Camera types may influence features of images or videos
in ways that are relevant for extraction of information.}

On-the-ground field studies are limited in how many \rem{specific }locations
they can target, due to time and resources. In contrast, the wide coverage
across locations found in the YFCC100M \cite{Choi2014,Kalkowski2015} makes it
possible to \sht{focus on comparing}{compare} many different places---or, conversely,
to reduce location bias within a study area.

\rem{
However, metadata---especially when it is user-generated---has its limits. 
Complete and correct metadata cannot be expected.
A robust approach leveraging this wealth of metadata must \rem{therefore} work around incorrect, ambiguous, or missing annotations\sht{. We discuss this in detail in Section \ref{s:fwovw}.}{ (see Section \ref{s:fwovw}).}
}

\subsection{Extended Resources and Tools}

An important advantage of the YFCC100M dataset is that new resources are often
added by research groups working together in the Multimedia Commons. These
include carefully selected annotated subsets \cite[e.g.,][]{Bernd2015a} and
preprocessed datasets (or subsets) of commonly used types of features
\sht{\cite[e.g.,][]{Choi2014,Bernd2015,JiangEtAl2015,Popescu2015,AmatoEtAl2016}.}{
\cite[e.g.,][]{Choi2014,JiangEtAl2015,Popescu2015}.
} \rem{Having
precomputed features can significantly speed up processing.} The Multimedia
Commons also includes a set of automatically generated tags (autotags) for all
of the images and the first frame of each video, labeling them for $1500$
visual concepts (object classes) with $90\%$ precision \cite{Thomee2016}, and a
set of\rem{ automatically generated} multimodal video labels for $609$ concepts \cite{JiangEtAl2015}.

The existing subsets can reduce the amount of data that has to be parsed to
extract relevant examples for a study. As one example, the YLI-MED
video subset \cite{Bernd2015a} provides strong annotations for ten targeted
events (or no targeted event), along with attributes like
languages spoken and musical scores. Our objective in part is
to enable scientists to create new strongly annotated subsets \rem{(as described in
Section \ref{s:app}) }and---ideally---to contribute them in turn to the
Multimedia Commons.

\rem{
The current Multimedia Commons ecosystem includes an easy-to-use image browser, 
described in Kalkowski et al.\ 2015
\cite{Kalkowski2015}. 
This browser can use the image metadata to generate subsets according to a user's specifications,
provide statistics about that subset or the dataset as a whole, 
allow users to view the images,
and provide URLs to download images for further analysis.
}

\section{The MMBDS Framework}
\label{s:app}

\sht{Our MMBDS framework takes a similar approach to Kalkowski et al.'s data browser \cite{Kalkowski2015}.}{The Multimedia Commons ecosystem includes an image browser, described in Kalkowski et al.\ 2015 \cite{Kalkowski2015}. Our MMBDS framework takes a similar approach in some respects.}
However, our new framework is open source, 
enables more types of searches (e.g., feature-based), 
and provides more ways to interact with and refine
a dataset to achieve the desired \sht{result. 

This}{result. This} section outlines the MMBDS \sht{framework---including specifications based on our conversations with scientists and our case study on origami\rem{ (see Section \ref{s:origami})}---and describes our progress in implementing it.}{framework, including specifications based on our conversations with scientists and our case study on origami.}

\subsection{The Proposed Dataset-Building Process}
\label{s:fwovw}

Ideally, a scientist will want high-quality data with strong---i.e., consistent
and reliable---annotations. But user-supplied tags are generally inconsistent,
sometimes inaccurate, and often do not exist at all. In addition, scientists
will frequently be looking for characteristics that a user would not typically
think to tag because they are unremarkable or backgrounded (like ordinary trees
lining a street)---or would not tag in the same way (for example, listing each
species of tree). However, obviously, it would be very cumbersome for expert
annotators to go through all of the YFCC100M images and videos and label them by hand
for a given research project.

We propose an iterative, hybrid approach to take advantage of 
both content and metadata\sht{ in this large corpus.}{.}
A typical search process might start with a user selecting 
some terms and filters to gather an initial candidate pool. 
These filters can be of multiple types, including 
metadata filters and/or 
(weak) detectors. 
The use of automatic detectors 
means that all data is included in the search,
whether the user has supplied tags or not.
The search engine would enable the user to
prioritize the various filters to sort the data 
\sht{\cite[as in][]{Jiang2015, Xu2015}.}{\cite[as in][]{Jiang2015}.}
(Alternatively, the user could start with a similarity search, 
if they already had some relevant examples on hand.)

\rem{
\begin{figure}[thb]
  \centering
  \includegraphics[width=0.45\textwidth]{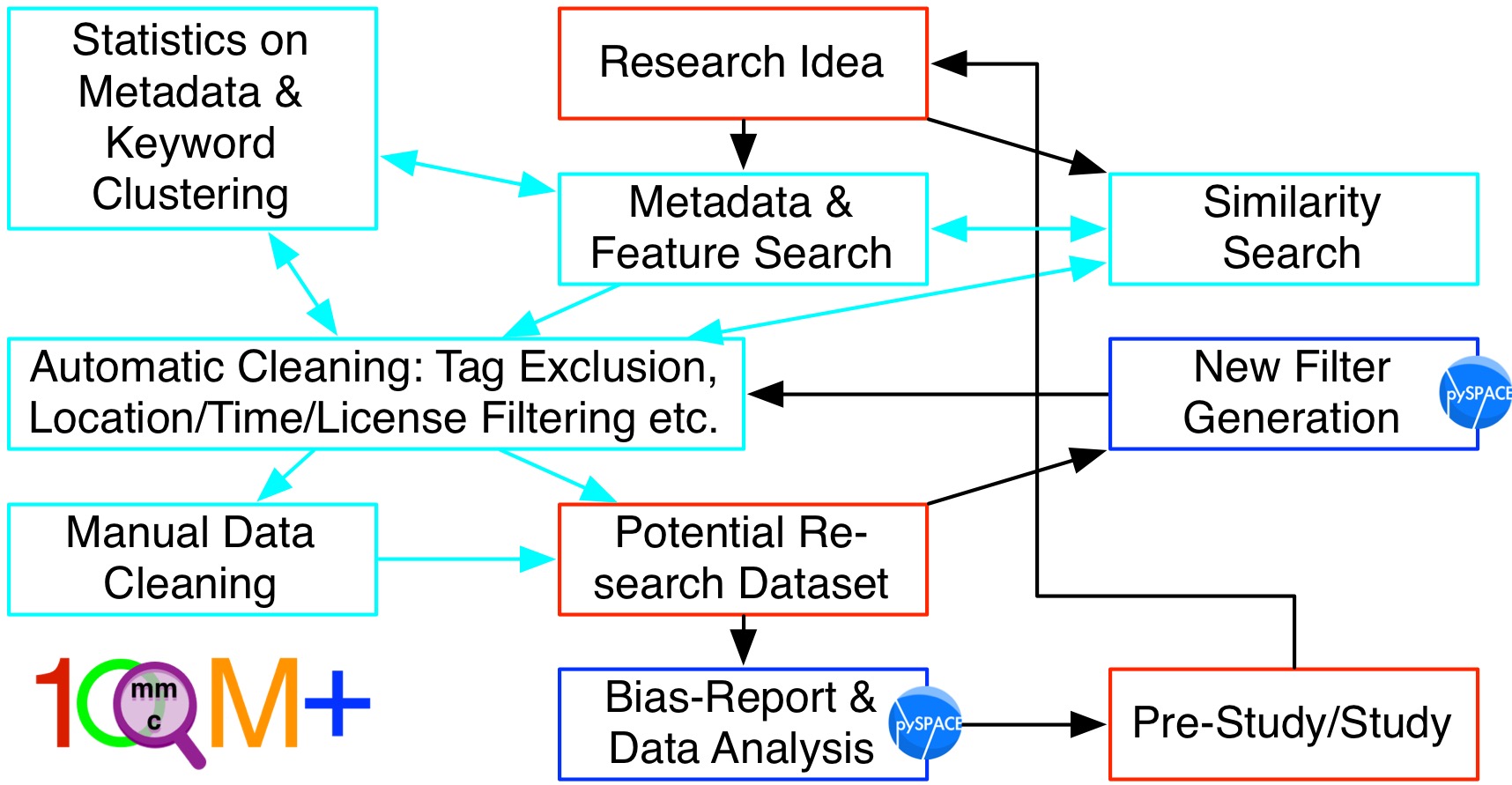}
  \caption{Schematic representation of the MMBDS framework,
  including ideas and products (red), 
  dataset extraction (cyan) handled by MMCS and scripts,
  and data analysis (blue) handled by {pySPACE}.}
  \label{f:framework}
\end{figure}
}

If the resulting set of candidates is sufficient and reasonably apropos,
the search could end there, with the expert (optionally) manually eliminating
the less relevant examples. \rem{The data browser may then be used to
add any additional annotations the researcher requires.}
If it is not satisfactory, the candidate pool could be automatically
narrowed\sht{. Narrowing could be done}{,} by adding or removing filters/detector types,
changing the filter parameters, adjusting their associated confidence bounds,
or selecting some relevant candidates and keeping only the most similar
examples among the rest.

Alternatively, if the candidate pool is too small or narrow, the next step
could be to expand it. This could be done by adding or removing
filters/detector types, adjusting filter parameters or confidence bounds, or
using similarity search based on the best candidates found so far. 
\rem{(Where
similarity search might be based on content/features or on metadata
clustering.)} The user could also examine the metadata \rem{(including autotags)}
associated with examples identified via automatic detectors, to inspire new
\rem{metadata }search \sht{terms.

Finally,}{terms. Finally,} the data processing framework (see Section \ref{s:procfw}) would allow
the user to create a new classifier/filter based on the candidate pool, and
re-apply it to a new search.

Each of these processes could be iterated until the field expert is
satisfied with the size and quality of the example dataset. Although the expert
would most likely have to do some manual reviewing and selecting, the automatic
filtering would make each step more manageable. In addition, any labels generated in the process could 
\rem{(if the user chose)}
be fed back into the system, to provide more reliable annotations for
future studies.
\rem{This process is summarized in \Reffig{f:framework}.}

\subsection{Implementation: Expanding Search Capabilities}
\label{s:engine}

There are, of course, a number of existing approaches to search.
But so far, none brings together all the features needed for MMBDS. For MMBDS, 
a scientist should be able to iteratively choose and customize searches on all
different types and aspects of multimedia data (text, textual metadata, images,
audio, etc.); retrieve data by labels, by content similarity, or by specifying
particular characteristics for detection; specify the fuzziness of the
parameters; and (if desired) retrieve \emph{all} of the matching examples. In
addition to browsing and selecting data, the framework should also allow for
annotation and for the creation of new filters within the same interface\rem{ (see
Section \ref{s:procfw})}.
\sht{The t}{T}esting of new
filters, as well as the goal of full customizability,
can be much better achieved when the framework is fully \sht{open source---unlike
the current YFCC100M browser~\cite{Kalkowski2015}.}{open source.}

We are therefore building a comprehensive new search engine, 
together with a web-based front end, called Multimedia Commons Search (MMCS). The open-source framework thus far
is built around 
a Solr-based search, connected via Flask to a React web page.
It is available at
\sht{\url{http://search.mmcommons.org/}}{\textit{[URL of search engine]}
(see also \Reffig{f:MMCS})}, on GitHub, and via Google Drive.
The search engine currently 
uses the YFCC100M metadata; Yahoo-supplied extensions such as geographical information and autotags; 
video labels~\cite{JiangEtAl2015}; and language labels~\cite{Koochali2016}.  
Confidence scores for the autotags and video tags can be used
to adjust the precision of the search.

\sht{
\Reffig{f:MMCS} shows the current state of the MMCS web interface.
The user can generate complex searches (optionally using Solr queries) 
and use metadata filters to exclude irrelevant results.
\begin{figure}
  \centering
  \includegraphics[width=0.45\textwidth]{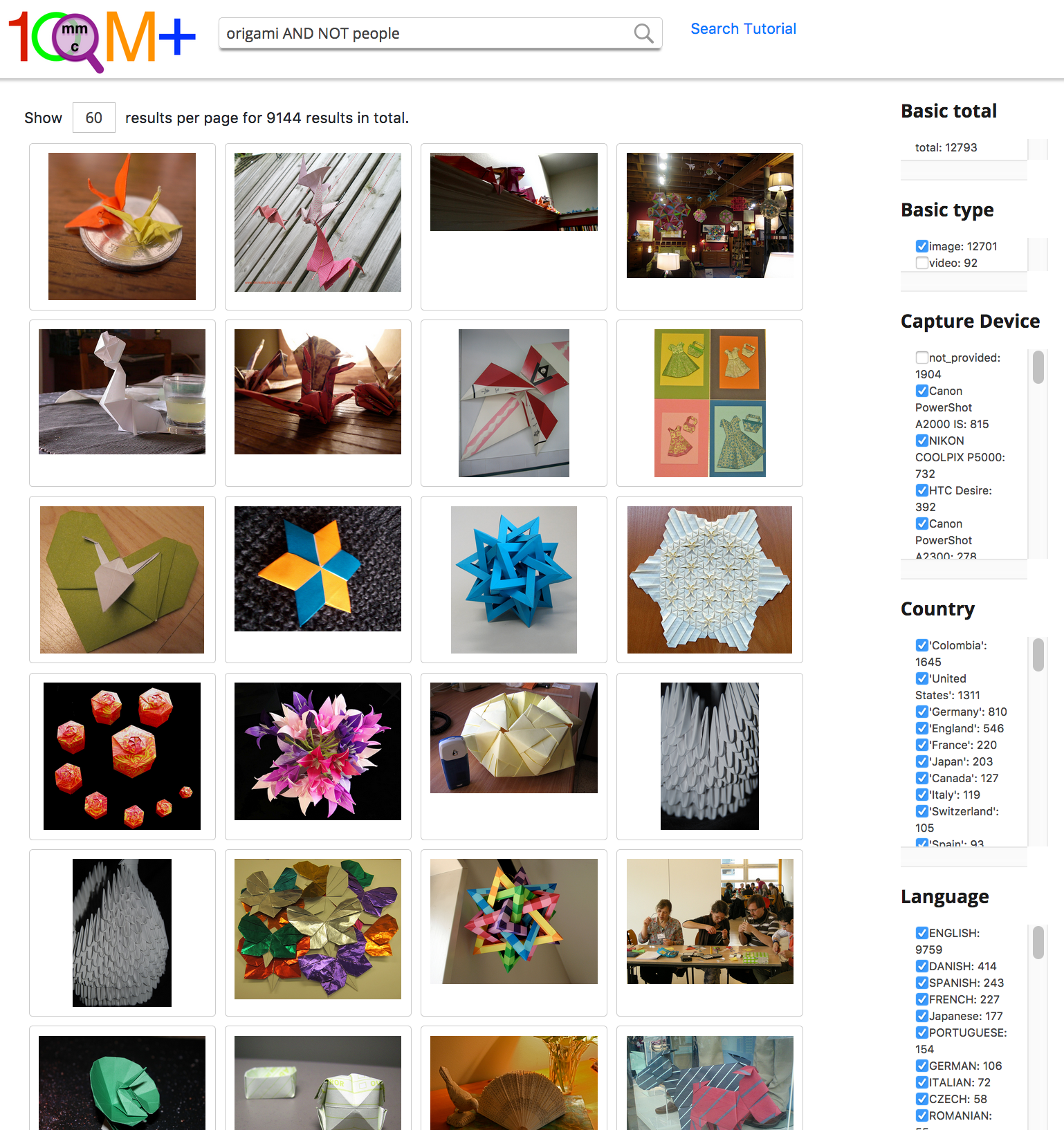}
  \caption{Screenshot of the Multimedia Commons Search (MMCS) interface.}
  \label{f:MMCS}
\end{figure}
}{
\begin{figure}
  \centering
  \includegraphics[width=0.39\textwidth]{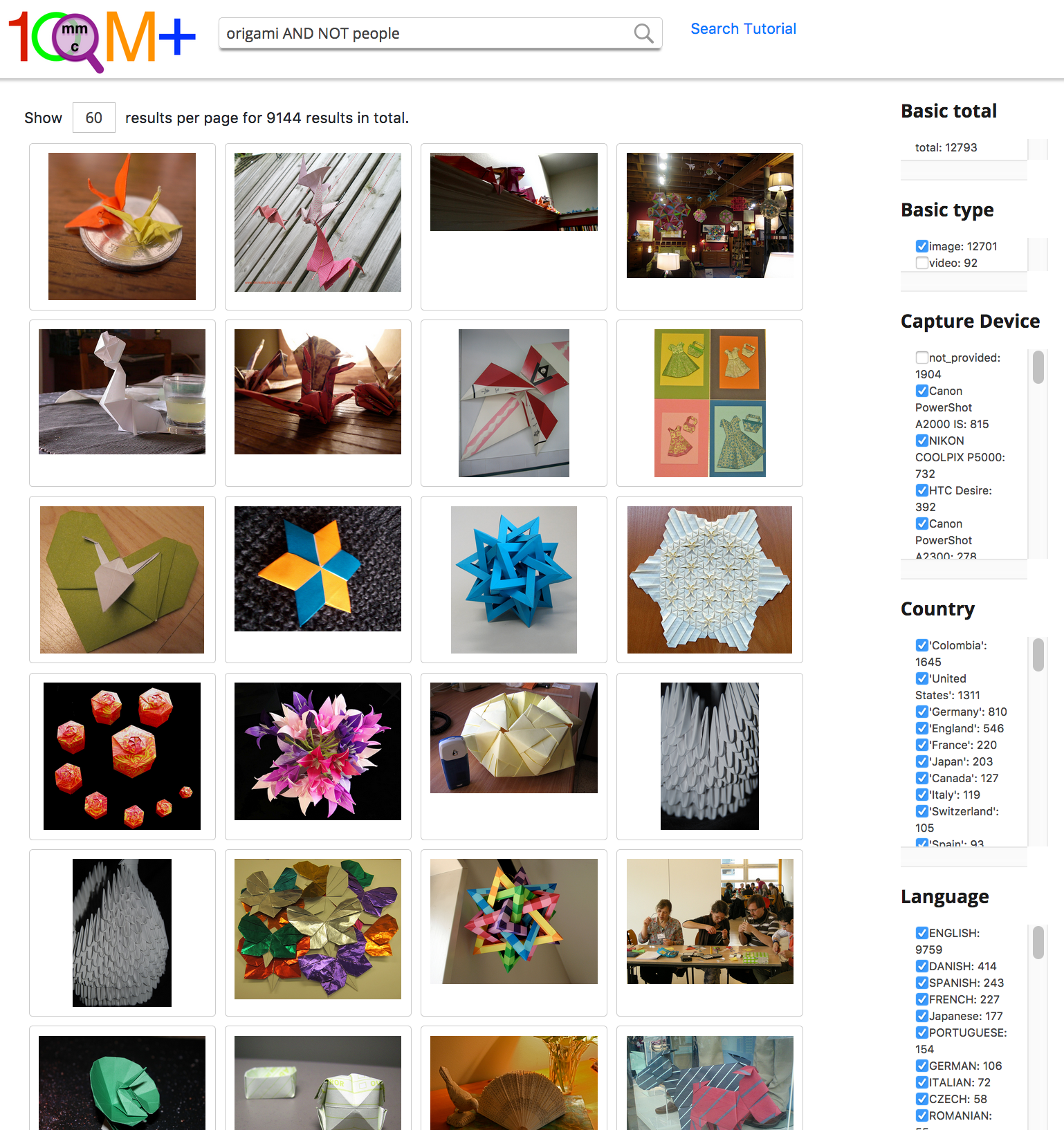}
  \caption{Screenshot of our new search interface.}
  \label{f:MMCS}
\end{figure}
}

\rem{We are continuing to add new filters and search types. }We next plan to add 
similarity search \rem{(e.g., based on LIRE~\cite{Lux2014}), }using
existing and/or new human-generated video labels \cite[e.g.,][]{Bernd2015a}\rem{, as well as adding some of the existing feature sets to aid in building new filters}.
\sht{We may also generate new autotags based on
high-frequency user-supplied tags, which can then generalize over the whole
dataset \cite{Thomee2016}. In addition, new autotags could be generated by
transferring classifiers trained on other datasets.

Several studies have already investigated how and to what degree the
user-supplied tags in the YFCC100M can be used to bootstrap annotations for
untagged media. For example, Izadinia et al.\ \cite{Izadinia2015} suggested a
new classifier to deal with noisy class labels. It
uses user-supplied tags (``wild tags'') to generate automatic (weak)
annotations for untagged data in the YFCC100M.
Popescu et al.\ \cite{Popescu2015} developed an evaluation scheme 
in which user-supplied tags can be used to evaluate new descriptors
(when enough user tags are available)---though again, field studies may require descriptors 
that do not tend to occur in tags.
As}{As} we expand\sht{ further,}{,} filters will incorporate any newly generated metadata, such as
estimated locations for non-geotagged media. 
\sht{Other additions may involve
automatically detecting}{We may also automatically detect} characteristics likely to be useful across a variety of
studies, such as 3D posture specifications.\rem{ Such estimation combines 
2D pose estimation \cite{Wei2016,Cao2016}
with a mapping from 2D to 3D, mostly based on existing 3D databases
of human 
poses.\footnote{However, addition of 3D postures may not be feasible for a while yet. In a
test of some state-of-the-art pose estimation tools, we realized that current capabilities are more limited than they
are often reported to be. The recognition quality was poor when we queried less common (or less straightforward)
postures involving, for example, crossed legs or rotated hips. We believe this discrepancy between the reported
performance and our results arises from standard issues with transfer from small datasets with a limited set of
targets (in this case, poses) to wild UGC data. In addition, in practical terms, pose estimation requires heavy
processing, which is rather slow on a dataset the size of the YFCC100M.}}
We also intend to \sht{}{bootstrap new autotags \cite{Popescu2015}, }add translation capabilities\sht{. 
And as we noted above,
users may \rem{choose to} contribute new classifiers \rem{they train} or new annotation
sets they create.

Finally, \rem{in addition to adding new metadata and new search capabilities,} we hope to }{, and }incorporate additional UGC corpora\rem{ beyond the YFCC100M, including text and pure audio corpora}.

\subsection{From Search to Data Processing}
\label{s:procfw}

To enable scientists to perform studies, 
developing search filters is not sufficient.
A user-friendly data-processing framework is also necessary, 
for annotating, correcting labels, extracting features from the data, 
and/or creating new classifiers\rem{ and filters} from search results or \rem{from target }images they \rem{already }have on hand.\sht{

In addition to improving MMCS, 
we will add a data processing component, for example by extending the signal processing and classification environment
pySPACE~\cite{Krell2013} to work on multimedia data.
To build new classifiers for images and video keyframes, 
the framework might incorporate  
CNN features from VGG16 trained on ImageNet~\cite{Simonyan2014}, 
train a simple classifier on the examples selected by the user,
and then use the classifier to retrieve additional images from the YFCC100M.\rem{ (Although this piece has 
not yet been incorporated in the publicly available version of the framework, Section
\ref{s:origami} describes a trial run of the process.)}
As we noted in Sections \ref{s:fwovw} and \ref{s:engine}, these classifiers and labels could 
then be incorporated\rem{ into the framework for future studies},
providing more prefab filtering options to new users.}{
Therefore, we will connect MMCS to an existing data-processing framework~\cite[e.g.,][]{Krell2013}
so it can easily configure a classifier---e.g., based on VGG16 features---using examples chosen by the 
user, then use that classifier to retrieve additional images.
}

\sht{\subsection{A Potential Issue: Selection Bias}
\label{s:select}

One issue the data-processing framework will need to help researchers address is selection bias.
Bias might arise from the filtering strategies or from the distribution of the dataset itself,
potentially affecting the results of a study.
Kordopatis-Zilos et al.\ \cite{Kordopatis-Zilos2016} 
analyzed several dimensions of bias in the YFCC100M with respect to the location estimation task:
\begin{itemize}[nolistsep,noitemsep]
\item Location bias: The YFCC100M is biased toward the U.S.\ and (to a lesser degree) Europe; 
people are more likely to take pictures in certain places (like tourist destinations);\footnote{In addition, filtering must account for non-unique place names like \textit{Richmond}.}
\item User bias: Some users contribute a much higher proportion of the data than others;
\item Text description bias: Some data comes with many tags and long descriptions, while some is not even titled;
\item Text diversity bias: Some tags and descriptions might be very similar (especially if uploaded together); and
\item Visual/audio content bias: Data may contain more or fewer of the particular visual or audio concepts 
targeted by automatic classifiers.
\end{itemize}

Other important dimensions might include language (of content or metadata), properties of the recording device,
time of day, or the gender, age, etc.\ of the contributors and subjects. 
For applied studies involving training and evaluating classification algorithms,
class imbalance can also be an issue 
\cite{Straube2014}.

For MMBDS filtering, we intend to build on the sampling strategy suggested by
Kordopatis-Zilos et al. 
In this approach, the percentage of the difference between a 
given metric computed on the target dataset compared to a metric computed
on a less biased reference dataset is reported (``volatility'' in their equation 6).
To generate one or more reference datasets, the system can apply strategies that mitigate
the aforementioned biases (like text diversity or geographical/user uniform sampling) or separate the biased dataset into several subgroups (like text-based, geographically focused, and ambiguity-based sampling).
The search engine can support creation of the reference datasets,
and then the data-processing framework can calculate the different performance metrics
in an evaluation setting.

In addition to trying to mitigate bias, the data
processing framework should make the user aware (e.g., via a visualization) of
possible residual biases that could influence their results.
}
{One issue the processing framework will need to help researchers address is selection bias. Bias might arise from the filtering strategies or from the distribution of the dataset itself,
potentially affecting the results of a study.
Kordopatis-Zilos et al.\ \cite{Kordopatis-Zilos2016} 
analyzed several dimensions of bias in the YFCC100M with respect to the
location estimation task, including location bias (towards the U.S.; towards
tourist destinations), bias towards high-contributing users, bias in quantity
and variety of text metadata, and bias in content targeted by automatic
classifiers. Other important dimensions might include language (of content or
metadata), properties of the recording device,
time of day, or the gender, age, etc.\ of the contributors and subjects. 
\rem{For applied studies involving training and evaluating classification algorithms,
class imbalance can also be an issue 
\cite{Straube2014}.}
Kordopatis-Zilos et al.\ suggested a sampling strategy to mitigate bias; in addition to building on this, 
the data processing framework should make 
the user aware (e.g., via a visualization) of possible residual biases. 
}

\sht{\section{Example Studies: Answering Scientific Questions With UGC}}{\section{Example Scientific Studies}}
\label{sec:sciexx}

\sht{A wide range of studies in natural science, social science, 
and the humanities could be performed or supplemented
using UGC \rem{media} data rather than controlled \sht{recording. 

In}{recording. In} some of the examples \sht{in this section}{here}, we describe existing studies 
and suggest how they could be reproduced or extended
with the YFCC100M\rem{ dataset}.
In other examples, we suggest studies that have not yet been performed at all.

Depending on the example, the UGC data might be the final object of study, or
it might act as a pilot.}{
A wide range of studies in natural science, social science, 
and the humanities could be performed or supplemented
using UGC data rather than controlled recording. In some cases, the UGC data might be the final object of study, 
and in some it might act as a pilot.}
 As a pilot or pre-study, it could help researchers get a handle on what variables they
most want to examine or isolate in controlled data-gathering, what other
variables they need to control for, how much data and how many camera angles
they need, etc. \sht{It can also alert them to additional factors or
possible variables of interest that they might not have expected \textit{a
priori}. Such a pilot could save a project significant time and money.}{It can also alert them to 
possible variables of interest they might not have expected \textit{a
priori}, as well as to potential difficulties. This can help researchers avoid mistakes and delays during
targeted data acquisition, saving time and money.}

\rem{
\subsection{Environmental Changes and Climate Indexes} 
\label{s:nature}
}

\sht{}{\textbf{\textit{Environmental Changes:}}} Researchers have already begun combining social-media 
images with data from other sources to analyze changes in the natural \sht{world.

In}{world. In} general, it is possible to extract or classify 
any specific kind of plant, tree, lake, mountain, river, cloud, etc.\  in images.
Since natural scenes are popular motifs in vacation images,
there is quite a bit of relevant data in the YFCC100M.
This data can be used to analyze changes in those features
across time and space.
\sht{For example, calculated features such as color scores can be used to create indexes, such as 
a snow index,
a pollution index, or an index representing the height of a river or creek.}{For example, calculated features 
such as color scores have been used to create indexes for environmental factors such as 
snow cover, cloud cover, and pollution, using targeted image datasets or combinations of targeted and UGC data.}

\rem{
In one recent case, Castelletti et al.\ \cite{Castelletti2016} used a combination of traditional and UGC data to
optimize a control policy for water management of Lake Como in Italy.
They calculated virtual snow indexes from webcam data 
and from Flickr photos.
Using data from even one webcam was already slightly better than using
satellite information, and combining the two showed large benefits.
However, they were not able to leverage the Flickr photos for similar gains because
their image dataset covered too short a time \sht{period.

By}{period. By} using the ten years of YFCC100M data with this approach, researchers could further improve on these results,
and even generalize to other regions of the world where YFCC100M coverage is dense (especially North America, Europe, the Middle East, and Australia). 
In related studies \cite{Wang2016,Zhang2012},
satellite images were taken as ground truth to estimate
worldwide snow and vegetation coverage using
(unfiltered, but geotagged and time-stamped) Flickr images.
Using a combination of these two approaches to enhance snow indexes all over 
the world would be very useful for climate analysis.

To estimate air pollution ($PM_{2.5}$ index) from images, recent approaches
have used image data from small, purpose-built datasets. For example, Liu et
al.\ 2016 \cite{Liu2016} correlated images with publicly available measurements
of the $PM_{2.5}$ index in Beijing, Shanghai, and Phoenix, using specific
features to construct estimators of the air pollution based purely on images.
Zhang et al.\ \cite{Zhang2016} used a CNN-based approach to the same task,
focusing on \sht{Beijing.
Extending}{Beijing. Extending} those studies to larger datasets
(different points of interest, different distances to observed objects,
and different times and seasons)
and more locales could enable the generation of air pollution
estimates for places where no sensors exist.
This could be achieved by correlating geotagged, time-stamped outdoor images in the YFCC100M dataset
with pollution measurements for locations where that data is publicly available,
then translating the results to locations without pollution sensors.

\rem{
Cloud-cover data is also highly relevant for longterm analysis of the natural
world~\cite{Eastman2013}. Some work has been done on automatically classifying
cloud types~\cite{Xia2015}. However, methods for globally complete cloud-cover
estimation have not been developed to extend localized automatic detection and
human-generated estimates, which often suffer from gaps.
Existing studies~\cite[e.g.,][]{Eastman2013} could be augmented by adding
YFCC100M image data to existing cloud-cover databases. This image data could be
gathered by using image segmentation \cite{Cimpoi2015} and/or classifiers to
pick out \sht{clouds, possibly augmented by incorporating PoseNet to determine
the direction of the camera~\cite{Kendall2015}. }{clouds.}
}

\rem{
In the related field of geography, there is already great interest in using UGC
to address scientific \rem{research} questions (\sht{a form of}{as} citizen science).
For example, crowdsourcing has  been used to gather data on
forest diseases~\cite{Connors2012}, and Flickr data has been used
to improve landcover maps~\cite{Estima2013}.
}
}

\rem{
\subsection{Human Language and Gesture Communication}
\label{sec:lxexx}
}

\sht{}{\textbf{\textit{Human Language and Gesture:}}} The YFCC100M contains a wealth of data on human interaction and communication \cite{Koochali2016}\sht{,
which could be quite valuable for linguistics, cognitive
science, anthropology, \sht{psychology, and other social sciences. }{and 
other social sciences. }

In}{. In} addition to
\rem{searching for and filtering relevant videos using }text metadata and location,
researchers could target specific situations using \rem{automatic classification
functions like }speech/non-speech detection, language identification,
\rem{emotion/affect recognition, }and pose recognition.\sht{ Language identifications on the YFCC100M 
metadata \cite{Koochali2016} have already been added to the MMBDS framework. 
Off-the-shelf speech/non-speech
detection\rem{ \cite[e.g.,][]{WebRTC-VAD}} 
and speech recognition
\sht{\cite[e.g.,][]{Povey_ASRU2011,hannun2014deep,lamere2003cmu},}{\cite[e.g.,][]{Povey_ASRU2011},}
written \cite[e.g.,][]{Lui2012}
and \sht{spoken \cite[e.g.,][]{Lang-ID}}{spoken} 
language identification, and 
speech-based emotion recognition 
\sht{\cite[e.g.,][]{EmoVoice,OpenEAR}}{\cite[e.g.,][]{OpenEAR}}
packages could also be incorporated.\footnote{Of course, no available off-the-shelf detector for any video or image characteristic will provide perfect accuracy. However, used in combination with other filters and given the ability to adjust the desired certainty, they can be a valuable tool for narrowing or broadening the search space.}
Extracting 3D models for pose recognition has been studied for images
\cite{Yasin2016,Bogo2016,Chen2017}, and will hopefully continue to improve and become
more efficient. If so, this work can be extended to video
\cite{Zhou_2016_CVPR}, in
combination with \sht{work on motion trajectories already being done with
YFCC100M videos \cite{CarranoCASIS}}{ongoing work on motion trajectories using the YFCC100M}. 
\rem{This aspect would be quite challenging, but very useful for several of the
examples described in this section and Section \ref{sec:applexx}.}

\rem{
An example of a study that could be expanded in this way compares how people
talk to pets, babies, and adults. Mitchell's (2001) analysis identified ways in which
people in the U.S.\ speak to their dogs as they would to infants, in terms of
content (short sentences) and acoustic features (higher pitch), and ways the two types of speech differ
\sht{\cite{Mitchell2001}.

With}{\cite{Mitchell2001}. With} MMBDS, 
the dataset \rem{for this study} could be 
\sht{broadened, and
comparisons made to how people talk to other pets and non-domestic animals,
along with comparing child-directed and animal-directed speech in other
cultures.}{broadened.}
For such a study, short videos \rem{like those on Flickr} would be sufficient.
\sht{A large number \rem{of videos} could be gathered using metadata searches, 
given the popularity of animals and children as video subjects in YFCC100M; 
in addition, some \rem{videos}}{Some videos in YFCC100M} already have strong annotations 
for interaction with animals \cite{Bernd2015a}. 
\sht{If necessary, t}{T}his could be supplemented with speech/non-speech 
detection and other feature-based filters 
to identify babies, children, and pets.
}

\sht{On a much wider scale, t}{T}}{ For example, t}here are many topics in child language acquisition 
that could be explored using UGC data\rem{, especially for high-frequency phenomena}. 
Existing corpora of children's \rem{speech }and child-directed speech 
\rem{\sht{(CHILDES \cite{childes} being the most widely used)}{\cite[e.g.,][]{childes}} }
\rem{usually }include some video data, but \rem{by far }the 
majority is audio-only or annotated transcripts. 
This limits researchers' ability to examine the relationship 
between children's utterances and the situational context \rem{for them }(\rem{for 
example, }what they might be trying to describe or achieve). 
Acquisition researchers therefore often spend a significant portion 
of their budget on video recording---and
\sht{a significant amount of time}{significant time} dealing with \sht{Institutional Review Boards'}{ethics boards'} 
requirements for data involving children.

\rem{
We describe here \sht{two}{one} among the many examples
where an important acquisition study could be extended using UGC data. 
\rem{In one example,} Choi and Bowerman \rem{(1991, 2003)} examined how English- and
Korean-speaking children conceptualized the relationships between two objects 
\cite{ChoiBowerman1991,BowermanChoi2003}. 
Different languages highlight different aspects of spatial relationships; 
for example, English \textit{put in} vs.\ 
\textit{put on} distinguish containment from surface attachment, 
while Korean \textit{kkita} vs.\ \textit{nehta} 
distinguish close-fitting from loose-fitting relationships. 
Choi and Bowerman used videotapes of both spontaneous speech and controlled
experiments to investigate how the difference in language affected 
children's spatial reasoning. 
They and other authors have since extended this work to, e.g., Dutch \cite{Bowerman1996} and Tzotzil \cite{deLeon2001}. 

The YFCC100M could be used to collect data for many more 
languages, using language identification, 
detectors for children or children's voices, 
and location metadata, perhaps combined with tag searches and/or 
object or pose detectors to identify particular target \sht{situations. 
A subfield of language acquisition explores 
how children's language learning is integrally 
related to learning about social behavior as a whole 
\sht{\cite{SchieffelinOchs1986,GarrettBL2002,OchsSchieffelin2012};}{\cite{OchsSchieffelin2012};} such studies
require a large amount of video data to get the necessary rich context. 
In one seminal 
study, Ochs and Schieffelin (1984) used data (including video) 
from several of their past projects to identify important differences 
in how caregivers in three cultures 
talk (or don't talk) to prelinguistic infants \cite{OchsSchieffelin1984}. 
These differences stem from varying assumptions 
about what kinds of communicative intentions 
an infant could have. 

For other researchers who are extending the findings 
about caregiver assumptions in large comparisons across many culture groups,
 in-depth data-gathering is of course necessary.
But to answer some preliminary questions and ascertain which cultural groups 
might follow which general patterns in addressing infants---i.e., to decide 
where to 
conduct that in-depth data-gathering---a pilot study using short, uncontrolled 
videos from a UGC corpus with worldwide coverage could be very helpful. In this 
case, 
location metadata could be combined with language identification and 
identification of babies and children in the videos (and/or broad tag searches) 
to find potential 
videos of interest.
Again, using such a pilot to prepare for more controlled, 
high-quality data gathering is especially helpful for studies involving 
children and conducted across 
national borders, given the added difficulties in 
scheduling data-gathering and getting proper permissions.}{situations.
Alternatively, MMBDS could be used as a pilot to prepare for more controlled, 
high-quality data gathering. Again, such pilots could be especially helpful for studies involving 
children---and particularly if they are conducted across 
national borders, given the added difficulties in 
scheduling data-gathering and getting proper permissions.}
}

\sht{
The related and growing field of gesture studies relies for obvious reasons on 
video-recorded data\sht{ \cite[e.g.,][]{Eliza2016,watson2013communicative,SOGM2005}.}{.}
Here, again, there are questions that can be answered using short 
videos or even images. \rem{\rem{Depending on the question,} UGC might provide the dataset for study or act as a pilot.} At the least, even uncontrolled, messy UGC data can give the researcher a preliminary sense of how frequent a particular phenomenon is and whether it is common across speakers or across a \sht{culture. 

But}{culture. But} in the case of gesture, tag-based search will likely produce 
little of value. \sht{Gesture researchers interested in systematic description of 
the ordinary 
hand movements, postures, and facial expressions that accompany normal 
conversation will not be able to find those ordinary gestures using tags}{Researchers will not necessarily be able to find the ordinary hand movements, postures, and facial expressions that accompany normal conversation}; after 
all, tags 
tend to point out the exceptional. 

To take a concrete example,}{
A related and growing field is gesture studies. To take a concrete example,
} a gesture researcher at one of our institutions 
wanted to study when people gesture with a pointed finger but without pointing at anything 
in specific 
(for example, how does it correlate with emphatic tone of voice?). However, 
when she tried a tag search for \textit{pointing} in UGC videos, of course, she 
found either 
extreme examples (shaking a pointed finger angrily) or examples where people 
were pointing at something or someplace, rather than the small hand movements 
she wanted to analyze. In this case, the researcher gave up on pursuing the 
question---but if she had been able to use feature-based query-by-example, or 
(better yet) 
initiate a search by specifying the 3D relationship between the hand and 
fingers, she could have had much better success.

\rem{
\subsection{Human Behavior: Emotion Examples}
\label{sec:psych}
}

\sht{}{\textbf{\textit{Emotion Expression and Recognition:}}} \sht{One area of behavioral research where m}{M}ultimedia data is vital \sht{is}{to} the 
study of how humans \sht{express emotion, and how we understand others' emotions. (And human 
emotion research in turn feeds into multimedia research on automatic emotion 
understanding; see Section \ref{sec:robointrxn} for examples.)

In particular,}{express and understand emotion. But} it can be difficult to obtain spontaneous recordings of a 
wide range of emotions in an experimental setting. Researchers can set up 
situations to try to elicit
emotional reactions (\rem{sometimes called }``induced emotion'')  \cite[e.g.,][]{HarrisAlvarado2005},
but there are
limits to this\rem{ practice}---especially given \sht{the ethical requirements to obtain
permission}{ethical considerations}.\rem{ UGC therefore has the potential
to fill a large gap in emotion \rem{and affect} research that is only beginning to be
addressed.}

\rem{
However, as with the gesture example discussed in Section \ref{sec:lxexx}, 
a simple approach to tag-based search---e.g., using emotion words like 
\textit{disappointment}---is not likely to yield scientifically useful results.
It will likely only turn up examples where the behavioral expression of that 
emotion is extreme, and/or 
where the uploader has some personal reason for commenting on it. Researchers 
targeting \rem{the expression of} specific emotions can find more representative 
samples by 
searching for situations likely to elicit those emotions, either using tags, existing event annotations, or 
detectors for events or other aspects of the situation.\footnote{\sht{While it is 
possible that off-the-shelf 
emotion/affect detectors could be used here, their utility}{Unfortunately, the utility of off-the-shelf emotion/affect detectors} is limited if the object of 
study is emotion expression or interpretation itself. Automatic 
classification of human 
behavior is necessarily always \rem{a few steps behind} what is known in 
behavioral science \cite{GunesHung2016,ZengEtAl2009}.
\rem{Creative approaches can \sht{at least partially account for}{help with} this \cite{MariEtAl2014}, but a simple system can only 
confidently identify the 
most prototypical, unambiguous examples\sht{, rather than the full range. (In 
addition, such detectors are often trained on acted emotion---see Section \ref{sec:robointrxn}.)}{.}}}
For example, sports events are associated 
with 
\rem{feelings of} excitement, suspense, triumph, and disappointment \cite{MatsumotoWillingham2009}. 

\rem{Alternatively, a researcher could start by searching for particular facial 
expressions, gestures, or tones of voice---either using feature-based 
query-by-example or by specifying 3D
postures, motion trajectories, pitch contours, etc.---then analyze the types of 
situations that lead up to those reactions and what, if anything, the 
participants say about them.} 

\sht{These avenues of}{Such} multimedia research could be very helpful in developing a 
fuller picture of the wide range of behaviors that can express any given 
emotion. Much has 
been done to identify the most prototypical facial expressions, vocal 
inflections, etc.\ associated with particular emotions
\sht{\cite{EkmanFriesen1971,EkmanFriesen1986,HaidtKeltner1999}}{\cite[e.g.,][]{EkmanFriesen1971,HaidtKeltner1999}}. However, one of the 
important questions in the field is how to 
get beyond those prototypical reactions to a more comprehensive 
understanding---especially as emotion expression is known to vary quite widely 
even within a single culture, 
much less across cultures \sht{\cite{Russell1994,ElfenbeinAmbady2002}.}{\cite[e.g.,][]{ElfenbeinAmbady2002}.} 
}

\sht{The flipside of research into emotion expression is research into how people 
interpret and categorize the emotions of others based on their behaviors. 
Image, audio, and video data are 
\sht{of course a mainstay for creating}{often used for} test stimuli in such experiments. 
However, }{Similarly, }
much of the most prominent research on emotion recognition 
\rem{\cite[e.g.,][]{EkmanFriesen1971,HaidtKeltner1999}}
has used acted \rem{rather than spontaneous }emotion, which 
\rem{(besides being unnatural)} tends to stick to prototypical cues.
Researchers are discovering the 
limits of this approach and the questions it can address \cite[e.g.,][]{NaabRussell2007}\sht{}{.}
\rem{(including for translation into automatic affect recognition; see Section 
\ref{sec:robointrxn}).} \sht{Hence, t}{T}he last few 
years have \sht{}{thus }seen a shift to recognizing the need for more spontaneous \sht{data, 
such as that found in the YFCC100M.}{data---a need UGC can help to meet.}
\rem{After all, humans can and do deal with quite messy data about the emotions of 
the people around them \cite{FeldmanBarrettEt2011}.}

One possible research project \rem{(to potentially be conducted by one of the authors) }
\sht{would be to}{could} use YFCC100M data to compare how speakers
of different languages conceptualize and talk about emotions. For example,
\sht{there is a large class of languages that express}{many Southeast Asian languages 
express} \sht{virtually all emotions}{most emotions} as
states of \rem{the experiencer's }body parts\rem{ \cite{Matisoff1986}}. \sht{\sht{This is
particularly common in Southeast Asia, as in the following example from the}{In the}
Hakha Lai language of Burma\sht{:}{,}}{In Hakha Lai,} \textit{ka-ha-thi na-thak}, literally `my
tooth-blood you itch'\sht{, meaning}{, means} `I can't stand you' \cite{VanBik1998}. Geotagged
videos from Southeast Asia could help to investigate whether other, non-verbal
aspects of emotion expression differ in related ways.

\rem{Another \sht{potential study by the same researcher}{study} might investigate how emotion categorizations and descriptions
are influenced by an understanding of context. A broadly representative set of
test stimuli could be compiled \rem{using the methods described above, along with
location metadata, existing strong subset annotations, and (ideally) language
identification, to find likely candidate images and videos} from different
languages and cultures. Researchers could \rem{then} compare how people described the
emotions of the people in them depending on whether they were shown the
\sht{preceding (or surrounding)}{surrounding} context, or only the snippet of the person
expressing the target emotion. \rem{An extended experiment could also compare
descriptions based on only the audio or only the visual stream from a video.}
}

\rem{
\subsection{Location-Based Comparisons}
\label{sec:LBS}
}

\sht{}{\textbf{\textit{Location-Based Comparisons:}}} In the YFCC100M, around $50\%$ of the data is geotagged, and location
estimations can be generated for many of the remaining images and videos,
especially ones recorded outdoors \sht{\cite{Choi2014,Choi2015a}.}{\cite{Choi2014}.} \rem{From this, it is
possible to approximately infer the location of many users' homes---or at least
their hometowns---as well as where they travel to. }Even where location
estimation for a given image is too difficult (especially indoors), information
can be gleaned at the user level based on the individual's other uploaded
images. \rem{For some studies, it may even be more helpful to know where the user is
from than where the picture was taken.}

One possible research question \rem{in this area} would be to compare
where people from different locales like to travel \sht{(and take pictures) \cite{Clements2010,Kadar2013}---do}{\cite[e.g.,][]{Kadar2013}---do} they go to
(other) urban areas? Do they go out into nature? At what times of year? A more
in-depth analysis could examine changes in those behaviors over time to
identify how preferred tourist spots change in response to world events.

\rem{
As another example, a researcher (e.g., in anthropology or marketing) 
could use geotagged indoor images and videos 
to identify patterns in the personal possessions of people
from different locations and backgrounds.
Using classifiers on labeled data, it would be possible to
determine the brands and values of at least some items\sht{.
In many cases, it might be possible to identify the objects and their
values by}{, possibly using} image comparisons, e.g., with an online store. 
Companies could also gather data about how their products are used \rem{in practice} 
in different cultures and countries,
and use this information to develop new services and products or
marketing strategies~\cite{sunderland2016doing}.
Such analysis could \sht{lead to a variety of automated applications as well}{also lead to automated applications}; see Section~\ref{sec:LBSapp}.
}

\rem{
As we noted in Sections \ref{sec:lxexx} and \ref{sec:psych}, location data can
also be used in cross-cultural studies. As a starting point, styles of
photography could themselves be compared across locations and over time. 

As
another example, comparisons of gender presentation and gender dynamics across
cultures are usually based on in-depth fieldwork on the ground. 
But
such studies could be supplemented with UGC data to provide cross-checking
against many more data points for a given culture, and to quickly gather at
least some data from many different locales without having to travel to all of
them. \sht{A researcher could identify such data using geotags 
and (optionally) inferred
locations, relevant user-supplied tags, person detectors, and possibly language
detectors.}{}\rem{\footnote{A potential filter could be built to exclude tourists'
contributions where they are unlikely to be apropos, using tags and inference
from the locations of other pictures from that user.} 
}
}

\rem{\subsection{Medical Studies} 
\label{s:medstud}
Wang et al.\ (2017) used a combination of machine-learning methods to attempt
to identify Flickr users who engage in deliberate self-harm \cite{Wang2017}.
They showed differences by text characteristics, user profile statistics,
activity patterns, and image features. Classification results were not as
accurate as is usual for more well-studied tasks, but were certainly accurate
enough to produce a good candidate set for a field expert to narrow down. Wang
et al.\ suggest that data gathered via such a detector could help researchers
enrich their understanding of the triggers and risk factors for self-harm,
along with studying the self-presentation and interactions of self-harmers on
social media per se.

That work had a specific topical focus (on content that might not be well
represented in Creative Commons media), and thus approached the problem
somewhat differently than we are proposing for a more multi-purpose search
interface. However, we consider the results to be a promising indicator of the
potential of such efforts.

We have not tried to quantitatively assess how much content can be found in the
YFCC100M to represent abnormal or pathological behavior or physical conditions.
However, even for cases where the YFCC100M does not contain many examples of a
targeted condition, it can still be quite useful to researchers studying that
condition: It can provide a quick and easy way to gather a baseline dataset to
compare to. (In fact, Wang et al.\ pulled their examples of non--self harm
content from the YFCC100M, though they did not target any specific behaviors for that control set
\cite{Wang2017}.)
}

\section{Example UGC-Based AI Applications}
\label{sec:applexx}

\sht{
As with the examples in Section \ref{sec:sciexx}, 
some of the possibilities we describe for applied research with UGC 
extend existing studies, while some of these 
areas have not been explored much at all.
}{As well as scientific studies, UGC can drive new applications.}

\rem{
\subsection{Movement Training for Robotics}
\label{HRI}
}

\sht{}{\textbf{\textit{Movement Training for Robotics:}}} Imitation learning in robotics requires recorded data 
of humans performing a target behavior or motion.
Movements can be learned and transferred via  
\sht{shadowing \cite[e.g.,][]{Argall2009,Metzen2013},}{
shadowing,} 
for example to a robot arm 
for grasping \rem{\cite{Yang2015} }or throwing\sht{, or an agent can learn from recordings
of a single human over a long time-span \cite{Jebara2002}.

T}{. T}aking throwing as an example, it would be possible to infer 
joint and object positions, velocities, and accelerations
from movements observed in 
data from a variety of sources, 
including targeted motion-tracking systems but also recordings of, e.g., ball games.
Within our framework, 3D human pose estimation would have to be applied to
single keyframe images from a video\sht{ \cite{Yasin2016,Bogo2016,Chen2017,Zhou_2016_CVPR}.

After}{. After} extracting the movements, the data can be segmented\rem{, for example
using velocity-based multiple change-point 
inference \cite{Senger2014}}.
The motion primitives can then be 
postprocessed and \sht{classified \cite{Gutzeit2016}.}{classified.}
Finally, the movement behavior can be optimized and \sht{transferred \cite{Gutzeit2017}.}{transferred.}

\rem{Such movement trajectories can be used not only
for transfer learning, but also for more general analyses 
of movement patterns and to build classifiers, for example to identify
movement disorders.}

\rem{
\sht{\subsection{Interaction Training for AI Systems}}{\subsection{Interaction and Affect for AI Systems}}
\label{sec:robointrxn}
}

\sht{}{\textbf{\textit{Interaction Training for AI:}}} \sht{As robots, dialogue systems,}{Robots,} AI assistants, and other AI-based services \sht{improve
in sophistication and interactivity, they }{}need to be able to recognize and
categorize \sht{not just speech, but }{}human emotions, attentional cues, and other
cues that can help in interpreting intent.\sht{\footnote{However, it is worth
pointing out that UGC data also has the potential to aid in automatic speech
recognition (ASR) in arbitrarily noisy environments, especially in recognizing
children's speech. Again because of the additional necessary permissions and
precautions, ASR researchers have a much smaller pool of recorded corpus data
to draw from for recognizing the speech of children, in comparison to the
available resources for adult speech \cite{claus2013survey}.}}{} 
There is therefore a major drive (for example, in affective computing) toward
automatic recognition of emotion\rem{\footnote{We use \textit{emotion (expression)}
and \textit{affect} somewhat interchangeably in talking about the whole field;
for particular applications, researchers often draw finer distinctions.}} 
in
multimedia data, \sht{including facial expressions \cite[surveys
in][]{PanticRoth2000,SariyanidiEtAl2015,GunesSchuller2013}, 
gesture/posture \cite[survey in][]{KleinBianchi2013},
vocal cues \cite[surveys in][]{VogtEtAl2008,BussoEtAl2013}, and biological signals \cite[surveys
in][]{GunesSchuller2013,ArroyoRomano2008}---and combinations of those modes \cite[surveys
in][]{CastellanoEtAl2015,ZengEtAl2009,DMelloKory2012}.}{including facial expressions, gesture, posture, vocal cues, and biological signals---and combinations of those modes.}  \rem{

However, as we noted in Section \ref{sec:psych}, it can be difficult to get truly
spontaneous, naturalistic data for emotion expression}\sht{---to use 
as training data for automatic systems, as much as for the scientific purposes
mentioned above 
\cite{CastellanoEtAl2015,ZengEtAl2009,VogtEtAl2008}}{ 
}\sht{.

For this reason, a}{However, a}utomatic affect recognition researchers 
\sht{have often used datasets of acted emotion \cite[e.g.,][]{CK,GEMEP,MMI}.}{often use acted datasets,
which does not necessarily help systems recognize human emotion ``in the wild'' \cite[e.g.,][]{MowerEtAl2009}.}
\rem{While the situation has improved in recent years, with several annotated video 
datasets of ``spontaneous'' emotion expression being released, much of that 
data has in fact been induced emotion, collected under contrived conditions 
\cite[e.g.,][]{MMI,RECOLA,SEMAINE}, 
or at best from television interviews 
\cite[e.g.,][]{VaM,Belfast}.}
{Even ``spontaneous'' datasets are usually collected under contrived conditions, either induced  
or (at best) interviews.} \sht{(In addition, \sht{such data}{they} are often collected under ideal conditions, 
in terms of lighting, head angle, etc.)}{}
\rem{Available annotated datasets collected in more truly naturalistic situations 
with rich context are fewer\, and are often audio-only
\cite[e.g.,][]{CREST,HardyEtAl2002}
 (where it is easier to minimize the effects of recording 
\cite{DevilEtAl2005}).}\rem{\footnote{By ``truly naturalistic'', we mean occurring
naturally in the course of everyday life\rem{, rather than induced by a researcher}. 
We do not (necessarily) mean what is sometimes called in behavioral research 
``biologically driven'' (or, even more confusingly, ``spontaneous'') emotion 
expression\sht{ \cite[e.g.,][]{EkmanEt1988}, which is then opposed to some general category of unnatural or non-spontaneous behavior that lumps together learned or socially driven emotion expression with acted or deliberately false emotion. \\}{.}
\rem{\setlength\parindent{10pt}{UGC image and video data will include a large proportion of emotion expression
driven by communicative purposes and manifested according to learned cultural
conventions (and varying in how closely it represents what the person is
``really'' feeling), but it is nonetheless
a response to a naturally arising context \cite{RussellEt2003,MatsumotoWillingham2009}.
For an AI system, it will be important to be able to interpret 
(and emulate \cite{EndrassEtAl2014}) both cues that can be consciously mediated (such as vocal pitch) and cues that (largely) cannot (such as pulse rate)---and to use the two different types of information appropriately in interaction. 
(Here, we set aside the effects of being recorded in the first place, which are 
nearly unminimizable for any kind of video data \cite{DevilEtAl2005}.)}}
}
The value of UGC data for this purpose is therefore coming to be recognized with new datasets \cite[e.g.,][]{BQ16},
but none as yet have strong human-generated annotations.

\sht{
Comparisons show that \sht{acted and induced or spontaneous emotion expression can 
differ in ways that have 
consequences for recognition}{differences even between acted and induced emotion expression can 
have consequences for recognition} 
\cite[e.g.,][]{CohnSchmidt2004,SchullerEtAl2007}---\sht{fundamental enough that}{in fact,} 
different features may be more discriminatory 
\sht{\cite[e.g.,][]{VogtAndre2005,SariyanidiEtAl2013,CaoEtAl2015}.}{
\cite[e.g.,][]{VogtAndre2005,CaoEtAl2015}}
\sht{Most tellingly, c}{C}ross-testing \rem{between datasets} shows that training 
systems to recognize \sht{acted, prototypical, or larger-than-life 
examples of affective cues}{acted or prototypical examples} does not \rem{necessarily} prepare them well 
for \rem{their actual task:} recognizing what humans are doing ``in the wild'' 
\sht{\cite{SchullerEtAl2010,ValstarPantic2012,LefterEtAl2010,MowerEtAl2009}
(though model adaptation can help
\cite{AbdelBusso2015})---nor vice versa, for that matter \cite{VidrascuDevillers2008}.\footnote{Many of these comparisons are between acted and induced data. 
    Extending to three-way comparisons with UGC data (of comparable quality) 
    is one obvious starting point for research.} 
}{\cite{ValstarPantic2012,MowerEtAl2009}.}
Natural emotion expression may be much subtler, 
and cues may be ambiguous\sht{---either because there are multiple emotions 
that cue commonly expresses or because the person is expressing 
an emotional state that does not fall neatly into one category}{} 
\sht{\cite[e.g.,][]{FernandezDolsCrivelli2013,KleinsmithEtAl2011,GunesSchuller2013,CowieCowieCox2005,MowerEtAl2009,CowieEtAl2005}.}{\cite{GunesSchuller2013,MowerEtAl2009}.}
\rem{
\sht{In addition, e}{E}ven dimensional \rem{(non-categorical)} analysis systems run into 
problems because of the wide variation between individuals 
\sht{\cite{GasparEsteves2012,GunesEtAl2011,PanticRoth2004,CaoEtAl2015}.}{\cite[e.g.,][]{CaoEtAl2015}.}
}

To take just one example of an intriguing study that invites replication with 
naturalistic data,
Metallinou et al.\ (2012) were able to improve emotion recognition \rem{in video} by
taking into account prior and following emotion estimations, i.e., by modeling
the emotional structure of an interaction at the same time as the individual
cues \cite{MetallEtAl2012}. However, they used dialogues improvised by actors
rather than naturally occurring emotional \sht{interactions.

An}{interactions. An} important question is whether they would have achieved similar results using
non-acted data\sht{. In other words,}{---i.e.,} is automatic recognition of emotions in the wild helped
in the same way \rem{or to the same degree} by taking into account prior and
following judgments, or is there something special about the unity of
presentation in an acted situation?
To investigate this, researchers could collect and annotate data from the
YFCC100M videos that depict \rem{interactional} situations similar to those
improvised by the actors, using tag search, speech detectors, and recognizers
for particular activities. For example, the existing autotags and video event
labels could be used to filter for targets like \sht{\textit{people}, }{}\textit{sport}, \textit{wedding},
\textit{hospital}, \textit{fight}, or \textit{love}.

For humans, context and world knowledge can help to disambiguate or clarify 
confusing emotional cues \cite{FeldmanBarrettEt2011}, 
but context-sensitive automatic affect interpretation is still 
a fairly young field\sht{ \cite[e.g.,][]{CastellanoEtAl2012,MetallEtAl2012,LeeEtAl2009,CowieCowieCox2005}.}{.}
As Castellano et al.\ \sht{2015}{\cite{CastellanoEtAl2015}} \sht{(among others)}{(e.g.)} point out, AI systems require both
training and testing data specific to the types of interactional situations
they are likely to face\sht{, to ensure they will interpret cues correctly and react
appropriately \cite{CastellanoEtAl2015}. They thus point out the need in
affective computing for more datasets drawn from recordings of specific natural
contexts.

The}{. The}}}{The} YFCC100M offers \sht{an opportunity to gather corpora of }emotion expression
occurring in natural situations and recorded under non-ideal conditions, with a
wide variety of contexts and drawn from \sht{a variety of}{many} cultures. \rem{These corpora
could then be annotated and used to train systems to interpret emotion relative
to situational and cultural context, and to produce interactional styles
matched to those contexts and to the culture\rem{ \cite{MascaraEtAl2016}}.
\rem{The fact that a given Flickr user will often have videos of the same people
(such as the user's family members) across many different situations can also
provide an opportunity to isolate individual variation from contextual
variation.
}

\rem{
To demonstrate the importance of context-specificity, Castellano and colleagues
collected data from children playing chess---in this case, with a robot---to
train the robot to interact socially while playing the game
\cite{CastellanoEtAl2017,CastellanoEtAl2013}. 
They found that including context features in the affect recognition component
increased children's engagement with the robot, and that game-specific context
features had a bigger effect than general social context \sht{features.

However,}{features. However,} playing chess with children is obviously only one of thousands or
millions of situations AIs might be called upon to interact in, each with its
own norms and scenarios that might be broken down into context features.
\sht{Whether one is building many purpose-specific AIs or a more general-purpose AI
system, it}{It} would quickly become prohibitively expensive to try to collect a
similarly controlled emotion dataset for every situation.
}

A researcher could use the MMBDS framework to
gather video data on situations of interest, combining tag and location
search with feature-based query-by-example, existing strong labels (e.g., event
labels), and recognizers for, e.g., human faces, speech/non-speech, and
particular relevant objects. 
For common situations such as tourist
interactions, researchers may be able to find enough high-quality videos from
useful angles to constitute a training dataset in itself, or at least enough to
identify which context features to target (i.e., to develop a codebook). 
\rem{For
other situations, YFCC100M data could help researchers prioritize what kind of
controlled data to collect, identify the range of conditions 
they might need to get a representative sample, and find alternatives or
proxies for situations where controlling data collection may be too difficult.}
}

\rem{
\subsection{Construction Site Navigation}

For autonomous driving systems, construction sites are very challenging
\cite{Balali2017}.
Each site looks different, with large differences between countries. 
In addition, it is difficult to get sufficient data.
For relatively predictable features of sites such as warning signs, existing classifiers \cite{Sermanet2011}
could be applied or extended
relatively easily. In the MMBDS framework, more
sophisticated approaches to data-gathering could be implemented to address more challenging and heterogeneous characteristics.

Using archived traffic reports or government data to locate construction sites,
a researcher could then filter out potentially relevant videos and images from the YFCC100M dataset using timestamps and geotags.
This data could be quickly reviewed to verify its relevance and perhaps add additional tags.
\rem{This approach would extract a much bigger dataset
than could be obtained solely using warning-sign classifiers.}
The next step would be to train classifiers and segmentation algorithms
to recognize and characterize construction sites (even if they are unmarked), and to assess the likelihood of various complications and risks.

\rem{As a side note,
this is a good example of a case where maximal applicability requires having human-interpretable algorithms,
so the knowledge gained can be best integrated 
into autonomous driving applications.}
}

\rem{
\subsection{Other Location-Based Applications}
\label{sec:LBSapp}
}

\sht{}{\textbf{\textit{Location-Based Applications:}}} \sht{In addition to studies \sht{of how people differ by location and cultural 
background, such as}{like} those we suggested in \sht{Section~\ref{sec:LBS},}{Section~\ref{sec:sciexx},}
the}{The} YFCC100M location information can \sht{be used in combination
with other data extracted from images/videos for}{aid} several AI applications.
For example, a number of studies have looked at using 
social-media content---including YFCC100M data---to automatically 
generate tourist guides\sht{
\cite{Chen2013,Kofler2011}.}{.} 
\sht{}{UGC data could also supplement training 
for autonomous driving systems, for example in recognizing construction sites.}

\rem{
The studies of \rem{people's} possessions proposed in Section~\ref{sec:LBS} could be extended to 
automatic applications, for example for targeted advertising and market research.
Data on possessions could also be used to generate features and
train an estimator for housing values based on public property-value data, 
then to transfer this classification to
regions that do not have publicly available data on housing values.

Conversely, the dangers posed by multimedia analysis techniques
like automatic valuation and location estimation are an important topic in online privacy,
where researchers are examining and measuring how much private information
users give away when they post, e.g., images and videos \cite{ChoiEtAl2017}.
The potential for new applications like location-aware classification of
\rem{people's} possessions increases those dangers. \sht{After all, such}{Such}
techniques---especially when combined with information from other \rem{online}
sources---could be used not only by marketers
but by criminals, for example to plan a robbery \sht{(called \textit{cybercasing} 
\cite{Friedland2010}).}{\cite{Friedland2010}.}
}

\section{Implementation Case Study: Origami}
\label{s:origami}

\sht{This section describes a practical case study we conducted to analyze the
requirements for the MMBDS framework. To our knowledge, this is the first time
that machine learning has been used in the study of origami.

\subsection{Background: Origami in Science}

\textit{Origami} is the art of paperfolding. 
The term arises specifically from the Japanese tradition, 
but that tradition has been practiced around the world for
several centuries now~\cite{Hatori2011}. In addition to being a recreational
art form, in recent years, origami has increasingly often been
incorporated into the work and
research of mathematicians and engineers. 

For example, in the 1970s, to solve the problem of packing large, flat
membrane structures to be sent into space, Koryo Miura used origami: he
developed a method for collapsing a large flat sheet to a much smaller area so
that collapsing or expanding the sheet would only require pushing or pulling at
the opposite corners of the sheet \cite{Miura1994}.
Similarly, Robert Lang
is using computational origami-based research to help the 
Lawrence Livermore National Laboratory 
design a space telescope lens (``Eyeglass'') that can collapse 
down from $100$ meters in diameter to a size that will fit 
in a rocket roughly
$4$ meters in diameter~\cite{Hyde2002,lang2008flapping}. 
In the field of mathematics, Thomas Hull is investigating enumeration 
of the valid ways 
to fold along a crease pattern (i.e., a diagram containing all the creases 
needed to create a model) 
such that it will lie flat. 
He uses approaches from coloring
in graph theory to solve the problem~\cite{Hull2015}.
In medicine, Kuribayashi et al.\ used
origami to design a metallic heart stent that can easily be threaded
through an artery before expanding where 
it needs to be deployed~\cite{Kuribayashi2006}.

In other words, origami is becoming an established part of science.
To support research on origami,
we decided to generate a large origami dataset, building around data in the 
YFCC100M.}{\sht{This section describes 
a practical case study we conducted}{We conducted a practical case study} to analyze the
requirements for the MMBDS framework. 
\textit{Origami} is the art of paperfolding\sht{. 
The term arises from the Japanese tradition, 
but that tradition has been}{, which is} practiced around the world\rem{ for
several centuries now}. \rem{In addition to being a recreational
art form, in recent years, origami has often been
incorporated into mathematics and engineering.}
To support research on origami,
we \sht{decided to generate}{generated} a large \rem{origami} dataset, building around YFCC100M data. To our knowledge, this is the first time
machine learning has been used \sht{in the study of}{to study} origami.}

\sht{\subsection{Potential Research Question: Regional Differences} 

As our test case for the requirements analysis, we gathered a dataset from the YFCC100M that could be used to answer questions about regional variation, such as:  
What are the differences between countries\sht{/regions}{} in terms of what styles and
subject matter are most popular? How do \sht{those}{the} differences interact with
other paperfolding traditions?

Some traditional approaches to this question might be to look at books or
informational websites about origami from different places, or to contact and
interview experts in those well-known places. However, the books and websites
are unlikely to be comprehensive across regions, and experts might not know
about \rem{(or be concerned about)} origami in all the places it is practiced.
Alternatively, one could travel the world, visiting local communities to gather
data about origami practices, but this would be very expensive and
time-consuming\sht{ (if one could even get funded to do it).}{.}

However, origami can also fruitfully be studied using UGC media. 
\sht{People are often proud of their origami art, 
especially when it is of high difficulty and quality. 
It is therefore the kind of thing that people take pictures of, and upload them
to social media like Flickr.

Using}{Using} the MMBDS framework to target location-specific images of origami, 
data for such a field study could be gathered in a day.
}
{As our test case, we gathered a dataset that could be used to answer questions about regional variation, such as:  
What are the differences between countries in terms of what styles and
subject matter are most popular? How do those differences interact with
other paperfolding traditions? Using the MMBDS framework to target location-specific images of origami, 
data for a field study across many countries could be gathered in a day.}

\subsection{The Limitations of Text-Based Search}
\label{s:browserlimits}

We began by assessing what could be gathered via a simple text metadata search,
using the YFCC100M browser \cite{Kalkowski2015}.
The keyword \textit{origami} netted
more than $13,000$ hits.
However, only about half of the returned images were geotagged, and
we identified several issues with the remainder.

Most obviously, more than $30\%$ of the images did not contain origami.
In addition, the spatial distribution map did not match where
common sense tells us origami should be prevalent. 
The most uploads came from Colombia, 
followed by the U.S.\ and Germany. Japan was in seventh place, and
there were only two examples from China.

This unexpected distribution likely had several causes.
First, there is a general
bias towards the U.S.\ in the YFCC100M \cite{Kalkowski2015}.
\rem{ It was gathered from Flickr,
which is most popular in the U.S.\ and Europe, and less popular elsewhere (and in the case of China, 
was blocked for part of the target period).}
Second, subset
geographical skew can also stem from user bias; in this case, nearly $1,000$ images
were uploaded by one artist \rem{(Jorge Jamarillo)} from Colombia. Finally, a search
on \textit{origami} would not catch examples tagged in Japanese characters.
Media whose metadata is in a different language than the researcher is
searching in will not be included, putting the burden of \rem{language-guessing and }
translation on the researcher.
(Though many Flickr users do include English tags, 
whatever other languages they use \cite{Koochali2016}.) Multilingual search is also limited by
character encoding issues\rem{; at present, a researcher could not search using
Japanese at all}.
This highlights the \rem{general }problem described in Section \ref{s:app}\sht{, that}{:} searching the text metadata 
will miss many examples, most obviously those 
that do not have text metadata at all.

These limitations show that we need a more comprehensive search engine 
that can consider multimedia content. 
As we described in Section \ref{s:engine}, 
the user should be able to create new filters by selecting good examples. 
For a location-based study like this one, 
location estimation could expand the dataset.
Furthermore, the search needs to incorporate translation, 
either of the search terms or the media metadata.
Finally, a science-ready search engine should allow a researcher 
to quantify and visualize bias (such as user bias)
and have a configurable filtering tool
for reducing bias \sht{(see Section \ref{s:select}).}{(see 
Section \ref{s:procfw}).}

\subsection{\sht{Data Processing and Filter Generation}{Data Processing, Filter Generation, and Application}}
\label{s:origamifilter}

\sht{For an effective search, s}{S}tatistics over the metadata \sht{are required}{help}
to \rem{quickly} generate ideas for additional
search terms to include or exclude.
In our case, we narrowed our search 
by using the prominent tags \textit{papiroflexia} (Spanish for
\textit{origami}) and \textit{origamiforum}, and found they were more reliable
than \textit{origami}. We used these terms, plus minimal hand-cleaning, to
collect $1,938$ geotagged origami images for the first part of our
filter-training dataset.\rem{\footnote{Because of our choice of terms, around $25\%$
of this \rem{training} dataset consisted of Colombian images. Again, \rem{ideally,} a
search engine should include sampling tools to \sht{easily ensure that the resulting
detector was not skewed towards a specific country.}{easily reduce bias.}}}

That process highlighted some considerations for the selection and
data-processing framework\rem{ (as described in Section \ref{s:procfw})---and why it
needs to be part of the same system}. \sht{For example, to create a training dataset
for filter generation, we wanted to use images where the origami object was
dominant (rather than, for instance, a person holding \sht{an origami
object).}{origami).} Instead of having to hand-prune the initial results,}{For example, to limit the data
to images where the origami object was
dominant (rather than, for instance, a person holding origami),} it
is much faster to begin with automatic annotations for, e.g., people\rem{
(such as the existing autotags, which we added later)}, or with similarity-based filters.

\sht{We began with our extracted dataset of $1,938$ examples
to analyze the requirements for the process of generating new filters. 
First we applied a VGG16 neural network \cite{Simonyan2014},
trained on ImageNet with $1,000$ common classes 
(not including \textit{origami}).
The top-1 predictions were spread across $263$ different classes, and the
top-5 predictions were spread across $529$ classes.

The most common classes \rem{we found} are summarized in Table~\ref{t:top15}. Our
ground truth origami images were quite often classified as \textit{pinwheel},
\textit{envelope}, \textit{carton}, \textit{paper towel}, \textit{packet}, or
\textit{handkerchief}---all visually (and conceptually) similar to
origami/paperfolding in involving paper. We also noted that \rem{(beyond
\textit{pinwheel})} some images were classified as containing the real-world
objects the origami was supposed to represent, such as bugs or candles---and in
some cases, like flowers, the origami was often difficult to distinguish from
the real object even for a human.

We therefore believe that assigning ImageNet labels to the whole YFCC100M
dataset via VGG16 could be a first step in improving the search function by
allowing multiple filter types. For example, searching for data tagged as
\textit{envelope} by the VGG16 net and \textit{origami} in the text metadata
would probably deliver cleaner results than just searching for
\textit{origami}\sht{. 

However,}{. However,} since VGG16 trained on ImageNet apparently includes
many classes that are at least visually similar to origami, VGG16 features (not
classes) would seem to be the better basis for constructing a \textit{new}
classifier/filter for origami. In general, features from deep learning networks
are quite powerful for image classification \cite[e.g.,][]{Murthy2015}, so are
good candidates for use in our framework.

\sht{\begin{table}[t]
\renewcommand{\arraystretch}{0.9}
\tabcolsep=0.11cm
\small
\caption{Top-i predictions for our extracted YFCC100M subset ($n = $ number of occurrences).}
\label{t:top15}
\begin{center}
\begin{tabular}{|l|r||l|r|c|c|c|c|c||c|c|}
\hline
\multicolumn{2}{|c||}{Top-1} &\multicolumn{2}{c|}{Top-5}\\ 
\multicolumn{1}{|l|}{name}&\multicolumn{1}{c||}{n}
&\multicolumn{1}{c|}{name} &\multicolumn{1}{c|}{n} 
\\ 
\hline 
pinwheel & 243 & envelope & 788\\
envelope & 243 & pinwheel & 518\\
carton & 117 & carton & 499\\
paper towel & 76 & packet & 328\\
honeycomb & 71 & handkerchief & 314\\
lampshade & 41 & paper towel & 302\\
rubber eraser & 39 & rubber eraser & 238\\
handkerchief & 37 & candle & 218\\
pencil sharpener & 35 & lampshade & 192\\
shower cap & 34 & wall clock & 168\\
\hline
\end{tabular}
\end{center}
\end{table}}{
\begin{table}[t]
\renewcommand{\arraystretch}{0.9}
\tabcolsep=0.11cm
\small
\caption{Top-i predictions for our extracted \rem{YFCC100M} subset\rem{ ($n = $ number of occurrences)}.}
\label{t:top15}
\begin{center}
\begin{tabular}{|l|r||l|r|c|c|c|c|c||c|c|}
\hline
\multicolumn{2}{|c||}{Top-1} &\multicolumn{2}{c|}{Top-5}\\ 
\multicolumn{1}{|l|}{name}&\multicolumn{1}{c||}{n}
&\multicolumn{1}{c|}{name} &\multicolumn{1}{c|}{n} 
\\ 
\hline 
pinwheel & 243 & envelope & 788\\
envelope & 243 & pinwheel & 518\\
carton & 117 & carton & 499\\
paper towel & 76 & packet & 328\\
honeycomb & 71 & handkerchief & 314\\
\hline
\end{tabular}
\end{center}
\end{table}
}

\sht{As we pointed out in Section \ref{s:procfw}, if a researcher already has some target images on hand, they can be used t}{T}o improve the filter\rem{. In this case}, we added additional data\rem{,} by scraping
images from two origami-specific databases~\cite{oriwiki,giladorigami}. 
After some
minor cleaning by hand (to remove instructions and \sht{placeholder images}{placeholders}), these
databases yielded $3,934$ and $2,140$ additional origami images. To construct a
non-origami class, we used the ILSVRC2011 validation
data~\cite{Russakovsky2014}.\footnote{We could not use the regular ImageNet
data for the non-origami class because that is what the VGG16 net was trained
on.} We used the first $8$ examples for each label, excluding
\textit{pinwheel}, \textit{envelope}, \textit{carton}, \textit{paper towel},
\textit{packet}, and \textit{handkerchief}. In all, we had $8,011$ images with
\rem{origami} and $7,976$ images without origami.

For features, we used the VGG16 output before the last layer.
VGG16 features are already available for the YFCC100M as part 
of the Multimedia Commons
\cite{Popescu2015}; such precomputed features are essential to speed up processing.
\rem{We generated features for the rest of the data using
MXNet.\footnote{\url{http://mxnet.io}}}
We evaluated the classifier using the pySPACE framework \cite{Krell2013}, 
with a logistic regression 
\rem{implemented in scikit-learn~\cite{Pedregosa2011}}
(default settings) with $5$-fold cross-validation and $5$ repetitions.
The classifier achieved a $97.4\%\pm0.3$ 
balanced accuracy \sht{(BA)~\cite{Straube2014}.}{(BA).} 
}
{To analyze the requirements for generating new filters, we applied a VGG16 net trained on ImageNet. 
We found that the class labels for paper objects could indirectly help narrow the dataset, but exactly because
those classes were so visually similar to origami, we decided that VGG16 features (not classes) would be a
better basis for constructing a \textit{new} classifier.
We used already available VGG16 features for the YFCC100M \cite{Popescu2015}, taking the output before 
the last layer. We also generated VGG16 features for images from two origami-specific databases
(OriWiki and Gilad's Origami). We trained a logistic regression and obtained a $97.4\%\pm0.3$ balanced accuracy (BA) for the resulting classifier.}

\sht{\subsection{Using the New Filter to Gather More Data}
\label{s:useclass}

Applying a new trained neural network to all the YFCC100M images would require 
a tremendous amount of processing. 
On the other hand, our approach---using a simple classifier model and 
precomputed features---enabled the transfer on
a very simple computing instance without a GPU.

In total, our classifier identified $1,960,303$ images as \textit{origami}.
The histogram distribution of the 
classification probability scores was
$[86.9, 5.3, 3.1, 1.5, 1.1, 0.8, 0.7, 0.3, 0.2, 0.1]$ (as \sht{percentages).

Visual}{percentages). Visual} inspection of the highest ranked $87$ origami images (scores $>0.99999$) 
showed only
$2$ incorrect identifications. 
But looking at the lower-scoring images, we found much worse performance.
Even for classification probabilities between $0.99$ and $0.9$, 
visual inspection\rem{ of a subset}
revealed that less than $50\%$ \rem{of the images} contained origami.

This discrepancy shows that it is crucial to allow the user to adjust
the decision boundary for a given filter. It also shows that, at this scale, filtering
will generally need to be above $99\%$ BA in quality.
Having an error of $1\%$ for the 
non-target class
can easily result in millions of misclassifications,
effectively swamping a low number of relevant examples.

Visual inspection also showed that many images were not
photos; this suggests that a pre-supplied photo/non-photo filter would allow for quick weeding.
This could be done using EXIF data (released as a YFCC100M extension) to 
select images that have camera information.
In addition, once the new \textit{origami} filter had been applied, autotags (or other annotations)
could be used to remove types of images that were commonly misclassified as origami\sht{ (in 
this case, people, animals, vehicles, and food)}{.}

\subsection{From Images to Field Study}
\label{s:answers}

For this case study, we extracted only those origami/paperfolding images with location information ($39\%$ of those found)\sht{. 

To}{. To} compare regional variations in origami styles and subjects, the researcher would want to begin by dividing the geotagged data into geographic units. 
For this, we used the YFCC100M Places extension, which has place names based on the GPS coordinates.

We looked at the country distribution of the top $5,167$ images (those scoring higher than $0.99$). 
In total, this high-scoring dataset included images from $178$ countries, with $93$ of those countries having at least $10$ examples for an origami researcher to work with. The top $12$ countries all had more than $100$ images each.
}{
Applying a new trained neural network to all the YFCC100M images would require 
a tremendous amount of processing. 
On the other hand, our approach---using a simple classifier model and 
features that are precomputed for the YFCC100M---enabled the transfer on
a very simple computing instance without a GPU. Adding other features might, of course, lead to even better performance.

Analyzing the results of applying the newly generated filter to the whole YFCC100M, we
identified several additional considerations.
First, the user must be able to adjust the decision boundary for a filter, deciding what classification probability
score.
Second, a BA of at least $99\%$ is required; having an error of more than $1\%$ for the 
non-target class
can easily result in millions of misclassifications,
effectively swamping a low number of relevant examples.
Finally, data cleaning must be supported with strong filters to remove irrelevant images, for example based
on autotags (as we mentioned above) or using EXIF data to remove non-photos.

In the end, the set of 5,167 images with the highest classification probabilities was distributed over 178 countries. Of those, 93 had at least 10 examples for an origami researcher to work with; the top 12 countries all had more than 100 each.
}

\section{Outlook and Call to Action}
\label{s:con}

In this paper, we introduced a cross-disciplinary framework for multimedia big
data studies (MMBDS) and gave a number of motivating examples of past or
potential real-world field studies that could be conducted, replicated,
piloted, or extended cheaply and easily with user-generated multimedia content.
We also described Multimedia Commons Search, the first open-source search for the
YFCC100M and the Multimedia Commons. We encourage researchers to add their contributions
to make the framework even more powerful.

\sht{Scientists (including some of the authors of this paper) are integral to building 
a resource like this. Our discussions}{Our discussions with scientists} about the research topics described in Sections
\ref{sec:sciexx} and \ref{sec:applexx} indicate that there is a high level of interest
in having an MMBDS framework like the one we are developing. These discussions
are already informing the design, and we will continue to involve these and
other scientists to ensure maximum utility and usability.
We view these kinds of discussions as essential 
to shifting the focus of the field from potential impact to actual impact. 
We encourage more multimedia scientists to get in contact
with scientists from other disciplines---from environmental science to 
linguistics to robotics---and vice versa, to build new on-the-ground MMBDS 
collaborations.

\rem{
\section*{Acknowledgments}
We would like to thank 
Per Pascal Grube for his assistance with the AWS search server and
Rick Jaffe and John Lowe for helping us make the 
Solr search engine open source.
We would also like to thank
Angjoo Kanazawa for sharing expertise on 3D pose estimation,
Roman Fedorov on snow index detection,
David M. Romps on cloud coverage estimation, and \sht{}{Michael Ellsworth 
and }Elise Stickles on language and human behavior. Thanks also to
Alan Woodley and anonymous reviewers for providing feedback on the paper.
Finally, we thank all the people who are providing the datasets 
and annotations being integrated into our MMBDS framework.

This work was supported by a fellowship from the FITweltweit program 
of the German Academic Exchange Service (DAAD),
by the Undergraduate Research Apprenticeship Program (URAP) 
at University of California, Berkeley, by grants from the U.S.\ National Science 
Foundation (1251276 and 1629990),
and by a collaborative Laboratory Directed Research \& Development grant led by 
Lawrence Livermore National Laboratory (U.S.\ Dept.\ of Energy contract 
DE-AC52-07NA27344). (Findings and conclusions are those of the authors, and do not necessarily
represent the views of the funders.)
}

\sht{
\bibliographystyle{ACM-Reference-Format}
\bibliography{library,add_library}
}{
\bibliographystyle{IEEEbib} 
\bibliography{shortform_library}


\begin{thebibliography}{00}


\ifx \showCODEN    \undefined \def \showCODEN     #1{\unskip}     \fi
\ifx \showDOI      \undefined \def \showDOI       #1{{\tt DOI:}\penalty0{#1}\ }
  \fi
\ifx \showISBNx    \undefined \def \showISBNx     #1{\unskip}     \fi
\ifx \showISBNxiii \undefined \def \showISBNxiii  #1{\unskip}     \fi
\ifx \showISSN     \undefined \def \showISSN      #1{\unskip}     \fi
\ifx \showLCCN     \undefined \def \showLCCN      #1{\unskip}     \fi
\ifx \shownote     \undefined \def \shownote      #1{#1}          \fi
\ifx \showarticletitle \undefined \def \showarticletitle #1{#1}   \fi
\ifx \showURL      \undefined \def \showURL       #1{#1}          \fi
\providecommand\bibfield[2]{#2}
\providecommand\bibinfo[2]{#2}
\providecommand\natexlab[1]{#1}
\providecommand\showeprint[2][]{arXiv:#2}

\bibitem[\protect\citeauthoryear{Abdelwahab and Busso}{Abdelwahab and
  Busso}{2015}]%
        {AbdelBusso2015}
\bibfield{author}{\bibinfo{person}{Mohammed Abdelwahab} {and}
  \bibinfo{person}{Carlos Busso}.} \bibinfo{year}{2015}\natexlab{}.
\newblock \showarticletitle{Supervised domain adaptation for emotion
  recognition from speech}. In \bibinfo{booktitle}{{\em 2015 {IEEE}
  International Conference on Acoustics, Speech and Signal Processing
  ({ICASSP})}}. \bibinfo{pages}{5058--5062}.
\newblock
\showDOI{%
\url{http://dx.doi.org/10.1109/ICASSP.2015.7178934}}


\bibitem[\protect\citeauthoryear{Aharoni}{Aharoni}{2017}]%
        {giladorigami}
\bibfield{author}{\bibinfo{person}{Gilad Aharoni}.}
  \bibinfo{year}{2017}\natexlab{}.
\newblock \bibinfo{title}{http://www.giladorigami.com/}.
\newblock   (\bibinfo{year}{2017}).
\newblock


\bibitem[\protect\citeauthoryear{Amato, Falchi, Gennaro, and Rabitti}{Amato
  et~al\mbox{.}}{2016}]%
        {AmatoEtAl2016}
\bibfield{author}{\bibinfo{person}{Giuseppe Amato}, \bibinfo{person}{Fabrizio
  Falchi}, \bibinfo{person}{Claudio Gennaro}, {and} \bibinfo{person}{Fausto
  Rabitti}.} \bibinfo{year}{2016}\natexlab{}.
\newblock \showarticletitle{{YFCC100M} {HybridNet} {Fc6} Deep Features for
  Content-Based Image Retrieval}. In \bibinfo{booktitle}{{\em Proceedings of
  the 2016 {ACM} Workshop on the Multimedia {COMMONS} ({MMCommons} '16)}}.
  \bibinfo{publisher}{ACM}, \bibinfo{address}{New York, NY, USA},
  \bibinfo{pages}{11--18}.
\newblock
\showDOI{%
\url{http://dx.doi.org/10.1145/2983554.2983557}}


\bibitem[\protect\citeauthoryear{Argall, Chernova, Veloso, and Browning}{Argall
  et~al\mbox{.}}{2009}]%
        {Argall2009}
\bibfield{author}{\bibinfo{person}{Brenna~D. Argall}, \bibinfo{person}{Sonia
  Chernova}, \bibinfo{person}{Manuela Veloso}, {and} \bibinfo{person}{Brett
  Browning}.} \bibinfo{year}{2009}\natexlab{}.
\newblock \showarticletitle{A survey of robot learning from demonstration}.
\newblock \bibinfo{journal}{{\em Robotics and Autonomous Systems\/}}
  \bibinfo{volume}{57}, \bibinfo{number}{5} (\bibinfo{date}{May}
  \bibinfo{year}{2009}), \bibinfo{pages}{469--483}.
\newblock
\showISSN{09218890}
\showDOI{%
\url{http://dx.doi.org/10.1016/j.robot.2008.10.024}}


\bibitem[\protect\citeauthoryear{Arroyo-Palacios and Romano}{Arroyo-Palacios
  and Romano}{2008}]%
        {ArroyoRomano2008}
\bibfield{author}{\bibinfo{person}{Jorge Arroyo-Palacios} {and}
  \bibinfo{person}{D.M. Romano}.} \bibinfo{year}{2008}\natexlab{}.
\newblock \showarticletitle{Towards a standardization in the use of
  physiological signals for affective recognition systems}. In
  \bibinfo{booktitle}{{\em Proceedings of Measuring Behavior 2008}}.
\newblock


\bibitem[\protect\citeauthoryear{Balali, Jahangiri, and Machiani}{Balali
  et~al\mbox{.}}{2017}]%
        {Balali2017}
\bibfield{author}{\bibinfo{person}{Vahid Balali}, \bibinfo{person}{Arash
  Jahangiri}, {and} \bibinfo{person}{Sahar~Ghanipoor Machiani}.}
  \bibinfo{year}{2017}\natexlab{}.
\newblock \showarticletitle{Multi-class {US} traffic signs {3D} recognition and
  localization via image-based point cloud model using color candidate
  extraction and texture-based recognition}.
\newblock \bibinfo{journal}{{\em Advanced Engineering Informatics\/}}
  \bibinfo{volume}{32} (\bibinfo{year}{2017}), \bibinfo{pages}{263--274}.
\newblock
\showISSN{1474-0346}
\showDOI{%
\url{http://dx.doi.org/https://doi.org/10.1016/j.aei.2017.03.006}}


\bibitem[\protect\citeauthoryear{{B\"{a}nziger}, Mortillaro, and
  Scherer}{{B\"{a}nziger} et~al\mbox{.}}{2012}]%
        {GEMEP}
\bibfield{author}{\bibinfo{person}{Tanja {B\"{a}nziger}},
  \bibinfo{person}{Marcello Mortillaro}, {and} \bibinfo{person}{Klaus~R.
  Scherer}.} \bibinfo{year}{2012}\natexlab{}.
\newblock \showarticletitle{Introducing the {G}eneva Multimodal expression
  corpus for experimental research on emotion perception}.
\newblock \bibinfo{journal}{{\em Emotion\/}} \bibinfo{volume}{12},
  \bibinfo{number}{5} (\bibinfo{date}{October} \bibinfo{year}{2012}),
  \bibinfo{pages}{1161--1179}.
\newblock


\bibitem[\protect\citeauthoryear{Benitez-Quiroz, Srinivasan, and
  Martinez}{Benitez-Quiroz et~al\mbox{.}}{2016}]%
        {BQ16}
\bibfield{author}{\bibinfo{person}{C.~Fabian Benitez-Quiroz},
  \bibinfo{person}{Ramprakash Srinivasan}, {and} \bibinfo{person}{Aleix~M.
  Martinez}.} \bibinfo{year}{2016}\natexlab{}.
\newblock \showarticletitle{{EmotioNet}: An Accurate, Real-Time Algorithm for
  the Automatic Annotation of a Million Facial Expressions in the Wild}. In
  \bibinfo{booktitle}{{\em The {IEEE} Conference on Computer Vision and Pattern
  Recognition ({CVPR})}}. \bibinfo{pages}{5562--5570}.
\newblock


\bibitem[\protect\citeauthoryear{Bernd, Borth, Carrano, Choi, Elizalde,
  Friedland, Gottlieb, Ni, Pearce, Poland, Ashraf, Shamma, and Thomee}{Bernd
  et~al\mbox{.}}{2015a}]%
        {Bernd2015}
\bibfield{author}{\bibinfo{person}{Julia Bernd}, \bibinfo{person}{Damian
  Borth}, \bibinfo{person}{Carmen Carrano}, \bibinfo{person}{Jaeyoung Choi},
  \bibinfo{person}{Benjamin Elizalde}, \bibinfo{person}{Gerald Friedland},
  \bibinfo{person}{Luke Gottlieb}, \bibinfo{person}{Karl Ni},
  \bibinfo{person}{Roger Pearce}, \bibinfo{person}{Doug Poland},
  \bibinfo{person}{Khalid Ashraf}, \bibinfo{person}{David~A. Shamma}, {and}
  \bibinfo{person}{Bart Thomee}.} \bibinfo{year}{2015}\natexlab{a}.
\newblock \showarticletitle{Kickstarting the {C}ommons: The {YFCC100M} and the
  {YLI} Corpora}. In \bibinfo{booktitle}{{\em Proceedings of the 2015 Workshop
  on Community-Organized Multimodal Mining: Opportunities for Novel Solutions
  (MMCommons '15)}}. \bibinfo{publisher}{ACM}, \bibinfo{pages}{1--6}.
\newblock
\showDOI{%
\url{http://dx.doi.org/10.1145/2814815.2816986}}


\bibitem[\protect\citeauthoryear{Bernd, Borth, Elizalde, Friedland, Gallagher,
  Gottlieb, Janin, Karabashlieva, Takahashi, and Won}{Bernd
  et~al\mbox{.}}{2015b}]%
        {Bernd2015a}
\bibfield{author}{\bibinfo{person}{Julia Bernd}, \bibinfo{person}{Damian
  Borth}, \bibinfo{person}{Benjamin Elizalde}, \bibinfo{person}{Gerald
  Friedland}, \bibinfo{person}{Heather Gallagher}, \bibinfo{person}{Luke
  Gottlieb}, \bibinfo{person}{Adam Janin}, \bibinfo{person}{Sara
  Karabashlieva}, \bibinfo{person}{Jocelyn Takahashi}, {and}
  \bibinfo{person}{Jennifer Won}.} \bibinfo{year}{2015}\natexlab{b}.
\newblock \bibinfo{booktitle}{{\em The {YLI-MED} Corpus: Characteristics,
  Procedures, and Plans}}.
\newblock \bibinfo{type}{{T}echnical {R}eport} {TR}-15-001.
  \bibinfo{institution}{International Computer Science Institute}.
\newblock
\showeprint[arxiv]{1503.04250}
\showURL{%
\url{http://arxiv.org/abs/1503.04250}}
\newblock
\shownote{{arXiv}:1503.04250.}


\bibitem[\protect\citeauthoryear{Bogo, Kanazawa, Lassner, Gehler, Romero, and
  Black}{Bogo et~al\mbox{.}}{2016}]%
        {Bogo2016}
\bibfield{author}{\bibinfo{person}{Federica Bogo}, \bibinfo{person}{Angjoo
  Kanazawa}, \bibinfo{person}{Christoph Lassner}, \bibinfo{person}{Peter
  Gehler}, \bibinfo{person}{Javier Romero}, {and} \bibinfo{person}{Michael~J.
  Black}.} \bibinfo{year}{2016}\natexlab{}.
\newblock \showarticletitle{{Keep it SMPL : Automatic Estimation of 3D Human
  Pose and Shape from a Single Image}}. In \bibinfo{booktitle}{{\em Computer
  Vision -- ECCV 2016}}. \bibinfo{publisher}{Springer International
  Publishing}, \bibinfo{pages}{34--36}.
\newblock
\showISBNx{9783319464534}
\showISSN{16113349}
\showDOI{%
\url{http://dx.doi.org/10.1007/978-3-319-46454-1_34}}
\showeprint[arxiv]{1607.08128}


\bibitem[\protect\citeauthoryear{Bowerman}{Bowerman}{1996}]%
        {Bowerman1996}
\bibfield{author}{\bibinfo{person}{Melissa Bowerman}.}
  \bibinfo{year}{1996}\natexlab{}.
\newblock \showarticletitle{Learning how to structure space for language: A
  crosslinguistic perspective}.
\newblock In \bibinfo{booktitle}{{\em Language and Space}},
  \bibfield{editor}{\bibinfo{person}{Paul Bloom}} (Ed.).
  \bibinfo{publisher}{MIT Press}, \bibinfo{pages}{385--436}.
\newblock


\bibitem[\protect\citeauthoryear{Bowerman and Choi}{Bowerman and Choi}{2003}]%
        {BowermanChoi2003}
\bibfield{author}{\bibinfo{person}{Melissa Bowerman} {and}
  \bibinfo{person}{Soonja Choi}.} \bibinfo{year}{2003}\natexlab{}.
\newblock \showarticletitle{Space under construction: Language-specific spatial
  categorization in first language acquisition}.
\newblock In \bibinfo{booktitle}{{\em Language in Mind: Advances in the Study
  of Language and Thought}}, \bibfield{editor}{\bibinfo{person}{Dedre Gentner}
  {and} \bibinfo{person}{Susan Goldin-Meadow}} (Eds.). \bibinfo{publisher}{MIT
  Press}, \bibinfo{address}{Cambridge, MA}, \bibinfo{pages}{387--428}.
\newblock


\bibitem[\protect\citeauthoryear{Busso, Bulut, and Narayanan}{Busso
  et~al\mbox{.}}{2013}]%
        {BussoEtAl2013}
\bibfield{author}{\bibinfo{person}{Carlos Busso}, \bibinfo{person}{Murtaza
  Bulut}, {and} \bibinfo{person}{Shrikanth Narayanan}.}
  \bibinfo{year}{2013}\natexlab{}.
\newblock \showarticletitle{Toward effective automatic recognition systems of
  emotion in speech}.
\newblock In \bibinfo{booktitle}{{\em Social Emotions in Nature and Artifact:
  Emotions in human and human-computer interaction}},
  \bibfield{editor}{\bibinfo{person}{Jonathan Gratch} {and}
  \bibinfo{person}{Stacy Marsella}} (Eds.). \bibinfo{publisher}{Oxford
  University Press}, \bibinfo{pages}{110--127}.
\newblock


\bibitem[\protect\citeauthoryear{Campbell}{Campbell}{2014}]%
        {CREST}
\bibfield{author}{\bibinfo{person}{Nick Campbell}.}
  \bibinfo{year}{2014}\natexlab{}.
\newblock \showarticletitle{Databases of Expressive Speech}.
\newblock \bibinfo{journal}{{\em Journal of Chinese Language and Computing\/}}
  \bibinfo{volume}{14}, \bibinfo{number}{4} (\bibinfo{year}{2014}).
\newblock


\bibitem[\protect\citeauthoryear{Cao, Verma, and Nenkova}{Cao
  et~al\mbox{.}}{2015}]%
        {CaoEtAl2015}
\bibfield{author}{\bibinfo{person}{Houwei Cao}, \bibinfo{person}{Ragini Verma},
  {and} \bibinfo{person}{Ani Nenkova}.} \bibinfo{year}{2015}\natexlab{}.
\newblock \showarticletitle{Speaker-sensitive emotion recognition via ranking:
  Studies on acted and spontaneous speech}.
\newblock \bibinfo{journal}{{\em Computer Speech \& Language\/}}
  \bibinfo{volume}{29}, \bibinfo{number}{1} (\bibinfo{year}{2015}),
  \bibinfo{pages}{186--202}.
\newblock
\showDOI{%
\url{http://dx.doi.org/https://doi.org/10.1016/j.csl.2014.01.003}}


\bibitem[\protect\citeauthoryear{Cao, Simon, Wei, and Sheikh}{Cao
  et~al\mbox{.}}{2016}]%
        {Cao2016}
\bibfield{author}{\bibinfo{person}{Zhe Cao}, \bibinfo{person}{Tomas Simon},
  \bibinfo{person}{Shih-En Wei}, {and} \bibinfo{person}{Yaser Sheikh}.}
  \bibinfo{year}{2016}\natexlab{}.
\newblock \bibinfo{title}{{Realtime Multi-Person 2D Pose Estimation using Part
  Affinity Fields}}.
\newblock   (\bibinfo{date}{Nov} \bibinfo{year}{2016}).
\newblock
\showeprint[arxiv]{1611.08050}
\showURL{%
\url{http://arxiv.org/abs/1611.08050}}


\bibitem[\protect\citeauthoryear{Carrano and Poland}{Carrano and
  Poland}{2015}]%
        {CarranoCASIS}
\bibfield{author}{\bibinfo{person}{Carmen Carrano} {and} \bibinfo{person}{Doug
  Poland}.} \bibinfo{year}{2015}\natexlab{}.
\newblock \showarticletitle{Ad Hoc Video Query and Retrieval Using Multi-Modal
  Feature Graphs}. In \bibinfo{booktitle}{{\em Proceedings of the {19th Annual
  Signal \& Image Sciences Workshop} ({CASIS Workshop})}}.
\newblock
\showURL{%
\url{https://casis.llnl.gov/content/pages/casis-2015/docs/poster/Carrano_CASIS_2015.pdf}}


\bibitem[\protect\citeauthoryear{Castellano, Gunes, Peters, and
  Schuller}{Castellano et~al\mbox{.}}{2015}]%
        {CastellanoEtAl2015}
\bibfield{author}{\bibinfo{person}{Ginevra Castellano}, \bibinfo{person}{Hatice
  Gunes}, \bibinfo{person}{Christopher Peters}, {and}
  \bibinfo{person}{Bj\"{o}rn Schuller}.} \bibinfo{year}{2015}\natexlab{}.
\newblock \showarticletitle{Multimodal Affect Recognition for Naturalistic
  Human-Computer and Human-Robot Interactions}.
\newblock In \bibinfo{booktitle}{{\em The {O}xford Handbook of Affective
  Computing}}, \bibfield{editor}{\bibinfo{person}{Rafael Calvo},
  \bibinfo{person}{Sidney D'Mello}, \bibinfo{person}{Jonathan Gratch}, {and}
  \bibinfo{person}{Arvid Kappas}} (Eds.). \bibinfo{publisher}{Oxford University
  Press}, \bibinfo{address}{New York}, \bibinfo{pages}{246--257}.
\newblock
\showDOI{%
\url{http://dx.doi.org/10.1093/oxfordhb/9780199942237.013.026}}


\bibitem[\protect\citeauthoryear{Castellano, Leite, and Paiva}{Castellano
  et~al\mbox{.}}{2017}]%
        {CastellanoEtAl2017}
\bibfield{author}{\bibinfo{person}{Ginevra Castellano},
  \bibinfo{person}{Iolanda Leite}, {and} \bibinfo{person}{Ana Paiva}.}
  \bibinfo{year}{2017}\natexlab{}.
\newblock \showarticletitle{Detecting perceived quality of interaction with a
  robot using contextual features}.
\newblock \bibinfo{journal}{{\em Autonomous Robots\/}} \bibinfo{volume}{41},
  \bibinfo{number}{5} (\bibinfo{year}{2017}), \bibinfo{pages}{1245--1261}.
\newblock
\showDOI{%
\url{http://dx.doi.org/10.1007/s10514-016-9592-y}}


\bibitem[\protect\citeauthoryear{Castellano, Leite, Pereira, Martinho, Paiva,
  and McOwan}{Castellano et~al\mbox{.}}{2012}]%
        {CastellanoEtAl2012}
\bibfield{author}{\bibinfo{person}{Ginevra Castellano},
  \bibinfo{person}{Iolanda Leite}, \bibinfo{person}{Andr\'{e} Pereira},
  \bibinfo{person}{Carlos Martinho}, \bibinfo{person}{Ana Paiva}, {and}
  \bibinfo{person}{Peter~W. McOwan}.} \bibinfo{year}{2012}\natexlab{}.
\newblock \showarticletitle{Detecting Engagement in {HRI}: An Exploration of
  Social and Task-Based Context}. In \bibinfo{booktitle}{{\em 2012
  International Conference on Privacy, Security, Risk, and Trust and 2012
  International Confernece on Social Computing}}. \bibinfo{pages}{421--428}.
\newblock
\showDOI{%
\url{http://dx.doi.org/10.1109/SocialCom-PASSAT.2012.51}}


\bibitem[\protect\citeauthoryear{Castellano, Leite, Pereira, Martinho, Paiva,
  and McOwan}{Castellano et~al\mbox{.}}{2013}]%
        {CastellanoEtAl2013}
\bibfield{author}{\bibinfo{person}{Ginevra Castellano},
  \bibinfo{person}{Iolanda Leite}, \bibinfo{person}{Andre Pereira},
  \bibinfo{person}{Carlos Martinho}, \bibinfo{person}{Ana Paiva}, {and}
  \bibinfo{person}{Peter~W. McOwan}.} \bibinfo{year}{2013}\natexlab{}.
\newblock \showarticletitle{Multimodal affect modelling and recognition for
  empathic robot companions}.
\newblock \bibinfo{journal}{{\em International Journal of Humanoid Robotics\/}}
  \bibinfo{volume}{10}, \bibinfo{number}{1} (\bibinfo{year}{2013}),
  \bibinfo{pages}{1--23}.
\newblock


\bibitem[\protect\citeauthoryear{Castelletti, Fedorov, Fraternali, and
  Giuliani}{Castelletti et~al\mbox{.}}{2016}]%
        {Castelletti2016}
\bibfield{author}{\bibinfo{person}{Andrea Castelletti}, \bibinfo{person}{Roman
  Fedorov}, \bibinfo{person}{Piero Fraternali}, {and} \bibinfo{person}{Matteo
  Giuliani}.} \bibinfo{year}{2016}\natexlab{}.
\newblock \showarticletitle{Multimedia on the Mountaintop: Using Public Snow
  Images to Improve Water Systems Operation}. In \bibinfo{booktitle}{{\em
  Proceedings of the 2016 ACM Conference on Multimedia - MM '16}}.
  \bibinfo{publisher}{ACM Press}, \bibinfo{address}{New York, New York, USA},
  \bibinfo{pages}{948--957}.
\newblock
\showISBNx{9781450336031}
\showDOI{%
\url{http://dx.doi.org/10.1145/2964284.2976759}}


\bibitem[\protect\citeauthoryear{Chen and Ramanan}{Chen and Ramanan}{2017}]%
        {Chen2017}
\bibfield{author}{\bibinfo{person}{Ching-Hang Chen} {and} \bibinfo{person}{Deva
  Ramanan}.} \bibinfo{year}{2017}\natexlab{}.
\newblock \bibinfo{title}{{3D Human Pose Estimation = 2D Pose Estimation +
  Matching}}.
\newblock   (\bibinfo{date}{Dec} \bibinfo{year}{2017}).
\newblock
\showeprint[arxiv]{1612.06524}
\showURL{%
\url{http://arxiv.org/abs/1612.06524}}


\bibitem[\protect\citeauthoryear{Chen, Cheng, and Hsu}{Chen
  et~al\mbox{.}}{2013}]%
        {Chen2013}
\bibfield{author}{\bibinfo{person}{Yan-Ying Chen}, \bibinfo{person}{An-Jung
  Cheng}, {and} \bibinfo{person}{Winston~H. Hsu}.}
  \bibinfo{year}{2013}\natexlab{}.
\newblock \showarticletitle{Travel Recommendation by Mining People Attributes
  and Travel Group Types From Community-Contributed Photos}.
\newblock \bibinfo{journal}{{\em IEEE Transactions on Multimedia\/}}
  \bibinfo{volume}{15}, \bibinfo{number}{6} (\bibinfo{date}{Oct}
  \bibinfo{year}{2013}), \bibinfo{pages}{1283--1295}.
\newblock
\showISSN{1520-9210}
\showDOI{%
\url{http://dx.doi.org/10.1109/TMM.2013.2265077}}


\bibitem[\protect\citeauthoryear{Choi and Friedland}{Choi and
  Friedland}{2015}]%
        {Choi2015a}
\bibfield{editor}{\bibinfo{person}{Jaeyoung Choi} {and} \bibinfo{person}{Gerald
  Friedland}} (Eds.). \bibinfo{year}{2015}\natexlab{}.
\newblock \bibinfo{booktitle}{{\em Multimodal Location Estimation of Videos and
  Images}}.
\newblock \bibinfo{publisher}{Springer International Publishing},
  \bibinfo{address}{Cham}.
\newblock
\showISBNx{9783319098616}
\showDOI{%
\url{http://dx.doi.org/10.1007/978-3-319-09861-6}}


\bibitem[\protect\citeauthoryear{Choi, Larson, Li, Li, Friedland, and
  Hanjalic}{Choi et~al\mbox{.}}{2017}]%
        {ChoiEtAl2017}
\bibfield{author}{\bibinfo{person}{Jaeyoung Choi}, \bibinfo{person}{Martha
  Larson}, \bibinfo{person}{Xinchao Li}, \bibinfo{person}{Kevin Li},
  \bibinfo{person}{Gerald Friedland}, {and} \bibinfo{person}{Alan Hanjalic}.}
  \bibinfo{year}{2017}\natexlab{}.
\newblock \showarticletitle{The Geo-Privacy Bonus of Popular Photo
  Enhancements}. In \bibinfo{booktitle}{{\em Proceedings of the 2017 {ACM}
  International Conference on Multimedia Retrieval ({ICMR} '17)}}.
  \bibinfo{publisher}{ACM}, \bibinfo{address}{New York, NY, USA},
  \bibinfo{pages}{84--92}.
\newblock
\showDOI{%
\url{http://dx.doi.org/10.1145/3078971.3080543}}


\bibitem[\protect\citeauthoryear{Choi, Pearce, Poland, Thomee, Friedland, Cao,
  Ni, Borth, Elizalde, Gottlieb, and Carrano}{Choi et~al\mbox{.}}{2014}]%
        {Choi2014}
\bibfield{author}{\bibinfo{person}{Jaeyoung Choi}, \bibinfo{person}{Roger
  Pearce}, \bibinfo{person}{Doug Poland}, \bibinfo{person}{Bart Thomee},
  \bibinfo{person}{Gerald Friedland}, \bibinfo{person}{Liangliang Cao},
  \bibinfo{person}{Karl Ni}, \bibinfo{person}{Damian Borth},
  \bibinfo{person}{Benjamin Elizalde}, \bibinfo{person}{Luke Gottlieb}, {and}
  \bibinfo{person}{Carmen Carrano}.} \bibinfo{year}{2014}\natexlab{}.
\newblock \showarticletitle{The {P}lacing {T}ask: A Large-Scale Geo-Estimation
  Challenge for Social-Media Videos and Images}. In \bibinfo{booktitle}{{\em
  Proceedings of the 3rd {ACM Multimedia} Workshop on Geotagging and Its
  Applications in Multimedia ({GeoMM}'14)}}. \bibinfo{publisher}{ACM Press},
  \bibinfo{address}{New York}, \bibinfo{pages}{27--31}.
\newblock
\showISBNx{9781450331272}
\showDOI{%
\url{http://dx.doi.org/10.1145/2661118.2661125}}


\bibitem[\protect\citeauthoryear{Choi and Bowerman}{Choi and Bowerman}{1991}]%
        {ChoiBowerman1991}
\bibfield{author}{\bibinfo{person}{Soonja Choi} {and} \bibinfo{person}{Melissa
  Bowerman}.} \bibinfo{year}{1991}\natexlab{}.
\newblock \showarticletitle{Learning to express motion events in {E}nglish and
  {K}orean: The influence of language-specific lexicalization patterns}.
\newblock \bibinfo{journal}{{\em Cognition\/}}  \bibinfo{volume}{41}
  (\bibinfo{year}{1991}), \bibinfo{pages}{83--121}.
\newblock


\bibitem[\protect\citeauthoryear{Cimpoi, Maji, and Vedaldi}{Cimpoi
  et~al\mbox{.}}{2015}]%
        {Cimpoi2015}
\bibfield{author}{\bibinfo{person}{Mircea Cimpoi}, \bibinfo{person}{Subhransu
  Maji}, {and} \bibinfo{person}{Andrea Vedaldi}.}
  \bibinfo{year}{2015}\natexlab{}.
\newblock \showarticletitle{Deep filter banks for texture recognition and
  segmentation}. In \bibinfo{booktitle}{{\em 2015 IEEE Conference on Computer
  Vision and Pattern Recognition (CVPR)}}. \bibinfo{publisher}{IEEE},
  \bibinfo{pages}{3828--3836}.
\newblock
\showISBNx{978-1-4673-6964-0}
\showDOI{%
\url{http://dx.doi.org/10.1109/CVPR.2015.7299007}}


\bibitem[\protect\citeauthoryear{Claus, Rosales, Petrick, Hain, and
  Hoffmann}{Claus et~al\mbox{.}}{2013}]%
        {claus2013survey}
\bibfield{author}{\bibinfo{person}{Felix Claus},
  \bibinfo{person}{Hamurabi~Gamboa Rosales}, \bibinfo{person}{Rico Petrick},
  \bibinfo{person}{Horst-Udo Hain}, {and} \bibinfo{person}{R{\"u}diger
  Hoffmann}.} \bibinfo{year}{2013}\natexlab{}.
\newblock \showarticletitle{A survey about databases of children's speech}. In
  \bibinfo{booktitle}{{\em Proceedings of INTERSPEECH}}.
  \bibinfo{pages}{2410--2414}.
\newblock


\bibitem[\protect\citeauthoryear{Clements, Serdyukov, de~Vries, and
  Reinders}{Clements et~al\mbox{.}}{2010}]%
        {Clements2010}
\bibfield{author}{\bibinfo{person}{Maarten Clements}, \bibinfo{person}{Pavel
  Serdyukov}, \bibinfo{person}{Arjen~P. de Vries}, {and}
  \bibinfo{person}{Marcel~J.T. Reinders}.} \bibinfo{year}{2010}\natexlab{}.
\newblock \showarticletitle{Using {F}lickr geotags to predict user travel
  behaviour}. In \bibinfo{booktitle}{{\em Proceedings of the 33rd International
  {ACM SIGIR} Conference on Research and Development in Information Retrieval
  ({SIGIR} '10)}}. \bibinfo{publisher}{ACM Press}, \bibinfo{address}{New York,
  New York, USA}, \bibinfo{pages}{851}.
\newblock
\showISBNx{9781450301534}
\showDOI{%
\url{http://dx.doi.org/10.1145/1835449.1835648}}


\bibitem[\protect\citeauthoryear{Cohn and Schmidt}{Cohn and Schmidt}{2004}]%
        {CohnSchmidt2004}
\bibfield{author}{\bibinfo{person}{Jeffrey Cohn} {and} \bibinfo{person}{Karen
  Schmidt}.} \bibinfo{year}{2004}\natexlab{}.
\newblock \showarticletitle{The timing of facial motion in posed and
  spontaneous smiles}.
\newblock \bibinfo{journal}{{\em International Journal of Wavelets,
  Multiresolution, and Information Processing\/}} \bibinfo{volume}{2},
  \bibinfo{number}{2} (\bibinfo{date}{March} \bibinfo{year}{2004}),
  \bibinfo{pages}{1--12}.
\newblock
\showURL{%
\url{http://www.worldscientific.com/doi/abs/10.1142/S021969130400041X?journalCode=ijwmip}}


\bibitem[\protect\citeauthoryear{Congdon, Novack, and Goldin-Meadow}{Congdon
  et~al\mbox{.}}{2016}]%
        {Eliza2016}
\bibfield{author}{\bibinfo{person}{Eliza~L. Congdon},
  \bibinfo{person}{Miriam~A. Novack}, {and} \bibinfo{person}{Susan
  Goldin-Meadow}.} \bibinfo{year}{2016}\natexlab{}.
\newblock \showarticletitle{Gesture in Experimental Studies}.
\newblock \bibinfo{journal}{{\em Organizational Research Methods\/}}
  (\bibinfo{year}{2016}), \bibinfo{pages}{1094428116654548}.
\newblock
\showDOI{%
\url{http://dx.doi.org/10.1177/1094428116654548}}
\showeprint{http://dx.doi.org/10.1177/1094428116654548}


\bibitem[\protect\citeauthoryear{Connors, Lei, and Kelly}{Connors
  et~al\mbox{.}}{2012}]%
        {Connors2012}
\bibfield{author}{\bibinfo{person}{John~Patrick Connors},
  \bibinfo{person}{Shufei Lei}, {and} \bibinfo{person}{Maggi Kelly}.}
  \bibinfo{year}{2012}\natexlab{}.
\newblock \showarticletitle{Citizen Science in the Age of Neogeography:
  Utilizing Volunteered Geographic Information for Environmental Monitoring}.
\newblock \bibinfo{journal}{{\em Annals of the Association of American
  Geographers\/}} \bibinfo{volume}{102}, \bibinfo{number}{6}
  (\bibinfo{date}{Nov} \bibinfo{year}{2012}), \bibinfo{pages}{1267--1289}.
\newblock
\showISSN{0004-5608}
\showDOI{%
\url{http://dx.doi.org/10.1080/00045608.2011.627058}}


\bibitem[\protect\citeauthoryear{Cowie, Douglas-Cowie, and Cox}{Cowie
  et~al\mbox{.}}{2005}]%
        {CowieCowieCox2005}
\bibfield{author}{\bibinfo{person}{Roddy Cowie}, \bibinfo{person}{Ellen
  Douglas-Cowie}, {and} \bibinfo{person}{Cate Cox}.}
  \bibinfo{year}{2005}\natexlab{}.
\newblock \showarticletitle{Beyond emotion archetypes: Databases for emotion
  modelling using neural networks}.
\newblock \bibinfo{journal}{{\em Neural Networks\/}} \bibinfo{volume}{18},
  \bibinfo{number}{4} (\bibinfo{year}{2005}), \bibinfo{pages}{371--388}.
\newblock
\showDOI{%
\url{http://dx.doi.org/https://doi.org/10.1016/j.neunet.2005.03.002}}


\bibitem[\protect\citeauthoryear{de~Le\'{o}n}{de~Le\'{o}n}{2001}]%
        {deLeon2001}
\bibfield{author}{\bibinfo{person}{Lourdes de Le\'{o}n}.}
  \bibinfo{year}{2001}\natexlab{}.
\newblock \showarticletitle{Finding the richest path: Language and cognition in
  the acquisition of verticality in {T}zotzil ({M}ayan)}.
\newblock In \bibinfo{booktitle}{{\em Language Acquisition and Conceptual
  Development}}, \bibfield{editor}{\bibinfo{person}{Melissa Bowerman} {and}
  \bibinfo{person}{Stephen Levinson}} (Eds.). \bibinfo{publisher}{Cambridge
  University Press}, \bibinfo{pages}{544--565}.
\newblock
\showDOI{%
\url{http://dx.doi.org/10.1017/CBO9780511620669.020}}


\bibitem[\protect\citeauthoryear{Devillers, Vidrascu, and Lamel}{Devillers
  et~al\mbox{.}}{2005}]%
        {DevilEtAl2005}
\bibfield{author}{\bibinfo{person}{Laurence Devillers},
  \bibinfo{person}{Laurence Vidrascu}, {and} \bibinfo{person}{Lori Lamel}.}
  \bibinfo{year}{2005}\natexlab{}.
\newblock \showarticletitle{Challenges in real-life emotion annotation and
  machine learning based detection}.
\newblock \bibinfo{journal}{{\em Neural Networks\/}} \bibinfo{volume}{18},
  \bibinfo{number}{4} (\bibinfo{year}{2005}), \bibinfo{pages}{407--422}.
\newblock
\showDOI{%
\url{http://dx.doi.org/10.1016/j.neunet.2005.03.007}}
\newblock
\shownote{Emotion and Brain.}


\bibitem[\protect\citeauthoryear{D'Mello and Kory}{D'Mello and Kory}{2012}]%
        {DMelloKory2012}
\bibfield{author}{\bibinfo{person}{Sidney D'Mello} {and}
  \bibinfo{person}{Jacqueline Kory}.} \bibinfo{year}{2012}\natexlab{}.
\newblock \showarticletitle{Consistent but Modest: A Meta-analysis on Unimodal
  and Multimodal Affect Detection Accuracies from 30 Studies}. In
  \bibinfo{booktitle}{{\em Proceedings of the 14th {ACM} International
  Conference on Multimodal Interaction ({ICMI} '12)}}.
  \bibinfo{publisher}{ACM}, \bibinfo{address}{New York, NY, USA},
  \bibinfo{pages}{31--38}.
\newblock
\showISBNx{978-1-4503-1467-1}
\showDOI{%
\url{http://dx.doi.org/10.1145/2388676.2388686}}


\bibitem[\protect\citeauthoryear{Douglas-Cowie, Campbell, Cowie, and
  Roach}{Douglas-Cowie et~al\mbox{.}}{2003}]%
        {Belfast}
\bibfield{author}{\bibinfo{person}{Ellen Douglas-Cowie}, \bibinfo{person}{Nick
  Campbell}, \bibinfo{person}{Roddy Cowie}, {and} \bibinfo{person}{Peter
  Roach}.} \bibinfo{year}{2003}\natexlab{}.
\newblock \showarticletitle{Emotional speech: Towards a new generation of
  databases}.
\newblock \bibinfo{journal}{{\em Speech Communication\/}} \bibinfo{volume}{40},
  \bibinfo{number}{1--2} (\bibinfo{year}{2003}), \bibinfo{pages}{33--60}.
\newblock
\showDOI{%
\url{http://dx.doi.org/https://doi.org/10.1016/S0167-6393(02)00070-5}}


\bibitem[\protect\citeauthoryear{Douglas-Cowie, Devillers, Martin, Cowie,
  Savvidou, Abrilian, and Cox}{Douglas-Cowie et~al\mbox{.}}{2005}]%
        {CowieEtAl2005}
\bibfield{author}{\bibinfo{person}{Ellen Douglas-Cowie},
  \bibinfo{person}{Laurence Devillers}, \bibinfo{person}{Jean-Claude Martin},
  \bibinfo{person}{Roddy Cowie}, \bibinfo{person}{Suzie Savvidou},
  \bibinfo{person}{Sarkis Abrilian}, {and} \bibinfo{person}{Cate Cox}.}
  \bibinfo{year}{2005}\natexlab{}.
\newblock \showarticletitle{Multimodal databases of everyday emotion: facing up
  to complexity}. In \bibinfo{booktitle}{{\em Proceedings of {INTERSPEECH}}}.
  \bibinfo{pages}{813--816}.
\newblock


\bibitem[\protect\citeauthoryear{Eastman, Warren, Eastman, and Warren}{Eastman
  et~al\mbox{.}}{2013}]%
        {Eastman2013}
\bibfield{author}{\bibinfo{person}{Ryan Eastman}, \bibinfo{person}{Stephen~G.
  Warren}, \bibinfo{person}{Ryan Eastman}, {and} \bibinfo{person}{Stephen~G.
  Warren}.} \bibinfo{year}{2013}\natexlab{}.
\newblock \showarticletitle{{A 39-Yr Survey of Cloud Changes from Land Stations
  Worldwide 1971--2009: Long-Term Trends, Relation to Aerosols, and Expansion
  of the Tropical Belt}}.
\newblock \bibinfo{journal}{{\em Journal of Climate\/}} \bibinfo{volume}{26},
  \bibinfo{number}{4} (\bibinfo{date}{Feb} \bibinfo{year}{2013}),
  \bibinfo{pages}{1286--1303}.
\newblock
\showISSN{0894-8755}
\showDOI{%
\url{http://dx.doi.org/10.1175/JCLI-D-12-00280.1}}


\bibitem[\protect\citeauthoryear{Ekman and Friesen}{Ekman and Friesen}{1971}]%
        {EkmanFriesen1971}
\bibfield{author}{\bibinfo{person}{Paul Ekman} {and} \bibinfo{person}{Wallace~V
  Friesen}.} \bibinfo{year}{1971}\natexlab{}.
\newblock \showarticletitle{Constants across cultures in the face and emotion}.
\newblock \bibinfo{journal}{{\em Journal of Personality and Social
  Psychology\/}}  \bibinfo{volume}{17} (\bibinfo{year}{1971}),
  \bibinfo{pages}{124--129}.
\newblock


\bibitem[\protect\citeauthoryear{Ekman and Friesen}{Ekman and Friesen}{1986}]%
        {EkmanFriesen1986}
\bibfield{author}{\bibinfo{person}{Paul Ekman} {and} \bibinfo{person}{Wallace~V
  Friesen}.} \bibinfo{year}{1986}\natexlab{}.
\newblock \showarticletitle{A new pan-cultural facial expression of emotion}.
\newblock \bibinfo{journal}{{\em Motivation and Emotion\/}}
  \bibinfo{volume}{10} (\bibinfo{year}{1986}), \bibinfo{pages}{159--168}.
\newblock


\bibitem[\protect\citeauthoryear{Elfenbein and Ambady}{Elfenbein and
  Ambady}{2002}]%
        {ElfenbeinAmbady2002}
\bibfield{author}{\bibinfo{person}{Hillary~Anger Elfenbein} {and}
  \bibinfo{person}{Nalini Ambady}.} \bibinfo{year}{2002}\natexlab{}.
\newblock \showarticletitle{On the Universality and Cultural Specificity of
  Emotion Recognition: A Meta-Analysis}.
\newblock \bibinfo{journal}{{\em Psychological Bulletin\/}}
  \bibinfo{volume}{128}, \bibinfo{number}{2} (\bibinfo{year}{2002}),
  \bibinfo{pages}{203--235}.
\newblock


\bibitem[\protect\citeauthoryear{Endrass, Haering, Akila, and
  Andr{\'e}}{Endrass et~al\mbox{.}}{2014}]%
        {EndrassEtAl2014}
\bibfield{author}{\bibinfo{person}{Birgit Endrass}, \bibinfo{person}{Markus
  Haering}, \bibinfo{person}{Gasser Akila}, {and} \bibinfo{person}{Elisabeth
  Andr{\'e}}.} \bibinfo{year}{2014}\natexlab{}.
\newblock \showarticletitle{Simulating Deceptive Cues of Joy in Humanoid
  Robots}. In \bibinfo{booktitle}{{\em Proceedings of the 14th International
  Conference on Intelligent Virtual Agents ({IVA} 2014)}},
  \bibfield{editor}{\bibinfo{person}{Timothy Bickmore}, \bibinfo{person}{Stacy
  Marsella}, {and} \bibinfo{person}{Candace Sidner}} (Eds.).
  \bibinfo{publisher}{Springer International Publishing},
  \bibinfo{address}{Cham}, \bibinfo{pages}{174--177}.
\newblock
\showDOI{%
\url{http://dx.doi.org/10.1007/978-3-319-09767-1_20}}


\bibitem[\protect\citeauthoryear{Estima and Painho}{Estima and Painho}{2013}]%
        {Estima2013}
\bibfield{author}{\bibinfo{person}{Jacinto Estima} {and} \bibinfo{person}{Marco
  Painho}.} \bibinfo{year}{2013}\natexlab{}.
\newblock \showarticletitle{{F}lickr Geotagged and Publicly Available Photos:
  Preliminary Study of Its Adequacy for Helping Quality Control of Corine Land
  Cover}. In \bibinfo{booktitle}{{\em {ICCSA} 2013: Computational Science and
  Its Applications}}. \bibinfo{pages}{205--220}.
\newblock
\showDOI{%
\url{http://dx.doi.org/10.1007/978-3-642-39649-6_15}}


\bibitem[\protect\citeauthoryear{Eyben, W\"{o}llmer, and Schuller}{Eyben
  et~al\mbox{.}}{2009}]%
        {OpenEAR}
\bibfield{author}{\bibinfo{person}{Florian Eyben}, \bibinfo{person}{Martin
  W\"{o}llmer}, {and} \bibinfo{person}{Bj\"{o}rn Schuller}.}
  \bibinfo{year}{2009}\natexlab{}.
\newblock \showarticletitle{{OpenEAR}: Introducing the {M}unich open-source
  emotion and affect recognition toolkit}. In \bibinfo{booktitle}{{\em 2009 3rd
  International Conference on Affective Computing and Intelligent Interaction
  and Workshops}}. \bibinfo{pages}{1--6}.
\newblock
\showISSN{2156-8103}
\showDOI{%
\url{http://dx.doi.org/10.1109/ACII.2009.5349350}}


\bibitem[\protect\citeauthoryear{Feldman~Barrett, Mesquita, and
  Gendron}{Feldman~Barrett et~al\mbox{.}}{2011}]%
        {FeldmanBarrettEt2011}
\bibfield{author}{\bibinfo{person}{Lisa Feldman~Barrett},
  \bibinfo{person}{Batja Mesquita}, {and} \bibinfo{person}{Maria Gendron}.}
  \bibinfo{year}{2011}\natexlab{}.
\newblock \showarticletitle{Context in Emotion Perception}.
\newblock \bibinfo{journal}{{\em Current Directions in Psychological
  Science\/}}  \bibinfo{volume}{20} (\bibinfo{year}{2011}),
  \bibinfo{pages}{286--290}.
\newblock


\bibitem[\protect\citeauthoryear{Fern\'andez-Dols and
  Crivelli}{Fern\'andez-Dols and Crivelli}{2013}]%
        {FernandezDolsCrivelli2013}
\bibfield{author}{\bibinfo{person}{Jos\'e-Miguel Fern\'andez-Dols} {and}
  \bibinfo{person}{Carlos Crivelli}.} \bibinfo{year}{2013}\natexlab{}.
\newblock \showarticletitle{Emotion and Expression: Naturalistic Studies}.
\newblock \bibinfo{journal}{{\em Emotion Review\/}} \bibinfo{volume}{5},
  \bibinfo{number}{1} (\bibinfo{year}{2013}), \bibinfo{pages}{24--29}.
\newblock


\bibitem[\protect\citeauthoryear{Friedland and Sommer}{Friedland and
  Sommer}{2010}]%
        {Friedland2010}
\bibfield{author}{\bibinfo{person}{Gerald Friedland} {and}
  \bibinfo{person}{Robin Sommer}.} \bibinfo{year}{2010}\natexlab{}.
\newblock \showarticletitle{Cybercasing the joint: On the privacy implications
  of geo-tagging}. In \bibinfo{booktitle}{{\em Proceedings of the 5th {USENIX}
  Conference on Hot Topics in Security ({HotSec})}}. \bibinfo{publisher}{USENIX
  Association}, \bibinfo{pages}{1--8}.
\newblock


\bibitem[\protect\citeauthoryear{Garrett and Baquedano-L\'{o}pez}{Garrett and
  Baquedano-L\'{o}pez}{2002}]%
        {GarrettBL2002}
\bibfield{author}{\bibinfo{person}{Paul~B. Garrett} {and}
  \bibinfo{person}{Patricia Baquedano-L\'{o}pez}.}
  \bibinfo{year}{2002}\natexlab{}.
\newblock \showarticletitle{Language socialization: Reproduction and
  continuity, transformation and change}.
\newblock \bibinfo{journal}{{\em Annual Review of Anthropology\/}}
  \bibinfo{volume}{31} (\bibinfo{year}{2002}), \bibinfo{pages}{339--361}.
\newblock


\bibitem[\protect\citeauthoryear{Gaspar and Esteves}{Gaspar and
  Esteves}{2012}]%
        {GasparEsteves2012}
\bibfield{author}{\bibinfo{person}{Augusta Gaspar} {and}
  \bibinfo{person}{Francisco~G. Esteves}.} \bibinfo{year}{2012}\natexlab{}.
\newblock \showarticletitle{Preschooler's faces in spontaneous emotional
  contexts---How well do they match adult facial expression prototypes?}
\newblock \bibinfo{journal}{{\em International Journal of Behavioral
  Development\/}} \bibinfo{volume}{36}, \bibinfo{number}{5}
  (\bibinfo{year}{2012}), \bibinfo{pages}{348--357}.
\newblock


\bibitem[\protect\citeauthoryear{Grimm, Kroschel, and Narayanan}{Grimm
  et~al\mbox{.}}{2008}]%
        {VaM}
\bibfield{author}{\bibinfo{person}{Michael Grimm}, \bibinfo{person}{Kristian
  Kroschel}, {and} \bibinfo{person}{Shrikanth Narayanan}.}
  \bibinfo{year}{2008}\natexlab{}.
\newblock \showarticletitle{The {Vera am Mittag} {G}erman audio-visual
  emotional speech database}. In \bibinfo{booktitle}{{\em 2008 {IEEE}
  International Conference on Multimedia and Expo}}. \bibinfo{pages}{865--868}.
\newblock
\showISSN{1945-7871}
\showDOI{%
\url{http://dx.doi.org/10.1109/ICME.2008.4607572}}


\bibitem[\protect\citeauthoryear{Gunes and Hung}{Gunes and Hung}{2016}]%
        {GunesHung2016}
\bibfield{author}{\bibinfo{person}{Hatice Gunes} {and} \bibinfo{person}{Hayley
  Hung}.} \bibinfo{year}{2016}\natexlab{}.
\newblock \showarticletitle{Is automatic facial expression recognition of
  emotions coming to a dead end? The rise of the new kids on the block}.
\newblock \bibinfo{journal}{{\em Image and Vision Computing\/}}
  \bibinfo{volume}{55}, \bibinfo{number}{1} (\bibinfo{year}{2016}),
  \bibinfo{pages}{6--8}.
\newblock


\bibitem[\protect\citeauthoryear{Gunes and Schuller}{Gunes and
  Schuller}{2013}]%
        {GunesSchuller2013}
\bibfield{author}{\bibinfo{person}{Hatice Gunes} {and}
  \bibinfo{person}{Bj\"{o}rn Schuller}.} \bibinfo{year}{2013}\natexlab{}.
\newblock \showarticletitle{Categorical and dimensional affect analysis in
  continuous input: Current trends and future directions}.
\newblock \bibinfo{journal}{{\em Image and Vision Computing\/}}
  \bibinfo{volume}{31}, \bibinfo{number}{2} (\bibinfo{year}{2013}),
  \bibinfo{pages}{120--136}.
\newblock
\showISSN{0262-8856}
\showDOI{%
\url{http://dx.doi.org/10.1016/j.imavis.2012.06.016}}


\bibitem[\protect\citeauthoryear{Gunes, Schuller, Pantic, and Cowie}{Gunes
  et~al\mbox{.}}{2011}]%
        {GunesEtAl2011}
\bibfield{author}{\bibinfo{person}{Hatice Gunes}, \bibinfo{person}{Bj\"{o}rn
  Schuller}, \bibinfo{person}{Maja Pantic}, {and} \bibinfo{person}{Roddy
  Cowie}.} \bibinfo{year}{2011}\natexlab{}.
\newblock \showarticletitle{Emotion representation, analysis and synthesis in
  continuous space: A survey}. In \bibinfo{booktitle}{{\em Face and Gesture
  2011}}. \bibinfo{pages}{827--834}.
\newblock
\showDOI{%
\url{http://dx.doi.org/10.1109/FG.2011.5771357}}


\bibitem[\protect\citeauthoryear{Gutzeit, Fabisch, Otto, Metzen, Hansen,
  Kirchner, and Kirchner}{Gutzeit et~al\mbox{.}}{2017}]%
        {Gutzeit2017}
\bibfield{author}{\bibinfo{person}{Lisa Gutzeit}, \bibinfo{person}{Alexander
  Fabisch}, \bibinfo{person}{Marc Otto}, \bibinfo{person}{Jan~Hendrik Metzen},
  \bibinfo{person}{Jonas Hansen}, \bibinfo{person}{Frank Kirchner}, {and}
  \bibinfo{person}{Elsa~Andrea Kirchner}.} \bibinfo{year}{2017}\natexlab{}.
\newblock \showarticletitle{The {BesMan} Learning Platform for Automated Robot
  Skill Learning}.
\newblock \bibinfo{journal}{{\em Autonomous Robots\/}} (\bibinfo{year}{2017}).
\newblock
\newblock
\shownote{Submitted.}


\bibitem[\protect\citeauthoryear{Gutzeit and Kirchner}{Gutzeit and
  Kirchner}{2016}]%
        {Gutzeit2016}
\bibfield{author}{\bibinfo{person}{Lisa Gutzeit} {and}
  \bibinfo{person}{Elsa~Andrea Kirchner}.} \bibinfo{year}{2016}\natexlab{}.
\newblock \showarticletitle{{Automatic Detection and Recognition of Human
  Movement Patterns in Manipulation Tasks}}. In \bibinfo{booktitle}{{\em
  Proceedings of the 3rd International Conference on Physiological Computing
  Systems}}. \bibinfo{publisher}{SCITEPRESS - Science and Technology
  Publications}, \bibinfo{pages}{54--63}.
\newblock
\showISBNx{978-989-758-197-7}
\showDOI{%
\url{http://dx.doi.org/10.5220/0005946500540063}}


\bibitem[\protect\citeauthoryear{Haidt and Keltner}{Haidt and Keltner}{1999}]%
        {HaidtKeltner1999}
\bibfield{author}{\bibinfo{person}{Jonathan Haidt} {and}
  \bibinfo{person}{Dacher Keltner}.} \bibinfo{year}{1999}\natexlab{}.
\newblock \showarticletitle{Culture and facial expression: Open-ended methods
  find more expressions and a gradient of recognition}.
\newblock \bibinfo{journal}{{\em Cognition and Emotion\/}}
  \bibinfo{volume}{13} (\bibinfo{year}{1999}), \bibinfo{pages}{225--266}.
\newblock


\bibitem[\protect\citeauthoryear{Hannun, Case, Casper, Catanzaro, Diamos,
  Elsen, Prenger, Satheesh, Sengupta, Coates, and Ng}{Hannun
  et~al\mbox{.}}{2014}]%
        {hannun2014deep}
\bibfield{author}{\bibinfo{person}{Awni Hannun}, \bibinfo{person}{Carl Case},
  \bibinfo{person}{Jared Casper}, \bibinfo{person}{Bryan Catanzaro},
  \bibinfo{person}{Greg Diamos}, \bibinfo{person}{Erich Elsen},
  \bibinfo{person}{Ryan Prenger}, \bibinfo{person}{Sanjeev Satheesh},
  \bibinfo{person}{Shubho Sengupta}, \bibinfo{person}{Adam Coates}, {and}
  \bibinfo{person}{Andrew~Y. Ng}.} \bibinfo{year}{2014}\natexlab{}.
\newblock \bibinfo{title}{Deep speech: Scaling up end-to-end speech
  recognition}.
\newblock   (\bibinfo{year}{2014}).
\newblock
\newblock
\shownote{arXiv:1412.5567.}


\bibitem[\protect\citeauthoryear{Hardy, Baker, Devillers, Lamel, Rosset, Ursu,
  and Webb}{Hardy et~al\mbox{.}}{2002}]%
        {HardyEtAl2002}
\bibfield{author}{\bibinfo{person}{Hilda Hardy}, \bibinfo{person}{Kirk Baker},
  \bibinfo{person}{Laurence Devillers}, \bibinfo{person}{Lori Lamel},
  \bibinfo{person}{Sophie Rosset}, \bibinfo{person}{Cristian Ursu}, {and}
  \bibinfo{person}{Nick Webb}.} \bibinfo{year}{2002}\natexlab{}.
\newblock \showarticletitle{Multi-layer dialogue annotation for automated
  multilingual customer service}. In \bibinfo{booktitle}{{\em Proceedings of
  the International Standards for Language Engineering Workshop}}.
\newblock


\bibitem[\protect\citeauthoryear{Harris and Alvarado}{Harris and
  Alvarado}{2005}]%
        {HarrisAlvarado2005}
\bibfield{author}{\bibinfo{person}{Christine Harris} {and}
  \bibinfo{person}{Nancy Alvarado}.} \bibinfo{year}{2005}\natexlab{}.
\newblock \showarticletitle{Facial expressions, smile types, and self-report
  during humor, tickle, and pain}.
\newblock \bibinfo{journal}{{\em Cognition and Emotion\/}}
  \bibinfo{volume}{19}, \bibinfo{number}{5} (\bibinfo{year}{2005}),
  \bibinfo{pages}{655--669}.
\newblock


\bibitem[\protect\citeauthoryear{Harutyunyan, Fidanza, and
  Khachatrian}{Harutyunyan et~al\mbox{.}}{}]%
        {Lang-ID}
\bibfield{author}{\bibinfo{person}{Hrayr Harutyunyan},
  \bibinfo{person}{Guillaume Fidanza}, {and} \bibinfo{person}{Hrant
  Khachatrian}.}
\newblock \bibinfo{title}{Spoken language identification with deep learning}.
\newblock   (\bibinfo{year}{????}).
\newblock
\newblock
\shownote{\url{https://github.com/YerevaNN/Spoken-language-identification}.}


\bibitem[\protect\citeauthoryear{Hatori}{Hatori}{2011}]%
        {Hatori2011}
\bibfield{author}{\bibinfo{person}{Koshiro Hatori}.}
  \bibinfo{year}{2011}\natexlab{}.
\newblock \showarticletitle{History of Origami in the East and the West before
  Interfusion}. In \bibinfo{booktitle}{{\em Fifth International Meeting of
  Origami Science, Mathematics, and Education (Origami 5)}}.
  \bibinfo{publisher}{A K Peters/CRC Press}, \bibinfo{pages}{3--11}.
\newblock
\showISBNx{978-1-56881-714-9}
\showDOI{%
\url{http://dx.doi.org/10.1201/b10971-3}}
\newblock
\shownote{0.}


\bibitem[\protect\citeauthoryear{Hull}{Hull}{2015}]%
        {Hull2015}
\bibfield{author}{\bibinfo{person}{Thomas~C Hull}.}
  \bibinfo{year}{2015}\natexlab{}.
\newblock \showarticletitle{Coloring connections with counting mountain-valley
  assignments}.
\newblock \bibinfo{journal}{{\em Origami${}^6$: I. Mathematics\/}}
  (\bibinfo{year}{2015}), \bibinfo{pages}{3}.
\newblock


\bibitem[\protect\citeauthoryear{Hyde, Dixit, Weisberg, and Rushford}{Hyde
  et~al\mbox{.}}{2002}]%
        {Hyde2002}
\bibfield{author}{\bibinfo{person}{Roderick~A. Hyde},
  \bibinfo{person}{Shamasundar~N. Dixit}, \bibinfo{person}{Andrew~H. Weisberg},
  {and} \bibinfo{person}{Michael~C. Rushford}.}
  \bibinfo{year}{2002}\natexlab{}.
\newblock \showarticletitle{{E}yeglass: a very large aperture diffractive space
  telescope}. In \bibinfo{booktitle}{{\em SPIE 4849, Highly Innovative Space
  Telescope Concepts, 28}}, \bibfield{editor}{\bibinfo{person}{Howard~A.
  MacEwen}} (Ed.). \bibinfo{publisher}{International Society for Optics and
  Photonics}, \bibinfo{pages}{28}.
\newblock
\showDOI{%
\url{http://dx.doi.org/10.1117/12.460420}}


\bibitem[\protect\citeauthoryear{Izadinia, Russell, Farhadi, Hoffman, and
  Hertzmann}{Izadinia et~al\mbox{.}}{2015}]%
        {Izadinia2015}
\bibfield{author}{\bibinfo{person}{Hamid Izadinia}, \bibinfo{person}{Bryan~C.
  Russell}, \bibinfo{person}{Ali Farhadi}, \bibinfo{person}{Matthew~D.
  Hoffman}, {and} \bibinfo{person}{Aaron Hertzmann}.}
  \bibinfo{year}{2015}\natexlab{}.
\newblock \showarticletitle{{Deep Classifiers from Image Tags in the Wild}}. In
  \bibinfo{booktitle}{{\em Proceedings of the 2015 Workshop on
  Community-Organized Multimodal Mining: Opportunities for Novel Solutions
  (MMCommons '15)}}. \bibinfo{publisher}{ACM}, \bibinfo{pages}{13--18}.
\newblock
\showDOI{%
\url{http://dx.doi.org/10.1145/2814815.2814821}}


\bibitem[\protect\citeauthoryear{Jebara and Pentland}{Jebara and
  Pentland}{2002}]%
        {Jebara2002}
\bibfield{author}{\bibinfo{person}{T. Jebara} {and} \bibinfo{person}{A.
  Pentland}.} \bibinfo{year}{2002}\natexlab{}.
\newblock \showarticletitle{{Statistical imitative learning from perceptual
  data}}. In \bibinfo{booktitle}{{\em Proceedings of the 2nd International
  Conference on Development and Learning (ICDL 2002)}}.
  \bibinfo{publisher}{IEEE Comput. Soc}, \bibinfo{pages}{191--196}.
\newblock
\showISBNx{0-7695-1459-6}
\showDOI{%
\url{http://dx.doi.org/10.1109/DEVLRN.2002.1011859}}


\bibitem[\protect\citeauthoryear{Jiang, Yu, Meng, Mitamura, and
  Hauptmann}{Jiang et~al\mbox{.}}{2015a}]%
        {Jiang2015}
\bibfield{author}{\bibinfo{person}{Lu Jiang}, \bibinfo{person}{Shoou-I Yu},
  \bibinfo{person}{Deyu Meng}, \bibinfo{person}{Teruko Mitamura}, {and}
  \bibinfo{person}{Alexander~G. Hauptmann}.} \bibinfo{year}{2015}\natexlab{a}.
\newblock \showarticletitle{Bridging the Ultimate Semantic Gap}. In
  \bibinfo{booktitle}{{\em Proceedings of the 5th ACM on International
  Conference on Multimedia Retrieval - ICMR '15}}. \bibinfo{publisher}{ACM
  Press}, \bibinfo{address}{New York, New York, USA}, \bibinfo{pages}{27--34}.
\newblock
\showISBNx{9781450332743}
\showDOI{%
\url{http://dx.doi.org/10.1145/2671188.2749399}}


\bibitem[\protect\citeauthoryear{Jiang, Yu, Meng, Yang, Mitamura, and
  Hauptmann}{Jiang et~al\mbox{.}}{2015b}]%
        {JiangEtAl2015}
\bibfield{author}{\bibinfo{person}{Lu Jiang}, \bibinfo{person}{Shoou-I Yu},
  \bibinfo{person}{Deyu Meng}, \bibinfo{person}{Yi Yang},
  \bibinfo{person}{Teruko Mitamura}, {and} \bibinfo{person}{Alexander~G.
  Hauptmann}.} \bibinfo{year}{2015}\natexlab{b}.
\newblock \showarticletitle{Fast and accurate content-based semantic search in
  {100M} {I}nternet videos}. In \bibinfo{booktitle}{{\em Proceedings of the
  23rd {ACM} International Conference on Multimedia ({MM} '15)}}. ACM,
  \bibinfo{pages}{49--58}.
\newblock


\bibitem[\protect\citeauthoryear{K{\'{a}}d{\'{a}}r and Gede}{K{\'{a}}d{\'{a}}r
  and Gede}{2013}]%
        {Kadar2013}
\bibfield{author}{\bibinfo{person}{B{\'{a}}lint K{\'{a}}d{\'{a}}r} {and}
  \bibinfo{person}{M{\'{a}}ty{\'{a}}s Gede}.} \bibinfo{year}{2013}\natexlab{}.
\newblock \showarticletitle{Where Do Tourists Go? Visualizing and Analysing the
  Spatial Distribution of Geotagged Photography}.
\newblock \bibinfo{journal}{{\em Cartographica: The International Journal for
  Geographic Information and Geovisualization\/}} \bibinfo{volume}{48},
  \bibinfo{number}{2} (\bibinfo{date}{Jun} \bibinfo{year}{2013}),
  \bibinfo{pages}{78--88}.
\newblock
\showISSN{0317-7173}
\showDOI{%
\url{http://dx.doi.org/10.3138/carto.48.2.1839}}


\bibitem[\protect\citeauthoryear{Kalkowski, Schulze, Dengel, and
  Borth}{Kalkowski et~al\mbox{.}}{2015}]%
        {Kalkowski2015}
\bibfield{author}{\bibinfo{person}{Sebastian Kalkowski},
  \bibinfo{person}{Christian Schulze}, \bibinfo{person}{Andreas Dengel}, {and}
  \bibinfo{person}{Damian Borth}.} \bibinfo{year}{2015}\natexlab{}.
\newblock \showarticletitle{Real-time Analysis and Visualization of the
  {YFCC100m} Dataset}. In \bibinfo{booktitle}{{\em Proceedings of the 2015
  Workshop on Community-Organized Multimodal Mining: Opportunities for Novel
  Solutions (MMCommons '15)}}. \bibinfo{publisher}{ACM},
  \bibinfo{pages}{25--30}.
\newblock
\showDOI{%
\url{http://dx.doi.org/10.1145/2814815.2814820}}


\bibitem[\protect\citeauthoryear{Kanade, Cohn, and Tian}{Kanade
  et~al\mbox{.}}{2000}]%
        {CK}
\bibfield{author}{\bibinfo{person}{Takeo Kanade}, \bibinfo{person}{Jeffrey~F.
  Cohn}, {and} \bibinfo{person}{Yingli Tian}.} \bibinfo{year}{2000}\natexlab{}.
\newblock \showarticletitle{Comprehensive database for facial expression
  analysis}. In \bibinfo{booktitle}{{\em Proceedings of the Fourth {IEEE}
  International Conference on Automatic Face and Gesture Recognition (Cat. No.
  {PR}00580)}}. \bibinfo{pages}{46--53}.
\newblock
\showDOI{%
\url{http://dx.doi.org/10.1109/AFGR.2000.840611}}


\bibitem[\protect\citeauthoryear{Kendall, Grimes, and Cipolla}{Kendall
  et~al\mbox{.}}{2015}]%
        {Kendall2015}
\bibfield{author}{\bibinfo{person}{Alex Kendall}, \bibinfo{person}{Matthew
  Grimes}, {and} \bibinfo{person}{Roberto Cipolla}.}
  \bibinfo{year}{2015}\natexlab{}.
\newblock \showarticletitle{{PoseNet: A Convolutional Network for Real-Time
  6-DOF Camera Relocalization}}. In \bibinfo{booktitle}{{\em 2015 IEEE
  International Conference on Computer Vision (ICCV)}}.
  \bibinfo{publisher}{IEEE}, \bibinfo{pages}{2938--2946}.
\newblock
\showISBNx{978-1-4673-8391-2}
\showDOI{%
\url{http://dx.doi.org/10.1109/ICCV.2015.336}}


\bibitem[\protect\citeauthoryear{Kleinsmith and Bianchi-Berthouze}{Kleinsmith
  and Bianchi-Berthouze}{2013}]%
        {KleinBianchi2013}
\bibfield{author}{\bibinfo{person}{Andrea Kleinsmith} {and}
  \bibinfo{person}{Nadia Bianchi-Berthouze}.} \bibinfo{year}{2013}\natexlab{}.
\newblock \showarticletitle{Affective Body Expression Perception and
  Recognition: A Survey}.
\newblock \bibinfo{journal}{{\em {IEEE} Transactions on Affective Computing\/}}
  \bibinfo{volume}{4}, \bibinfo{number}{1} (\bibinfo{date}{Jan}
  \bibinfo{year}{2013}), \bibinfo{pages}{15--33}.
\newblock
\showISSN{1949-3045}
\showDOI{%
\url{http://dx.doi.org/10.1109/T-AFFC.2012.16}}


\bibitem[\protect\citeauthoryear{Kleinsmith, Bianchi-Berthouze, and
  Steed}{Kleinsmith et~al\mbox{.}}{2011}]%
        {KleinsmithEtAl2011}
\bibfield{author}{\bibinfo{person}{Andrea Kleinsmith}, \bibinfo{person}{Nadia
  Bianchi-Berthouze}, {and} \bibinfo{person}{Anthony Steed}.}
  \bibinfo{year}{2011}\natexlab{}.
\newblock \showarticletitle{Automatic Recognition of Non-Acted Affective
  Postures}.
\newblock \bibinfo{journal}{{\em {IEEE} Transactions on Systems, Man, and
  Cybernetics---{Part B}: Cybernetics\/}} \bibinfo{volume}{41},
  \bibinfo{number}{4} (\bibinfo{date}{August} \bibinfo{year}{2011}),
  \bibinfo{pages}{1027--1038}.
\newblock
\showURL{%
\url{http://ieeexplore.ieee.org/stamp/stamp.jsp?arnumber=5704207}}


\bibitem[\protect\citeauthoryear{Kofler, Caballero, Menendez, Occhialini, and
  Larson}{Kofler et~al\mbox{.}}{2011}]%
        {Kofler2011}
\bibfield{author}{\bibinfo{person}{Christoph Kofler}, \bibinfo{person}{Luz
  Caballero}, \bibinfo{person}{Maria Menendez}, \bibinfo{person}{Valentina
  Occhialini}, {and} \bibinfo{person}{Martha Larson}.}
  \bibinfo{year}{2011}\natexlab{}.
\newblock \showarticletitle{Near2me: an authentic and personalized social
  media-based recommender for travel destinations}. In \bibinfo{booktitle}{{\em
  Proceedings of the 3rd {ACM SIGMM} International Workshop on Social Media
  ({WSM} '11)}}. \bibinfo{publisher}{ACM}, \bibinfo{address}{New York, New
  York, USA}, \bibinfo{pages}{47}.
\newblock
\showISBNx{9781450309899}
\showDOI{%
\url{http://dx.doi.org/10.1145/2072609.2072624}}


\bibitem[\protect\citeauthoryear{Koochali, Kalkowski, Dengel, Borth, and
  Schulze}{Koochali et~al\mbox{.}}{2016}]%
        {Koochali2016}
\bibfield{author}{\bibinfo{person}{Alireza Koochali},
  \bibinfo{person}{Sebastian Kalkowski}, \bibinfo{person}{Andreas Dengel},
  \bibinfo{person}{Damian Borth}, {and} \bibinfo{person}{Christian Schulze}.}
  \bibinfo{year}{2016}\natexlab{}.
\newblock \showarticletitle{Which Languages Do People Speak on {F}lickr?: A
  Language and Geo-Location Study of the {YFCC100m} Dataset}. In
  \bibinfo{booktitle}{{\em Proceedings of the 2016 ACM Workshop on the
  Multimedia COMMONS (MMCommons '16)}}. ACM, \bibinfo{pages}{35--42}.
\newblock


\bibitem[\protect\citeauthoryear{Kordopatis-Zilos, Papadopoulos, and
  Kompatsiaris}{Kordopatis-Zilos et~al\mbox{.}}{2016}]%
        {Kordopatis-Zilos2016}
\bibfield{author}{\bibinfo{person}{Giorgos Kordopatis-Zilos},
  \bibinfo{person}{Symeon Papadopoulos}, {and} \bibinfo{person}{Yiannis
  Kompatsiaris}.} \bibinfo{year}{2016}\natexlab{}.
\newblock \showarticletitle{In-Depth Exploration of Geotagging Performance
  Using Sampling Strategies on {YFCC100M}}. In \bibinfo{booktitle}{{\em
  Proceedings of the 2016 ACM Workshop on the Multimedia COMMONS (MMCommons
  '16)}}. \bibinfo{publisher}{ACM Press}, \bibinfo{address}{New York, New York,
  USA}, \bibinfo{pages}{3--10}.
\newblock
\showISBNx{9781450345156}
\showDOI{%
\url{http://dx.doi.org/10.1145/2983554.2983558}}


\bibitem[\protect\citeauthoryear{Krell, Straube, Seeland, W{\"{o}}hrle, Teiwes,
  Metzen, Kirchner, and Kirchner}{Krell et~al\mbox{.}}{2013}]%
        {Krell2013}
\bibfield{author}{\bibinfo{person}{Mario~Michael Krell}, \bibinfo{person}{Sirko
  Straube}, \bibinfo{person}{Anett Seeland}, \bibinfo{person}{Hendrik
  W{\"{o}}hrle}, \bibinfo{person}{Johannes Teiwes},
  \bibinfo{person}{Jan~Hendrik Metzen}, \bibinfo{person}{Elsa~Andrea Kirchner},
  {and} \bibinfo{person}{Frank Kirchner}.} \bibinfo{year}{2013}\natexlab{}.
\newblock \showarticletitle{{pySPACE}---a signal processing and classification
  environment in {P}ython}.
\newblock \bibinfo{journal}{{\em Frontiers in Neuroinformatics\/}}
  \bibinfo{volume}{7}, \bibinfo{number}{40} (\bibinfo{date}{Dec}
  \bibinfo{year}{2013}), \bibinfo{pages}{1--11}.
\newblock
\showISSN{1662-5196}
\showDOI{%
\url{http://dx.doi.org/10.3389/fninf.2013.00040}}


\bibitem[\protect\citeauthoryear{Kuribayashi, Tsuchiya, You, Tomus, Umemoto,
  Ito, and Sasaki}{Kuribayashi et~al\mbox{.}}{2006}]%
        {Kuribayashi2006}
\bibfield{author}{\bibinfo{person}{Kaori Kuribayashi}, \bibinfo{person}{Koichi
  Tsuchiya}, \bibinfo{person}{Zhong You}, \bibinfo{person}{Dacian Tomus},
  \bibinfo{person}{Minoru Umemoto}, \bibinfo{person}{Takahiro Ito}, {and}
  \bibinfo{person}{Masahiro Sasaki}.} \bibinfo{year}{2006}\natexlab{}.
\newblock \showarticletitle{Self-deployable origami stent grafts as a
  biomedical application of {Ni-rich} {TiNi} shape memory alloy foil}.
\newblock \bibinfo{journal}{{\em Materials Science and Engineering: A\/}}
  \bibinfo{volume}{419}, \bibinfo{number}{1--2} (\bibinfo{date}{Mar}
  \bibinfo{year}{2006}), \bibinfo{pages}{131--137}.
\newblock
\showISSN{0921-5093}
\showDOI{%
\url{http://dx.doi.org/10.1016/j.msea.2005.12.016}}


\bibitem[\protect\citeauthoryear{Lamere, Kwok, Gouvea, Raj, Singh, Walker,
  Warmuth, and Wolf}{Lamere et~al\mbox{.}}{2003}]%
        {lamere2003cmu}
\bibfield{author}{\bibinfo{person}{Paul Lamere}, \bibinfo{person}{Philip Kwok},
  \bibinfo{person}{Evandro Gouvea}, \bibinfo{person}{Bhiksha Raj},
  \bibinfo{person}{Rita Singh}, \bibinfo{person}{William Walker},
  \bibinfo{person}{Manfred Warmuth}, {and} \bibinfo{person}{Peter Wolf}.}
  \bibinfo{year}{2003}\natexlab{}.
\newblock \showarticletitle{The {CMU SPHINX-4} speech recognition system}. In
  \bibinfo{booktitle}{{\em Proceedings of the IEEE International Conference on
  Acoustics, Speech and Signal Processing (ICASSP 2003), Hong Kong, April
  6--10, 2003}}, Vol.~\bibinfo{volume}{1}. \bibinfo{pages}{2--5}.
\newblock


\bibitem[\protect\citeauthoryear{Lang}{Lang}{2008}]%
        {lang2008flapping}
\bibfield{author}{\bibinfo{person}{Robert~J Lang}.}
  \bibinfo{year}{2008}\natexlab{}.
\newblock \showarticletitle{From flapping birds to space telescopes: The modern
  science of origami}. In \bibinfo{booktitle}{{\em 6th Symposium on
  Non-Photorealistic Animation and Rendering (NPAR)}}. \bibinfo{pages}{7}.
\newblock


\bibitem[\protect\citeauthoryear{Lee, Busso, Lee, and Narayanan}{Lee
  et~al\mbox{.}}{2009}]%
        {LeeEtAl2009}
\bibfield{author}{\bibinfo{person}{Chi-Chun Lee}, \bibinfo{person}{Carlos
  Busso}, \bibinfo{person}{Sungbok Lee}, {and} \bibinfo{person}{Shrikanth
  Narayanan}.} \bibinfo{year}{2009}\natexlab{}.
\newblock \showarticletitle{Modeling mutual influence of interlocutor emotion
  states in dyadic spoken interactions}. In \bibinfo{booktitle}{{\em
  Proceedings of {INTERSPEECH}}}. \bibinfo{pages}{1983--1986}.
\newblock


\bibitem[\protect\citeauthoryear{Lefter, Rothkrantz, Wiggers, and van
  Leeuwen}{Lefter et~al\mbox{.}}{2010}]%
        {LefterEtAl2010}
\bibfield{author}{\bibinfo{person}{Iulia Lefter}, \bibinfo{person}{Leon J.~M.
  Rothkrantz}, \bibinfo{person}{Pascal Wiggers}, {and}
  \bibinfo{person}{David~A. van Leeuwen}.} \bibinfo{year}{2010}\natexlab{}.
\newblock \showarticletitle{Emotion recognition from speech by combining
  databases and fusion of classifiers}. In \bibinfo{booktitle}{{\em Proceedings
  of the 13th International Conference on Text, Speech, and Dialogue ({TSD}
  2010), Brno, Czech Republic, September 6--10, 2010}} {\em
  (\bibinfo{series}{Lecture Notes in Computer Science})},
  \bibfield{editor}{\bibinfo{person}{Petr Sojka}, \bibinfo{person}{Ale\v{s}
  Hor\'{a}k}, \bibinfo{person}{Ivan Kope\v{c}ek}, {and} \bibinfo{person}{Karel
  Pala}} (Eds.). \bibinfo{publisher}{Springer}, \bibinfo{pages}{353--360}.
\newblock


\bibitem[\protect\citeauthoryear{Liu, Tsow, Zou, and Tao}{Liu
  et~al\mbox{.}}{2016}]%
        {Liu2016}
\bibfield{author}{\bibinfo{person}{Chenbin Liu}, \bibinfo{person}{Francis
  Tsow}, \bibinfo{person}{Yi Zou}, {and} \bibinfo{person}{Nongjian Tao}.}
  \bibinfo{year}{2016}\natexlab{}.
\newblock \showarticletitle{{Particle Pollution Estimation Based on Image
  Analysis}}.
\newblock \bibinfo{journal}{{\em PLOS ONE\/}} \bibinfo{volume}{11},
  \bibinfo{number}{2} (\bibinfo{date}{Feb} \bibinfo{year}{2016}),
  \bibinfo{pages}{e0145955}.
\newblock
\showISSN{1932-6203}
\showDOI{%
\url{http://dx.doi.org/10.1371/journal.pone.0145955}}


\bibitem[\protect\citeauthoryear{Lui and Baldwin}{Lui and Baldwin}{2012}]%
        {Lui2012}
\bibfield{author}{\bibinfo{person}{Marco Lui} {and} \bibinfo{person}{Timothy
  Baldwin}.} \bibinfo{year}{2012}\natexlab{}.
\newblock \showarticletitle{Langid.Py: An Off-the-shelf Language Identification
  Tool}. In \bibinfo{booktitle}{{\em Proceedings of the ACL 2012 System
  Demonstrations}} {\em (\bibinfo{series}{ACL '12})}.
  \bibinfo{publisher}{Association for Computational Linguistics},
  \bibinfo{address}{Stroudsburg, PA, USA}, \bibinfo{pages}{25--30}.
\newblock
\showURL{%
\url{http://dl.acm.org/citation.cfm?id=2390470.2390475}}


\bibitem[\protect\citeauthoryear{Lux and Macstravic}{Lux and
  Macstravic}{2014}]%
        {Lux2014}
\bibfield{author}{\bibinfo{person}{Mathias Lux} {and} \bibinfo{person}{Glenn
  Macstravic}.} \bibinfo{year}{2014}\natexlab{}.
\newblock \showarticletitle{The LIRE Request Handler: A Solr Plug-In for Large
  Scale Content Based Image Retrieval}. In \bibinfo{booktitle}{{\em MultiMedia
  Modeling: 20th Anniversary International Conference, MMM 2014, Dublin,
  Ireland, January 6-10, 2014, Proceedings, Part II}},
  \bibfield{editor}{\bibinfo{person}{Cathal Gurrin}, \bibinfo{person}{Frank
  Hopfgartner}, \bibinfo{person}{Wolfgang Hurst}, \bibinfo{person}{H{\aa}vard
  Johansen}, \bibinfo{person}{Hyowon Lee}, {and} \bibinfo{person}{Noel
  O'Connor}} (Eds.). \bibinfo{publisher}{Springer International Publishing},
  \bibinfo{address}{Cham}, \bibinfo{pages}{374--377}.
\newblock
\showISBNx{978-3-319-04117-9}
\showDOI{%
\url{http://dx.doi.org/10.1007/978-3-319-04117-9_39}}


\bibitem[\protect\citeauthoryear{MacWhinney}{MacWhinney}{2000}]%
        {childes}
\bibfield{author}{\bibinfo{person}{Brian MacWhinney}.}
  \bibinfo{year}{2000}\natexlab{}.
\newblock \bibinfo{booktitle}{{\em The {CHILDES} Project: Tools for Analyzing
  Talk\/} (\bibinfo{edition}{3rd} ed.)}.
\newblock \bibinfo{publisher}{Lawrence Erlbaum Associates}.
\newblock
\newblock
\shownote{\url{http://childes.talkbank.org/}.}


\bibitem[\protect\citeauthoryear{Mammucari, Caltagirone, Ekman, Friesen,
  Gainotti, Pizzamiglio, and Zoccolotti}{Mammucari et~al\mbox{.}}{1988}]%
        {EkmanEt1988}
\bibfield{author}{\bibinfo{person}{A Mammucari}, \bibinfo{person}{C
  Caltagirone}, \bibinfo{person}{P Ekman}, \bibinfo{person}{W Friesen},
  \bibinfo{person}{G Gainotti}, \bibinfo{person}{L Pizzamiglio}, {and}
  \bibinfo{person}{P Zoccolotti}.} \bibinfo{year}{1988}\natexlab{}.
\newblock \showarticletitle{Spontaneous Facial Expression of Emotions in
  Brain-damaged Patients}.
\newblock \bibinfo{journal}{{\em Cortex\/}}  \bibinfo{volume}{24}
  (\bibinfo{year}{1988}), \bibinfo{pages}{521--533}.
\newblock


\bibitem[\protect\citeauthoryear{Mariooryad, Lotfian, and Busso}{Mariooryad
  et~al\mbox{.}}{2014}]%
        {MariEtAl2014}
\bibfield{author}{\bibinfo{person}{Soroosh Mariooryad}, \bibinfo{person}{Reza
  Lotfian}, {and} \bibinfo{person}{Carlos Busso}.}
  \bibinfo{year}{2014}\natexlab{}.
\newblock \showarticletitle{Building a naturalistic emotional speech corpus by
  retrieving expressive behaviors from existing speech corpora}. In
  \bibinfo{booktitle}{{\em Proceedings of {INTERSPEECH}}}.
  \bibinfo{pages}{238--242}.
\newblock


\bibitem[\protect\citeauthoryear{Mascarenhas, Degens, Paiva, Prada, Hofstede,
  Beulens, and Aylett}{Mascarenhas et~al\mbox{.}}{2016}]%
        {MascaraEtAl2016}
\bibfield{author}{\bibinfo{person}{Samuel Mascarenhas}, \bibinfo{person}{Nick
  Degens}, \bibinfo{person}{Ana Paiva}, \bibinfo{person}{Rui Prada},
  \bibinfo{person}{Gert~Jan Hofstede}, \bibinfo{person}{Adrie Beulens}, {and}
  \bibinfo{person}{Ruth Aylett}.} \bibinfo{year}{2016}\natexlab{}.
\newblock \showarticletitle{Modeling culture in intelligent virtual agents}.
\newblock \bibinfo{journal}{{\em Autonomous Agents and Multi-Agent Systems\/}}
  \bibinfo{volume}{30}, \bibinfo{number}{5} (\bibinfo{year}{2016}),
  \bibinfo{pages}{931--962}.
\newblock
\showDOI{%
\url{http://dx.doi.org/10.1007/s10458-015-9312-6}}


\bibitem[\protect\citeauthoryear{Matisoff}{Matisoff}{1986}]%
        {Matisoff1986}
\bibfield{author}{\bibinfo{person}{James Matisoff}.}
  \bibinfo{year}{1986}\natexlab{}.
\newblock \showarticletitle{Hearts and minds in {S}outheast {A}sian languages
  and {E}nglish}.
\newblock \bibinfo{journal}{{\em Cahiers de Linguistique Asie Orientale\/}}
  \bibinfo{volume}{15}, \bibinfo{number}{1} (\bibinfo{year}{1986}),
  \bibinfo{pages}{5--57}.
\newblock


\bibitem[\protect\citeauthoryear{Matsumoto and Willingham}{Matsumoto and
  Willingham}{2009}]%
        {MatsumotoWillingham2009}
\bibfield{author}{\bibinfo{person}{David Matsumoto} {and} \bibinfo{person}{Bob
  Willingham}.} \bibinfo{year}{2009}\natexlab{}.
\newblock \showarticletitle{Spontaneous facial expressions of emotion of
  congenitally and noncongenitally blind individuals}.
\newblock \bibinfo{journal}{{\em Journal of Personal and Social Psychology\/}}
  \bibinfo{volume}{96}, \bibinfo{number}{1} (\bibinfo{year}{2009}),
  \bibinfo{pages}{1--10}.
\newblock
\showDOI{%
\url{http://dx.doi.org/10.1037/a0014037}}


\bibitem[\protect\citeauthoryear{McKeown, Valstar, Cowie, Pantic, and
  Schroder}{McKeown et~al\mbox{.}}{2012}]%
        {SEMAINE}
\bibfield{author}{\bibinfo{person}{Gary McKeown}, \bibinfo{person}{Michel
  Valstar}, \bibinfo{person}{Roddy Cowie}, \bibinfo{person}{Maja Pantic}, {and}
  \bibinfo{person}{Marc Schroder}.} \bibinfo{year}{2012}\natexlab{}.
\newblock \showarticletitle{The {SEMAINE} Database: Annotated Multimodal
  Records of Emotionally Colored Conversations Between a Person and a Limited
  Agent}.
\newblock \bibinfo{journal}{{\em {IEEE} Transactions on Affective Computing\/}}
  \bibinfo{volume}{3}, \bibinfo{number}{1} (\bibinfo{date}{Jan}
  \bibinfo{year}{2012}), \bibinfo{pages}{5--17}.
\newblock
\showDOI{%
\url{http://dx.doi.org/10.1109/T-AFFC.2011.20}}


\bibitem[\protect\citeauthoryear{Metallinou, Wollmer, Katsamanis, Eyben,
  Schuller, and Narayanan}{Metallinou et~al\mbox{.}}{2012}]%
        {MetallEtAl2012}
\bibfield{author}{\bibinfo{person}{Angeliki Metallinou},
  \bibinfo{person}{Martin Wollmer}, \bibinfo{person}{Athanasios Katsamanis},
  \bibinfo{person}{Florian Eyben}, \bibinfo{person}{Bjorn Schuller}, {and}
  \bibinfo{person}{Shrikanth Narayanan}.} \bibinfo{year}{2012}\natexlab{}.
\newblock \showarticletitle{Context-Sensitive Learning for Enhanced Audiovisual
  Emotion Classification}.
\newblock \bibinfo{journal}{{\em IEEE Transactions on Affective Computing\/}}
  \bibinfo{volume}{3}, \bibinfo{number}{2} (\bibinfo{date}{April}
  \bibinfo{year}{2012}), \bibinfo{pages}{184--198}.
\newblock
\showDOI{%
\url{http://dx.doi.org/10.1109/T-AFFC.2011.40}}


\bibitem[\protect\citeauthoryear{Metzen, Fabisch, Senger, {de Gea
  Fern{\'{a}}ndez}, and Kirchner}{Metzen et~al\mbox{.}}{2013}]%
        {Metzen2013}
\bibfield{author}{\bibinfo{person}{Jan~Hendrik Metzen},
  \bibinfo{person}{Alexander Fabisch}, \bibinfo{person}{Lisa Senger},
  \bibinfo{person}{Jos{\'{e}} {de Gea Fern{\'{a}}ndez}}, {and}
  \bibinfo{person}{Elsa~Andrea Kirchner}.} \bibinfo{year}{2013}\natexlab{}.
\newblock \showarticletitle{Towards Learning of Generic Skills for Robotic
  Manipulation}.
\newblock \bibinfo{journal}{{\em KI - K{\"{u}}nstliche Intelligenz\/}}
  \bibinfo{volume}{28}, \bibinfo{number}{1} (\bibinfo{date}{Dec}
  \bibinfo{year}{2013}), \bibinfo{pages}{15--20}.
\newblock
\showISSN{0933-1875}
\showDOI{%
\url{http://dx.doi.org/10.1007/s13218-013-0280-1}}


\bibitem[\protect\citeauthoryear{Mitchell}{Mitchell}{2001}]%
        {Mitchell2001}
\bibfield{author}{\bibinfo{person}{Robert~W. Mitchell}.}
  \bibinfo{year}{2001}\natexlab{}.
\newblock \showarticletitle{{Americans' Talk to Dogs: Similarities and
  Differences With Talk to Infants}}.
\newblock \bibinfo{journal}{{\em Research on Language {\&} Social
  Interaction\/}} \bibinfo{volume}{34}, \bibinfo{number}{2}
  (\bibinfo{date}{Apr} \bibinfo{year}{2001}), \bibinfo{pages}{183--210}.
\newblock
\showISSN{0835-1813}
\showDOI{%
\url{http://dx.doi.org/10.1207/S15327973RLSI34-2_2}}


\bibitem[\protect\citeauthoryear{Miura}{Miura}{1994}]%
        {Miura1994}
\bibfield{author}{\bibinfo{person}{Koryo Miura}.}
  \bibinfo{year}{1994}\natexlab{}.
\newblock \showarticletitle{Map fold a la {M}iura style, its physical
  characteristics and application to the space science}.
\newblock \bibinfo{journal}{{\em Research of Pattern Formation\/}}
  (\bibinfo{year}{1994}), \bibinfo{pages}{77--90}.
\newblock


\bibitem[\protect\citeauthoryear{Mower, Metallinou, Lee, Kazemzadeh, Busso,
  Lee, and Narayanan}{Mower et~al\mbox{.}}{2009}]%
        {MowerEtAl2009}
\bibfield{author}{\bibinfo{person}{Emily Mower}, \bibinfo{person}{Angeliki
  Metallinou}, \bibinfo{person}{Chi-Chun Lee}, \bibinfo{person}{Abe
  Kazemzadeh}, \bibinfo{person}{Carlos Busso}, \bibinfo{person}{Sungbok Lee},
  {and} \bibinfo{person}{Shrikanth Narayanan}.}
  \bibinfo{year}{2009}\natexlab{}.
\newblock \showarticletitle{Interpreting ambiguous emotional expressions}. In
  \bibinfo{booktitle}{{\em 2009 3rd International Conference on Affective
  Computing and Intelligent Interaction and Workshops}}. \bibinfo{pages}{1--8}.
\newblock
\showDOI{%
\url{http://dx.doi.org/10.1109/ACII.2009.5349500}}


\bibitem[\protect\citeauthoryear{Murthy, Maji, and Manmatha}{Murthy
  et~al\mbox{.}}{2015}]%
        {Murthy2015}
\bibfield{author}{\bibinfo{person}{Venkatesh~N. Murthy},
  \bibinfo{person}{Subhransu Maji}, {and} \bibinfo{person}{R. Manmatha}.}
  \bibinfo{year}{2015}\natexlab{}.
\newblock \showarticletitle{Automatic Image Annotation Using Deep Learning
  Representations}. In \bibinfo{booktitle}{{\em Proceedings of the 5th ACM
  International Conference on Multimedia Retrieval}} {\em
  (\bibinfo{series}{ICMR '15})}. \bibinfo{publisher}{ACM},
  \bibinfo{address}{New York, NY, USA}, \bibinfo{pages}{603--606}.
\newblock
\showISBNx{978-1-4503-3274-3}
\showDOI{%
\url{http://dx.doi.org/10.1145/2671188.2749391}}


\bibitem[\protect\citeauthoryear{Naab and Russell}{Naab and Russell}{2007}]%
        {NaabRussell2007}
\bibfield{author}{\bibinfo{person}{Pamela~J Naab} {and}
  \bibinfo{person}{James~A Russell}.} \bibinfo{year}{2007}\natexlab{}.
\newblock \showarticletitle{Judgements of Emotion From Spontaneous Facial
  Expressions of {N}ew {G}uineans}.
\newblock \bibinfo{journal}{{\em Emotion\/}} \bibinfo{volume}{7},
  \bibinfo{number}{4} (\bibinfo{year}{2007}), \bibinfo{pages}{736--744}.
\newblock


\bibitem[\protect\citeauthoryear{Ochs and Schieffelin}{Ochs and
  Schieffelin}{1984}]%
        {OchsSchieffelin1984}
\bibfield{author}{\bibinfo{person}{Elinor Ochs} {and} \bibinfo{person}{Bambi~B.
  Schieffelin}.} \bibinfo{year}{1984}\natexlab{}.
\newblock \showarticletitle{Language acquisition and socialization: Three
  developmental stories and their implications}.
\newblock In \bibinfo{booktitle}{{\em Culture Theory: Essays on Mind, Self, and
  Emotion}}, \bibfield{editor}{\bibinfo{person}{Richard~A. Shweder} {and}
  \bibinfo{person}{Robert~A. LeVine}} (Eds.). \bibinfo{publisher}{Cambridge
  University Press}, \bibinfo{address}{Cambridge, UK},
  \bibinfo{pages}{276--320}.
\newblock


\bibitem[\protect\citeauthoryear{Ochs and Schieffelin}{Ochs and
  Schieffelin}{2012}]%
        {OchsSchieffelin2012}
\bibfield{author}{\bibinfo{person}{Elinor Ochs} {and} \bibinfo{person}{Bambi~B.
  Schieffelin}.} \bibinfo{year}{2012}\natexlab{}.
\newblock \showarticletitle{The theory of language socialization}.
\newblock In \bibinfo{booktitle}{{\em The Handbook of Language Socialization}},
  \bibfield{editor}{\bibinfo{person}{Alessandro Duranti},
  \bibinfo{person}{Elinor Ochs}, {and} \bibinfo{person}{Bambi~B. Schieffelin}}
  (Eds.). \bibinfo{publisher}{Wiley-Blackwell}, \bibinfo{address}{Malden, MA},
  \bibinfo{pages}{1--21}.
\newblock


\bibitem[\protect\citeauthoryear{{{\"O}z\c{c}al{\i}\c{s}kan} and
  Goldin-Meadow}{{{\"O}z\c{c}al{\i}\c{s}kan} and Goldin-Meadow}{2005}]%
        {SOGM2005}
\bibfield{author}{\bibinfo{person}{{\c{S}eyda} {{\"O}z\c{c}al{\i}\c{s}kan}}
  {and} \bibinfo{person}{Susan Goldin-Meadow}.}
  \bibinfo{year}{2005}\natexlab{}.
\newblock \showarticletitle{Gesture is at the cutting edge of early language
  development}.
\newblock \bibinfo{journal}{{\em Cognition\/}} \bibinfo{volume}{96},
  \bibinfo{number}{3} (\bibinfo{year}{2005}), \bibinfo{pages}{B101--B113}.
\newblock
\showDOI{%
\url{http://dx.doi.org/https://doi.org/10.1016/j.cognition.2005.01.001}}


\bibitem[\protect\citeauthoryear{Pantic and Rothkrantz}{Pantic and
  Rothkrantz}{2000}]%
        {PanticRoth2000}
\bibfield{author}{\bibinfo{person}{Maja Pantic} {and}
  \bibinfo{person}{Leon~J.M. Rothkrantz}.} \bibinfo{year}{2000}\natexlab{}.
\newblock \showarticletitle{Automatic analysis of facial expressions: The state
  of the art}.
\newblock \bibinfo{journal}{{\em {IEEE} Transactions on Pattern Analysis and
  Machine Intelligence\/}} \bibinfo{volume}{22}, \bibinfo{number}{12}
  (\bibinfo{date}{Dec} \bibinfo{year}{2000}), \bibinfo{pages}{1424--1445}.
\newblock
\showISSN{0162-8828}


\bibitem[\protect\citeauthoryear{Pantic and Rothkrantz}{Pantic and
  Rothkrantz}{2004}]%
        {PanticRoth2004}
\bibfield{author}{\bibinfo{person}{Maja Pantic} {and}
  \bibinfo{person}{Leon~J.M. Rothkrantz}.} \bibinfo{year}{2004}\natexlab{}.
\newblock \showarticletitle{Case-based reasoning for user-profiled recognition
  of emotions from face images}. In \bibinfo{booktitle}{{\em 2004 {IEEE}
  International Conference on Multimedia and Expo {(ICME)}}},
  Vol.~\bibinfo{volume}{1}. \bibinfo{pages}{391--394}.
\newblock
\showDOI{%
\url{http://dx.doi.org/10.1109/ICME.2004.1394211}}


\bibitem[\protect\citeauthoryear{Pantic, Valstar, Rademaker, and Maat}{Pantic
  et~al\mbox{.}}{2005}]%
        {MMI}
\bibfield{author}{\bibinfo{person}{Maja Pantic}, \bibinfo{person}{Michel
  Valstar}, \bibinfo{person}{Ron Rademaker}, {and} \bibinfo{person}{Ludo
  Maat}.} \bibinfo{year}{2005}\natexlab{}.
\newblock \showarticletitle{Web-based database for facial expression analysis}.
  In \bibinfo{booktitle}{{\em 2005 {IEEE} International Conference on
  Multimedia and Expo}}.
\newblock
\showISSN{1945-7871}
\showDOI{%
\url{http://dx.doi.org/10.1109/ICME.2005.1521424}}


\bibitem[\protect\citeauthoryear{Pedregosa, Varoquaux, Gramfort, Michel,
  Thirion, Grisel, Blondel, Prettenhofer, Weiss, Dubourg, Vanderplas, Passos,
  Cournapeau, Brucher, Perrot, and Duchesnay}{Pedregosa et~al\mbox{.}}{2011}]%
        {Pedregosa2011}
\bibfield{author}{\bibinfo{person}{Fabian Pedregosa},
  \bibinfo{person}{Ga{\"{e}}l Varoquaux}, \bibinfo{person}{Alexandre Gramfort},
  \bibinfo{person}{Vincent Michel}, \bibinfo{person}{Bertrand Thirion},
  \bibinfo{person}{Olivier Grisel}, \bibinfo{person}{Mathieu Blondel},
  \bibinfo{person}{Peter Prettenhofer}, \bibinfo{person}{Ron Weiss},
  \bibinfo{person}{Vincent Dubourg}, \bibinfo{person}{Jake Vanderplas},
  \bibinfo{person}{Alexandre Passos}, \bibinfo{person}{David Cournapeau},
  \bibinfo{person}{Matthieu Brucher}, \bibinfo{person}{Matthieu Perrot}, {and}
  \bibinfo{person}{{\'{E}}douard Duchesnay}.} \bibinfo{year}{2011}\natexlab{}.
\newblock \showarticletitle{{Scikit-learn: Machine Learning in Python}}.
\newblock \bibinfo{journal}{{\em Journal of Machine Learning Research\/}}
  \bibinfo{volume}{12} (\bibinfo{date}{Feb} \bibinfo{year}{2011}),
  \bibinfo{pages}{2825--2830}.
\newblock
\showISSN{1532-4435}
\showURL{%
\url{http://dl.acm.org/citation.cfm?id=1953048.2078195}}


\bibitem[\protect\citeauthoryear{Popescu, Spyromitros-Xioufis, Papadopoulos,
  {Le Borgne}, and Kompatsiaris}{Popescu et~al\mbox{.}}{2015}]%
        {Popescu2015}
\bibfield{author}{\bibinfo{person}{Adrian Popescu},
  \bibinfo{person}{Eleftherios Spyromitros-Xioufis}, \bibinfo{person}{Symeon
  Papadopoulos}, \bibinfo{person}{Herv{\'{e}} {Le Borgne}}, {and}
  \bibinfo{person}{Ioannis Kompatsiaris}.} \bibinfo{year}{2015}\natexlab{}.
\newblock \showarticletitle{Toward an Automatic Evaluation of Retrieval
  Performance with Large Scale Image Collections}. In \bibinfo{booktitle}{{\em
  Proceedings of the 2015 Workshop on Community-Organized Multimodal Mining:
  Opportunities for Novel Solutions (MMCommons '15)}}.
  \bibinfo{publisher}{ACM}, \bibinfo{pages}{7--12}.
\newblock
\showDOI{%
\url{http://dx.doi.org/10.1145/2814815.2814819}}


\bibitem[\protect\citeauthoryear{Povey, Ghoshal, Boulianne, Burget, Glembek,
  Goel, Hannemann, Motlicek, Qian, Schwarz, Silovsky, Stemmer, and
  Vesely}{Povey et~al\mbox{.}}{2011}]%
        {Povey_ASRU2011}
\bibfield{author}{\bibinfo{person}{Daniel Povey}, \bibinfo{person}{Arnab
  Ghoshal}, \bibinfo{person}{Gilles Boulianne}, \bibinfo{person}{Lukas Burget},
  \bibinfo{person}{Ondrej Glembek}, \bibinfo{person}{Nagendra Goel},
  \bibinfo{person}{Mirko Hannemann}, \bibinfo{person}{Petr Motlicek},
  \bibinfo{person}{Yanmin Qian}, \bibinfo{person}{Petr Schwarz},
  \bibinfo{person}{Jan Silovsky}, \bibinfo{person}{Georg Stemmer}, {and}
  \bibinfo{person}{Karel Vesely}.} \bibinfo{year}{2011}\natexlab{}.
\newblock \showarticletitle{The {K}aldi speech recognition toolkit}. In
  \bibinfo{booktitle}{{\em Proceedings of the {IEEE} 2011 Workshop on Automatic
  Speech Recognition and Understanding}}. IEEE Signal Processing Society.
\newblock
\newblock
\shownote{IEEE Catalog No.: CFP11SRW-USB.}


\bibitem[\protect\citeauthoryear{Ringeval, Sonderegger, Sauer, and
  Lalanne}{Ringeval et~al\mbox{.}}{2013}]%
        {RECOLA}
\bibfield{author}{\bibinfo{person}{Fabien Ringeval}, \bibinfo{person}{Andreas
  Sonderegger}, \bibinfo{person}{Juergen Sauer}, {and} \bibinfo{person}{Denis
  Lalanne}.} \bibinfo{year}{2013}\natexlab{}.
\newblock \showarticletitle{Introducing the {RECOLA} multimodal corpus of
  remote collaborative and affective interactions}. In \bibinfo{booktitle}{{\em
  2013 10th {IEEE} International Conference and Workshops on Automatic Face and
  Gesture Recognition ({FG})}}. \bibinfo{pages}{1--8}.
\newblock
\showDOI{%
\url{http://dx.doi.org/10.1109/FG.2013.6553805}}


\bibitem[\protect\citeauthoryear{Russakovsky, Deng, Su, Krause, Satheesh, Ma,
  Huang, Karpathy, Khosla, Bernstein, Berg, and Fei-Fei}{Russakovsky
  et~al\mbox{.}}{2014}]%
        {Russakovsky2014}
\bibfield{author}{\bibinfo{person}{Olga Russakovsky}, \bibinfo{person}{Jia
  Deng}, \bibinfo{person}{Hao Su}, \bibinfo{person}{Jonathan Krause},
  \bibinfo{person}{Sanjeev Satheesh}, \bibinfo{person}{Sean Ma},
  \bibinfo{person}{Zhiheng Huang}, \bibinfo{person}{Andrej Karpathy},
  \bibinfo{person}{Aditya Khosla}, \bibinfo{person}{Michael Bernstein},
  \bibinfo{person}{Alexander~C. Berg}, {and} \bibinfo{person}{Li Fei-Fei}.}
  \bibinfo{year}{2014}\natexlab{}.
\newblock \bibinfo{title}{{ImageNet Large Scale Visual Recognition Challenge}}.
\newblock   (\bibinfo{date}{Sep} \bibinfo{year}{2014}).
\newblock
\showeprint[arxiv]{1409.0575}
\showURL{%
\url{http://arxiv.org/abs/1409.0575}}


\bibitem[\protect\citeauthoryear{Russell}{Russell}{1994}]%
        {Russell1994}
\bibfield{author}{\bibinfo{person}{James~A Russell}.}
  \bibinfo{year}{1994}\natexlab{}.
\newblock \showarticletitle{Is there universal recognition of emotion from
  facial expression?}
\newblock \bibinfo{journal}{{\em Psychological Bulletin\/}}
  \bibinfo{volume}{115}, \bibinfo{number}{1} (\bibinfo{year}{1994}),
  \bibinfo{pages}{102--141}.
\newblock


\bibitem[\protect\citeauthoryear{Russell, Bachorowski, and
  Fern\'andez-Dols}{Russell et~al\mbox{.}}{2003}]%
        {RussellEt2003}
\bibfield{author}{\bibinfo{person}{James~A Russell}, \bibinfo{person}{Jo-Anne
  Bachorowski}, {and} \bibinfo{person}{Jos\'e-Miguel Fern\'andez-Dols}.}
  \bibinfo{year}{2003}\natexlab{}.
\newblock \showarticletitle{Facial and vocal expressions of emotion}.
\newblock \bibinfo{journal}{{\em Annual Review of Psychology\/}}
  \bibinfo{volume}{54} (\bibinfo{year}{2003}), \bibinfo{pages}{329--349}.
\newblock


\bibitem[\protect\citeauthoryear{Sariyanidi, Gunes, and Cavallaro}{Sariyanidi
  et~al\mbox{.}}{2015}]%
        {SariyanidiEtAl2015}
\bibfield{author}{\bibinfo{person}{Evangelos Sariyanidi},
  \bibinfo{person}{Hatice Gunes}, {and} \bibinfo{person}{Andrea Cavallaro}.}
  \bibinfo{year}{2015}\natexlab{}.
\newblock \showarticletitle{Automatic Analysis of Facial Affect: A Survey of
  Registration, Representation, and Recognition}.
\newblock \bibinfo{journal}{{\em {IEEE} Transactions on Pattern Analysis and
  Machine Intelligence\/}} \bibinfo{volume}{36}, \bibinfo{number}{6}
  (\bibinfo{date}{June} \bibinfo{year}{2015}), \bibinfo{pages}{1113--1133}.
\newblock
\showISSN{0162-8828}


\bibitem[\protect\citeauthoryear{Sariyanidi, Gunes, {G\"{o}kmen}, and
  Cavallaro}{Sariyanidi et~al\mbox{.}}{2013}]%
        {SariyanidiEtAl2013}
\bibfield{author}{\bibinfo{person}{Evangelos Sariyanidi},
  \bibinfo{person}{Hatice Gunes}, \bibinfo{person}{Muhittin {G\"{o}kmen}},
  {and} \bibinfo{person}{Andrea Cavallaro}.} \bibinfo{year}{2013}\natexlab{}.
\newblock \showarticletitle{Local {Z}ernike Moment Representation for Facial
  Affect Recognition}. In \bibinfo{booktitle}{{\em Proceedings of the {B}ritish
  Machine Vision Conference ({BMVC})}}. \bibinfo{pages}{108.1--108.13}.
\newblock


\bibitem[\protect\citeauthoryear{Schieffelin and Ochs}{Schieffelin and
  Ochs}{1986}]%
        {SchieffelinOchs1986}
\bibfield{author}{\bibinfo{person}{Bambi~B. Schieffelin} {and}
  \bibinfo{person}{Elinor Ochs}.} \bibinfo{year}{1986}\natexlab{}.
\newblock \showarticletitle{Language socialization}.
\newblock \bibinfo{journal}{{\em Annual Review of Anthropology\/}}
  \bibinfo{volume}{15} (\bibinfo{year}{1986}), \bibinfo{pages}{163--191}.
\newblock


\bibitem[\protect\citeauthoryear{Schuller, Seppi, Batliner, Maier, and
  Steidl}{Schuller et~al\mbox{.}}{2007}]%
        {SchullerEtAl2007}
\bibfield{author}{\bibinfo{person}{Bj\"{o}rn Schuller}, \bibinfo{person}{Dino
  Seppi}, \bibinfo{person}{Anton Batliner}, \bibinfo{person}{Andreas Maier},
  {and} \bibinfo{person}{Stefan Steidl}.} \bibinfo{year}{2007}\natexlab{}.
\newblock \showarticletitle{Towards More Reality in the Recognition of
  Emotional Speech}. In \bibinfo{booktitle}{{\em 2007 {IEEE} International
  Conference on Acoustics, Speech and Signal Processing ({ICASSP} '07)}},
  Vol.~\bibinfo{volume}{4}. \bibinfo{pages}{IV.941--944}.
\newblock
\showISSN{1520-6149}
\showDOI{%
\url{http://dx.doi.org/10.1109/ICASSP.2007.367226}}


\bibitem[\protect\citeauthoryear{Schuller, Vlasenko, Eyben, Wollmer, Stuhlsatz,
  Wendemuth, and Rigoll}{Schuller et~al\mbox{.}}{2010}]%
        {SchullerEtAl2010}
\bibfield{author}{\bibinfo{person}{Bj\"{o}rn Schuller}, \bibinfo{person}{Bogdan
  Vlasenko}, \bibinfo{person}{Florian Eyben}, \bibinfo{person}{Martin Wollmer},
  \bibinfo{person}{Andre Stuhlsatz}, \bibinfo{person}{Andreas Wendemuth}, {and}
  \bibinfo{person}{Gerhard Rigoll}.} \bibinfo{year}{2010}\natexlab{}.
\newblock \showarticletitle{Cross-Corpus Acoustic Emotion Recognition:
  Variances and Strategies}.
\newblock \bibinfo{journal}{{\em {IEEE} Transactions on Affective Computing\/}}
  \bibinfo{volume}{1}, \bibinfo{number}{2} (\bibinfo{date}{July}
  \bibinfo{year}{2010}), \bibinfo{pages}{119--131}.
\newblock
\showISSN{1949-3045}
\showDOI{%
\url{http://dx.doi.org/10.1109/T-AFFC.2010.8}}


\bibitem[\protect\citeauthoryear{Senger, Schroer, Metzen, and Kirchner}{Senger
  et~al\mbox{.}}{2014}]%
        {Senger2014}
\bibfield{author}{\bibinfo{person}{Lisa Senger}, \bibinfo{person}{Martin
  Schroer}, \bibinfo{person}{Jan~Hendrik Metzen}, {and}
  \bibinfo{person}{Elsa~Andrea Kirchner}.} \bibinfo{year}{2014}\natexlab{}.
\newblock \showarticletitle{{Velocity-Based Multiple Change-Point Inference for
  Unsupervised Segmentation of Human Movement Behavior}}. In
  \bibinfo{booktitle}{{\em 2014 22nd International Conference on Pattern
  Recognition}}. \bibinfo{publisher}{IEEE}, \bibinfo{pages}{4564--4569}.
\newblock
\showISBNx{978-1-4799-5209-0}
\showDOI{%
\url{http://dx.doi.org/10.1109/ICPR.2014.781}}


\bibitem[\protect\citeauthoryear{Sermanet and LeCun}{Sermanet and
  LeCun}{2011}]%
        {Sermanet2011}
\bibfield{author}{\bibinfo{person}{Pierre Sermanet} {and} \bibinfo{person}{Yann
  LeCun}.} \bibinfo{year}{2011}\natexlab{}.
\newblock \showarticletitle{Traffic sign recognition with multi-scale
  Convolutional Networks}. In \bibinfo{booktitle}{{\em The 2011 International
  Joint Conference on Neural Networks}}. \bibinfo{pages}{2809--2813}.
\newblock
\showISSN{2161-4393}
\showDOI{%
\url{http://dx.doi.org/10.1109/IJCNN.2011.6033589}}


\bibitem[\protect\citeauthoryear{Simonyan and Zisserman}{Simonyan and
  Zisserman}{2015}]%
        {Simonyan2014}
\bibfield{author}{\bibinfo{person}{Karen Simonyan} {and}
  \bibinfo{person}{Andrew Zisserman}.} \bibinfo{year}{2015}\natexlab{}.
\newblock \showarticletitle{Very Deep Convolutional Networks for Large-Scale
  Image Recognition}. In \bibinfo{booktitle}{{\em International Conference on
  Learning Representations (ICLR 2015)}}.
\newblock
\showeprint[arxiv]{1409.1556}
\showURL{%
\url{http://arxiv.org/abs/1409.1556}}


\bibitem[\protect\citeauthoryear{Straube and Krell}{Straube and Krell}{2014}]%
        {Straube2014}
\bibfield{author}{\bibinfo{person}{Sirko Straube} {and}
  \bibinfo{person}{Mario~Michael Krell}.} \bibinfo{year}{2014}\natexlab{}.
\newblock \showarticletitle{{How to evaluate an agent's behaviour to infrequent
  events? -- Reliable performance estimation insensitive to class
  distribution}}.
\newblock \bibinfo{journal}{{\em Frontiers in Computational Neuroscience\/}}
  \bibinfo{volume}{8}, \bibinfo{number}{43} (\bibinfo{date}{Jan}
  \bibinfo{year}{2014}), \bibinfo{pages}{1--6}.
\newblock
\showISSN{1662-5188}
\showDOI{%
\url{http://dx.doi.org/10.3389/fncom.2014.00043}}


\bibitem[\protect\citeauthoryear{Sunderland and Denny}{Sunderland and
  Denny}{2016}]%
        {sunderland2016doing}
\bibfield{author}{\bibinfo{person}{Patricia~L. Sunderland} {and}
  \bibinfo{person}{Rita~M. Denny}.} \bibinfo{year}{2016}\natexlab{}.
\newblock \bibinfo{booktitle}{{\em Doing Anthropology in Consumer Research}}.
\newblock \bibinfo{publisher}{Routledge}.
\newblock


\bibitem[\protect\citeauthoryear{Thomee, Elizalde, Shamma, Ni, Friedland,
  Poland, Borth, and Li}{Thomee et~al\mbox{.}}{2016}]%
        {Thomee2016}
\bibfield{author}{\bibinfo{person}{Bart Thomee}, \bibinfo{person}{Benjamin
  Elizalde}, \bibinfo{person}{David~A. Shamma}, \bibinfo{person}{Karl Ni},
  \bibinfo{person}{Gerald Friedland}, \bibinfo{person}{Douglas Poland},
  \bibinfo{person}{Damian Borth}, {and} \bibinfo{person}{Li-Jia Li}.}
  \bibinfo{year}{2016}\natexlab{}.
\newblock \showarticletitle{{YFCC100M}: The New Data in Multimedia Research}.
\newblock \bibinfo{journal}{{\em {Communications of the ACM}\/}}
  \bibinfo{volume}{59}, \bibinfo{number}{2} (\bibinfo{date}{Jan}
  \bibinfo{year}{2016}), \bibinfo{pages}{64--73}.
\newblock
\showISSN{00010782}
\showDOI{%
\url{http://dx.doi.org/10.1145/2812802}}


\bibitem[\protect\citeauthoryear{Valstar and Pantic}{Valstar and
  Pantic}{2012}]%
        {ValstarPantic2012}
\bibfield{author}{\bibinfo{person}{Michel~F. Valstar} {and}
  \bibinfo{person}{Maja Pantic}.} \bibinfo{year}{2012}\natexlab{}.
\newblock \showarticletitle{Fully Automatic Recognition of the Temporal Phases
  of Facial Actions}.
\newblock \bibinfo{journal}{{\em {IEEE} Transactions on Systems, Man, and
  Cybernetics, Part {B} (Cybernetics)\/}} \bibinfo{volume}{42},
  \bibinfo{number}{1} (\bibinfo{date}{Feb} \bibinfo{year}{2012}),
  \bibinfo{pages}{28--43}.
\newblock
\showDOI{%
\url{http://dx.doi.org/10.1109/TSMCB.2011.2163710}}


\bibitem[\protect\citeauthoryear{Van-Bik}{Van-Bik}{1998}]%
        {VanBik1998}
\bibfield{author}{\bibinfo{person}{Kenneth Van-Bik}.}
  \bibinfo{year}{1998}\natexlab{}.
\newblock \showarticletitle{{L}ai Psycho-collocation}.
\newblock \bibinfo{journal}{{\em Linguistics of the {T}ibeto-{B}urman Area\/}}
  \bibinfo{volume}{21}, \bibinfo{number}{1} (\bibinfo{year}{1998}),
  \bibinfo{pages}{201--233}.
\newblock


\bibitem[\protect\citeauthoryear{Various}{Various}{2017}]%
        {oriwiki}
\bibfield{author}{\bibinfo{person}{Various}.} \bibinfo{year}{2017}\natexlab{}.
\newblock \bibinfo{title}{Origami Database: http://www.oriwiki.com/}.
\newblock   (\bibinfo{year}{2017}).
\newblock


\bibitem[\protect\citeauthoryear{Vidrascu and Devillers}{Vidrascu and
  Devillers}{2008}]%
        {VidrascuDevillers2008}
\bibfield{author}{\bibinfo{person}{Laurence Vidrascu} {and}
  \bibinfo{person}{Laurence Devillers}.} \bibinfo{year}{2008}\natexlab{}.
\newblock \showarticletitle{Anger detection performances based on prosodic and
  acoustic cues in several corpora}. In \bibinfo{booktitle}{{\em Proceedings of
  the Workshop on Corpora for Research on Emotion and Affect at the 2008
  International Conference on Language Resources and Evaluation ({LREC}
  2008)}}. \bibinfo{pages}{13--16}.
\newblock


\bibitem[\protect\citeauthoryear{Vogt and Andr{\'e}}{Vogt and
  Andr{\'e}}{2005}]%
        {VogtAndre2005}
\bibfield{author}{\bibinfo{person}{Thurid Vogt} {and}
  \bibinfo{person}{Elisabeth Andr{\'e}}.} \bibinfo{year}{2005}\natexlab{}.
\newblock \showarticletitle{Comparing Feature Sets for Acted and Spontaneous
  Speech in View of Automatic Emotion Recognition}. In \bibinfo{booktitle}{{\em
  2005 {IEEE} International Conference on Multimedia and Expo ({ICME} '05)}}.
  \bibinfo{pages}{474--477}.
\newblock
\showISSN{1945-7871}
\showDOI{%
\url{http://dx.doi.org/10.1109/ICME.2005.1521463}}


\bibitem[\protect\citeauthoryear{Vogt, Andr\'{e}, and Bee}{Vogt
  et~al\mbox{.}}{2008a}]%
        {EmoVoice}
\bibfield{author}{\bibinfo{person}{Thurid Vogt}, \bibinfo{person}{Elisabeth
  Andr\'{e}}, {and} \bibinfo{person}{Nikolaus Bee}.}
  \bibinfo{year}{2008}\natexlab{a}.
\newblock \showarticletitle{{EmoVoice}: A framework for online recognition of
  emotions from voice}. In \bibinfo{booktitle}{{\em Proceedings of Workshop on
  Perception and Interactive Technologies for Speech-Based Systems}}.
\newblock


\bibitem[\protect\citeauthoryear{Vogt, Andr{\'e}, and Wagner}{Vogt
  et~al\mbox{.}}{2008b}]%
        {VogtEtAl2008}
\bibfield{author}{\bibinfo{person}{Thurid Vogt}, \bibinfo{person}{Elisabeth
  Andr{\'e}}, {and} \bibinfo{person}{Johannes Wagner}.}
  \bibinfo{year}{2008}\natexlab{b}.
\newblock \showarticletitle{Automatic Recognition of Emotions from Speech: A
  Review of the Literature and Recommendations for Practical Realisation}.
\newblock In \bibinfo{booktitle}{{\em Affect and Emotion in Human-Computer
  Interaction: From Theory to Applications}},
  \bibfield{editor}{\bibinfo{person}{Christian Peter} {and}
  \bibinfo{person}{Russell Beale}} (Eds.). \bibinfo{publisher}{Springer},
  \bibinfo{address}{Berlin/Heidelberg}, \bibinfo{pages}{75--91}.
\newblock
\showISBNx{978-3-540-85099-1}
\showDOI{%
\url{http://dx.doi.org/10.1007/978-3-540-85099-1_7}}


\bibitem[\protect\citeauthoryear{Wang, Korayem, Blanco, and Crandall}{Wang
  et~al\mbox{.}}{2016}]%
        {Wang2016}
\bibfield{author}{\bibinfo{person}{Jingya Wang}, \bibinfo{person}{Mohammed
  Korayem}, \bibinfo{person}{Saul Blanco}, {and} \bibinfo{person}{David~J.
  Crandall}.} \bibinfo{year}{2016}\natexlab{}.
\newblock \showarticletitle{Tracking Natural Events through Social Media and
  Computer Vision}. In \bibinfo{booktitle}{{\em Proceedings of the 2016 ACM
  Conference on Multimedia - MM '16}}. \bibinfo{publisher}{ACM Press},
  \bibinfo{address}{New York, New York, USA}, \bibinfo{pages}{1097--1101}.
\newblock
\showISBNx{9781450336031}
\showDOI{%
\url{http://dx.doi.org/10.1145/2964284.2984067}}


\bibitem[\protect\citeauthoryear{Wang, Tang, Li, Li, Wan, Mellina, Hare, and
  Chang}{Wang et~al\mbox{.}}{2017}]%
        {Wang2017}
\bibfield{author}{\bibinfo{person}{Yilin Wang}, \bibinfo{person}{Jiliang Tang},
  \bibinfo{person}{Jundong Li}, \bibinfo{person}{Baoxin Li},
  \bibinfo{person}{Yali Wan}, \bibinfo{person}{Clayton Mellina},
  \bibinfo{person}{Neil~O Hare}, {and} \bibinfo{person}{Yi Chang}.}
  \bibinfo{year}{2017}\natexlab{}.
\newblock \showarticletitle{{Understanding and Discovering Deliberate Self-harm
  Content in Social Media}}. In \bibinfo{booktitle}{{\em International World
  Wide Web Conference (WWW)}}.
\newblock
\showISBNx{9781450349130}


\bibitem[\protect\citeauthoryear{Watson, Crais, Baranek, Dykstra, and
  Wilson}{Watson et~al\mbox{.}}{2013}]%
        {watson2013communicative}
\bibfield{author}{\bibinfo{person}{Linda~R. Watson},
  \bibinfo{person}{Elizabeth~R. Crais}, \bibinfo{person}{Grace~T. Baranek},
  \bibinfo{person}{Jessica~R. Dykstra}, {and} \bibinfo{person}{Kaitlyn~P.
  Wilson}.} \bibinfo{year}{2013}\natexlab{}.
\newblock \showarticletitle{Communicative gesture use in infants with and
  without autism: A retrospective home video study}.
\newblock \bibinfo{journal}{{\em American Journal of Speech-Language
  Pathology\/}} \bibinfo{volume}{22}, \bibinfo{number}{1}
  (\bibinfo{year}{2013}), \bibinfo{pages}{25--39}.
\newblock


\bibitem[\protect\citeauthoryear{Wei, Ramakrishna, Kanade, and Sheikh}{Wei
  et~al\mbox{.}}{2016}]%
        {Wei2016}
\bibfield{author}{\bibinfo{person}{Shih-En Wei}, \bibinfo{person}{Varun
  Ramakrishna}, \bibinfo{person}{Takeo Kanade}, {and} \bibinfo{person}{Yaser
  Sheikh}.} \bibinfo{year}{2016}\natexlab{}.
\newblock \showarticletitle{{Convolutional Pose Machines}}. In
  \bibinfo{booktitle}{{\em 2016 IEEE Conference on Computer Vision and Pattern
  Recognition (CVPR)}}.
\newblock
\showeprint[arxiv]{1602.00134}
\showURL{%
\url{http://arxiv.org/abs/1602.00134}}


\bibitem[\protect\citeauthoryear{Wiseman and Bondarenko}{Wiseman and
  Bondarenko}{}]%
        {WebRTC-VAD}
\bibfield{author}{\bibinfo{person}{John Wiseman} {and}
  \bibinfo{person}{Ivan~Yu. Bondarenko}.}
\newblock \bibinfo{title}{{P}ython interface to the {WebRTC} Voice Activity
  Detector}.
\newblock   (\bibinfo{year}{????}).
\newblock
\newblock
\shownote{\url{https://github.com/wiseman/py-webrtcvad}.}


\bibitem[\protect\citeauthoryear{Xia, Lu, Yang, Ma, Yao, and Zheng}{Xia
  et~al\mbox{.}}{2015}]%
        {Xia2015}
\bibfield{author}{\bibinfo{person}{Min Xia}, \bibinfo{person}{Weitao Lu},
  \bibinfo{person}{Jun Yang}, \bibinfo{person}{Ying Ma}, \bibinfo{person}{Wen
  Yao}, {and} \bibinfo{person}{Zichen Zheng}.} \bibinfo{year}{2015}\natexlab{}.
\newblock \showarticletitle{{A hybrid method based on extreme learning machine
  and k-nearest neighbor for cloud classification of ground-based visible cloud
  image}}.
\newblock \bibinfo{journal}{{\em Neurocomputing\/}}  \bibinfo{volume}{160}
  (\bibinfo{date}{Jul} \bibinfo{year}{2015}), \bibinfo{pages}{238--249}.
\newblock
\showISSN{09252312}
\showDOI{%
\url{http://dx.doi.org/10.1016/j.neucom.2015.02.022}}


\bibitem[\protect\citeauthoryear{Xu, Burger, Hauptmann, Li, Chang, Yu, Du, Li,
  Jiang, Mao, and Lan}{Xu et~al\mbox{.}}{2015}]%
        {Xu2015}
\bibfield{author}{\bibinfo{person}{Shicheng Xu}, \bibinfo{person}{Susanne
  Burger}, \bibinfo{person}{Alexander Hauptmann}, \bibinfo{person}{Huan Li},
  \bibinfo{person}{Xiaojun Chang}, \bibinfo{person}{Shoou-I Yu},
  \bibinfo{person}{Xingzhong Du}, \bibinfo{person}{Xuanchong Li},
  \bibinfo{person}{Lu Jiang}, \bibinfo{person}{Zexi Mao}, {and}
  \bibinfo{person}{Zhenzhong Lan}.} \bibinfo{year}{2015}\natexlab{}.
\newblock \showarticletitle{{Incremental Multimodal Query Construction for
  Video Search}}. In \bibinfo{booktitle}{{\em Proceedings of the 5th ACM on
  International Conference on Multimedia Retrieval - ICMR '15}}.
  \bibinfo{publisher}{ACM Press}, \bibinfo{address}{New York, New York, USA},
  \bibinfo{pages}{675--678}.
\newblock
\showISBNx{9781450332743}
\showDOI{%
\url{http://dx.doi.org/10.1145/2671188.2749413}}


\bibitem[\protect\citeauthoryear{Yang, Li, Ferm{\"{u}}ller, and Aloimonos}{Yang
  et~al\mbox{.}}{2015}]%
        {Yang2015}
\bibfield{author}{\bibinfo{person}{Yezhou Yang}, \bibinfo{person}{Yi Li},
  \bibinfo{person}{Cornelia Ferm{\"{u}}ller}, {and} \bibinfo{person}{Yiannis
  Aloimonos}.} \bibinfo{year}{2015}\natexlab{}.
\newblock \showarticletitle{Robot Learning Manipulation Action Plans by
  `Watching' Unconstrained Videos from the {W}orld {W}ide {W}eb}. In
  \bibinfo{booktitle}{{\em Twenty-Ninth AAAI Conference on Artificial
  Intelligence (AAAI-15)}}.
\newblock
\showISBNx{9781577357032}


\bibitem[\protect\citeauthoryear{Yasin, Iqbal, Kruger, Weber, and Gall}{Yasin
  et~al\mbox{.}}{2016}]%
        {Yasin2016}
\bibfield{author}{\bibinfo{person}{Hashim Yasin}, \bibinfo{person}{Umar Iqbal},
  \bibinfo{person}{Bjorn Kruger}, \bibinfo{person}{Andreas Weber}, {and}
  \bibinfo{person}{Juergen Gall}.} \bibinfo{year}{2016}\natexlab{}.
\newblock \showarticletitle{A Dual-Source Approach for {3D} Pose Estimation
  from a Single Image}. In \bibinfo{booktitle}{{\em 2016 IEEE Conference on
  Computer Vision and Pattern Recognition (CVPR)}}. \bibinfo{publisher}{IEEE},
  \bibinfo{pages}{4948--4956}.
\newblock
\showISBNx{978-1-4673-8851-1}
\showISSN{10636919}
\showDOI{%
\url{http://dx.doi.org/10.1109/CVPR.2016.535}}
\showeprint[arxiv]{1509.06720}


\bibitem[\protect\citeauthoryear{Zeng, Pantic, Roisman, and Huang}{Zeng
  et~al\mbox{.}}{2009}]%
        {ZengEtAl2009}
\bibfield{author}{\bibinfo{person}{Zhihong Zeng}, \bibinfo{person}{Maja
  Pantic}, \bibinfo{person}{Glenn~I. Roisman}, {and} \bibinfo{person}{Thomas~S.
  Huang}.} \bibinfo{year}{2009}\natexlab{}.
\newblock \showarticletitle{A Survey of Affect Recognition Methods: Audio,
  Visual, and Spontaneous Expressions}.
\newblock \bibinfo{journal}{{\em {IEEE} Transactions on Pattern Analysis and
  Machine Intelligence\/}} \bibinfo{volume}{31}, \bibinfo{number}{1}
  (\bibinfo{date}{Jan} \bibinfo{year}{2009}), \bibinfo{pages}{39--58}.
\newblock
\showISSN{0162-8828}
\showDOI{%
\url{http://dx.doi.org/10.1109/TPAMI.2008.52}}


\bibitem[\protect\citeauthoryear{Zhang, Yan, Li, Rui, Liu, and Bie}{Zhang
  et~al\mbox{.}}{2016}]%
        {Zhang2016}
\bibfield{author}{\bibinfo{person}{Chao Zhang}, \bibinfo{person}{Junchi Yan},
  \bibinfo{person}{Changsheng Li}, \bibinfo{person}{Xiaoguang Rui},
  \bibinfo{person}{Liang Liu}, {and} \bibinfo{person}{Rongfang Bie}.}
  \bibinfo{year}{2016}\natexlab{}.
\newblock \showarticletitle{On Estimating Air Pollution from Photos Using
  Convolutional Neural Network}. In \bibinfo{booktitle}{{\em Proceedings of the
  2016 ACM Conference on Multimedia (MM'16)}}. ACM, \bibinfo{pages}{297--301}.
\newblock


\bibitem[\protect\citeauthoryear{Zhang, Korayem, Crandall, and LeBuhn}{Zhang
  et~al\mbox{.}}{2012}]%
        {Zhang2012}
\bibfield{author}{\bibinfo{person}{Haipeng Zhang}, \bibinfo{person}{Mohammed
  Korayem}, \bibinfo{person}{David~J. Crandall}, {and}
  \bibinfo{person}{Gretchen LeBuhn}.} \bibinfo{year}{2012}\natexlab{}.
\newblock \showarticletitle{{Mining photo-sharing websites to study ecological
  phenomena}}. In \bibinfo{booktitle}{{\em Proceedings of the 21st
  International Conference on World Wide Web (WWW '12)}}.
  \bibinfo{publisher}{ACM Press}, \bibinfo{address}{New York, New York, USA},
  \bibinfo{pages}{749}.
\newblock
\showISBNx{9781450312295}
\showDOI{%
\url{http://dx.doi.org/10.1145/2187836.2187938}}


\bibitem[\protect\citeauthoryear{Zhou, Zhu, Leonardos, Derpanis, and
  Daniilidis}{Zhou et~al\mbox{.}}{2016}]%
        {Zhou_2016_CVPR}
\bibfield{author}{\bibinfo{person}{Xiaowei Zhou}, \bibinfo{person}{Menglong
  Zhu}, \bibinfo{person}{Spyridon Leonardos}, \bibinfo{person}{Konstantinos~G.
  Derpanis}, {and} \bibinfo{person}{Kostas Daniilidis}.}
  \bibinfo{year}{2016}\natexlab{}.
\newblock \showarticletitle{Sparseness Meets Deepness: {3D} Human Pose
  Estimation From Monocular Video}. In \bibinfo{booktitle}{{\em The {IEEE}
  Conference on Computer Vision and Pattern Recognition ({CVPR})}}.
\newblock


\end{thebibliography}
}

\end{document}